\documentclass[10pt,a4paper]{book}

\PassOptionsToPackage{table,svgnames,dvipsnames}{xcolor}

\usepackage[a4paper, portrait, margin=1.1811in]{geometry}
\usepackage[english]{babel}
\usepackage[utf8]{inputenc}
\usepackage[T1]{fontenc}
\usepackage{lmodern}%{helvet}
\usepackage{etoolbox}
\usepackage{graphicx}
\usepackage{titlesec}
\usepackage{caption}
\usepackage{booktabs}
\usepackage{xcolor} 
\usepackage[colorlinks, citecolor=cyan]{hyperref}
\usepackage{caption}
\graphicspath{ {./images/} }
\usepackage{scrextend}
\usepackage{fancyhdr}
\usepackage{graphicx}
\newcounter{lemma}

\newcounter{theorem}

\hypersetup{
    linkcolor=blue
}

\makeatletter
\patchcmd{\@maketitle}{\LARGE \@title}{\fontsize{16}{25}\selectfont\@title}{}{}
\makeatother

\usepackage{authblk}

\titlespacing\section{0pt}{12pt plus 4pt minus 2pt}{6pt}
\titlespacing\subsection{0pt}{12pt plus 4pt minus 2pt}{4pt}
\titlespacing\subsubsection{0pt}{12pt plus 4pt minus 2pt}{4pt}
\titlespacing\paragraph{12pt}{12pt plus 4pt minus 2pt}{4pt}

\usepackage{enumitem}
\setlist[itemize]{topsep=0pt} % ADDED TO REMOVE ADDITIONAL WHITESPACE BEFORE ITEMIZE

\titleformat{\section}{\normalfont\fontsize{14}{17}\bfseries}{\thesection.}{1em}{}
\titleformat{\subsection}{\normalfont\fontsize{12}{15}\bfseries}{\thesubsection.}{1em}{}
\titleformat{\subsubsection}{\normalfont\fontsize{10}{15}\bfseries}{\thesubsubsection.}{1em}{}

\titleformat{\author}{\normalfont\fontsize{10}{15}\bfseries}{\thesection}{1em}{}

\usepackage{parskip}
\newlength\paragraphmargin
\usepackage{caption}  %ADDED FOR SUBFIGURES
\usepackage{subcaption} %ADDED FOR SUBFIGURES
\usepackage{floatrow} % ADDED FOR ROWS OF SUBFIGURES
\usepackage{csquotes} % Added for quotations with indent formatting

\usepackage{amsmath,bm}
\usepackage{amsfonts}
\usepackage{fancyhdr}
\usepackage{amssymb}

\usepackage{upgreek} % ADDED FOR NON-ITALIC GREEK LETTERS
\usepackage[ nocomma ]{optidef} % ADDED FOR BETTER OPTIMIZATION PROBLEM FORMATTING 
\usepackage[mathscr]{euscript} % ADDED FOR SCRIPT SET LETTERS
\usepackage{wrapfig} % ADDED FOR FIGURE WRAPPING

\usepackage{natbib}
\newcommand{\citeay}[1]{(\citeauthor{#1}, \citeyear{#1})}
\usepackage{blindtext}

\usepackage{enumitem} % ADDED FOR LIST WITH MULTIPLE LEVELS
\usepackage{wrapfig}

\usepackage{longtable}
\usepackage{array} % ADDED TO HAVE A CENTERED P COLUMN (specified width and centered text)

\usepackage{acro} % PACKAGE USED FOR ABBREVIATIONS

\newlist{inparaenum}{enumerate}{2}% allow two levels of nesting in an enumerate-like environment
\setlist[inparaenum]{nosep}% compact spacing for all nesting levels
\setlist[inparaenum,1]{label=\bfseries\arabic*.}% labels for top level
\setlist[inparaenum,2]{label=\emph{\alph*})}% labels for second level

\definecolor{Prune}{RGB}{99,0,60} % l14-33 : couleurs de la charte graphique upsaclay
\definecolor{B1}{RGB}{49,62,72} 
\definecolor{C1}{RGB}{124,135,143}
\definecolor{D1}{RGB}{213,218,223}
\definecolor{A2}{RGB}{198,11,70}
\definecolor{B2}{RGB}{237,20,91}
\definecolor{C2}{RGB}{238,52,35}
\definecolor{D2}{RGB}{243,115,32}
\definecolor{A3}{RGB}{124,42,144}
\definecolor{B3}{RGB}{125,106,175}
\definecolor{C3}{RGB}{198,103,29}
\definecolor{D3}{RGB}{254,188,24}
\definecolor{A4}{RGB}{0,78,125}
\definecolor{B4}{RGB}{14,135,201}
\definecolor{C4}{RGB}{0,148,181}
\definecolor{D4}{RGB}{70,195,210}
\definecolor{A5}{RGB}{0,128,122}
\definecolor{B5}{RGB}{64,183,105}
\definecolor{C5}{RGB}{140,198,62}
\definecolor{D5}{RGB}{213,223,61}
\definecolor{SkyBlue}{RGB}{135,206,250}
\definecolor{MedAqua}{RGB}{102,205,170}
\definecolor{SteelBlue}{RGB}{70,130,180}
\definecolor{MyDarkGreen}{RGB}{0,153,0}
\definecolor{MyGreen}{RGB}{187,255,153}
\definecolor{Grey}{RGB}{192,192,192}
\definecolor{LightGrey}{RGB}{200,200,200}

\usepackage{graphicx}
\usepackage{pgfplots}
\usepackage{pgfplotstable}
\usepackage[final]{microtype}
\usepackage{eurosym}
\providecommand{\abs}[1]{\lvert#1\rvert}
\usepackage{bm}

\numberwithin{equation}{section}

\usepackage{algorithm}
\usepackage{algpseudocode}

\usepackage[capitalise]{cleveref}

\pgfplotsset{compat=1.18}

\newsavebox\affbox
\author[]{\textbf{Emily Little}}
\author[]{\textbf{Florent Cogen}}
\author[]{\textbf{Quentin Bustarret}}
\author[]{\textbf{Virginie Dussartre}}
\author[]{\textbf{Maxime Lâasri}}
\author[]{\textbf{Gabriel Kasmi}}
\author[]{\textbf{Marie Girod}}
\author[]{\textbf{Frederic Bienvenu}}
\author[]{\textbf{Maxime Fortin}}
\author[]{\textbf{Jean-Yves Bourmaud}}
\affil[]{\textbf{RTE} \\
7C Place du Dôme \\
La Défense \\France
}
\title{\textbf{\huge ATLAS: A Model of Short-term European Electricity Market Processes under Uncertainty}}
\date{\today}

\begin{document}

\pagestyle{headings}	
\newpage
\setcounter{page}{1}
\renewcommand{\thepage}{\arabic{page}}

\newgeometry{left=2cm,bottom=2cm, top=2cm, right=2cm}

\maketitle
	
%\noindent\rule{15cm}{0.5pt}
%	\begin{abstract}
%		\textbf{Abstract }consists of objectives, methods, findings, and research contributions in 150 to 250 words which contains the main conclusions and provides important information and is accompanied by \textbf{5 keywords}. Furthermore, the determination of keywords needs to pay attention to important words contained in the title and abstract, separated by a semicolon. \textbf{The novelty} in this paper briefly explains why no one else has adequately researched the question. Then \textbf{the results} are made a list of the empirical findings and write the discussion in one or two sentences. \\ \\
%		\let\thefootnote\relax\footnotetext{
%			\small $^{*}$\textbf{Corresponding author.} \textit{
%				\textit{E-mail address: \color{cyan}author4@email.com}}\\
%			\color{black} Received: xx xxxxx 20xx,\quad
%			Accepted: xx xxxxx 20xx and available online XX July 2022 \\
%			\color{cyan} https://doi.org/10.1016/j.compeleceng.2021.107553
%			
%		}
%		\textbf{\textit{Keywords}}: \textit{Keyword 1; keyword 2; keyword 3; keyword 4; keyword 5}
%	\end{abstract}
%\noindent\rule{15cm}{0.4pt}

\tableofcontents

\chapter{Introduction and Context} \label{ch:Introduction}

\begin{figure}[ht!]
    \centering
    \includegraphics[width=0.7\textwidth]{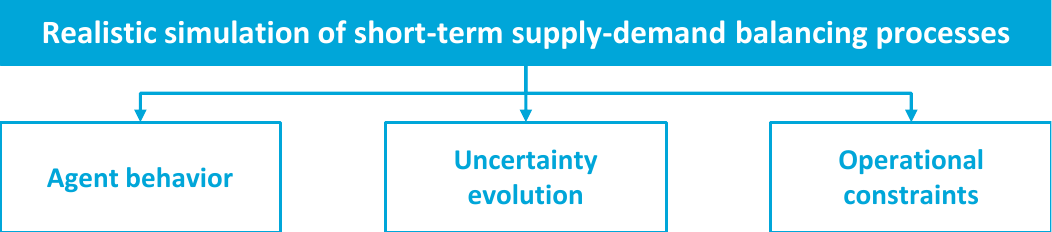}
    \caption{ATLAS Concepts}
    \label{fig:atlasConcepts}
\end{figure}

This document presents the ATLAS model, a model developed by RTE to simulate the electricity supply and demand system close to real-time. It was originally developed during the Optimate project [\cite{weber_optimate_2012}], followed by a phase of internal industrialization by RTE. The model was developed to address the impacts of specific concepts in complement with a large central planning optimization model (e.g. Antares\footnote{https://antares-simulator.org/}). It aims to realistically simulate the short-term processes (between day ahead and real-time) of the electricity system through three specific axes, as shown in Figure \ref{fig:atlasConcepts}. 

The model is mainly used to:  
\begin{itemize}
    \item simulate the impacts of various design choices in short-term electricity markets 
    \item understand the consequences of forecast error and strategic behavior on different market phases (for instance in terms of scarcity, market prices, revenue and risk etc.) 
\end{itemize}

The document focuses specifically on the day-ahead to intraday timeframe and does not include the different balancing processes. These are explained in detail in a second document [\cite{cogen_atlas_2024}]. The rest of this document is organized in the following way. This chapter presents the model environment and basic concepts, including a complete list of nomenclature. Chapter \ref{ch:ForecastModels} presents the different forecast models used currently within the ATLAS model. Note that as the model is quite modular, these are subject to change depending on the analysis required. Chapter \ref{ch:OptimalDispatch} shows the optimal dispatch problem that represents the core of several ATLAS modules. The different modules used to create the day ahead and intraday market orders are described in Chapter \ref{ch:OrderCreation}. The Market Clearing process is explicited in Chapter \ref{ch:MarketClearing}. A short summary of several studies previously performed using the ATLAS model, with links to the original works, are shown in Chapter \ref{ch:ExampleStudies}. Finally, Chapter \ref{ch:Conclusion} presents the future work of the model. 

\section{The ATLAS Environment}

% \begin{figure}[ht!]
%     \centering
%     \includegraphics[width=\textwidth]{images/ATLASmodules.pdf}
%     \caption{ATLAS Modules}
%     \label{fig:atlasModules}
% \end{figure}

\begin{figure}[ht!]
    \centering
    \includegraphics[width=0.8\textwidth]{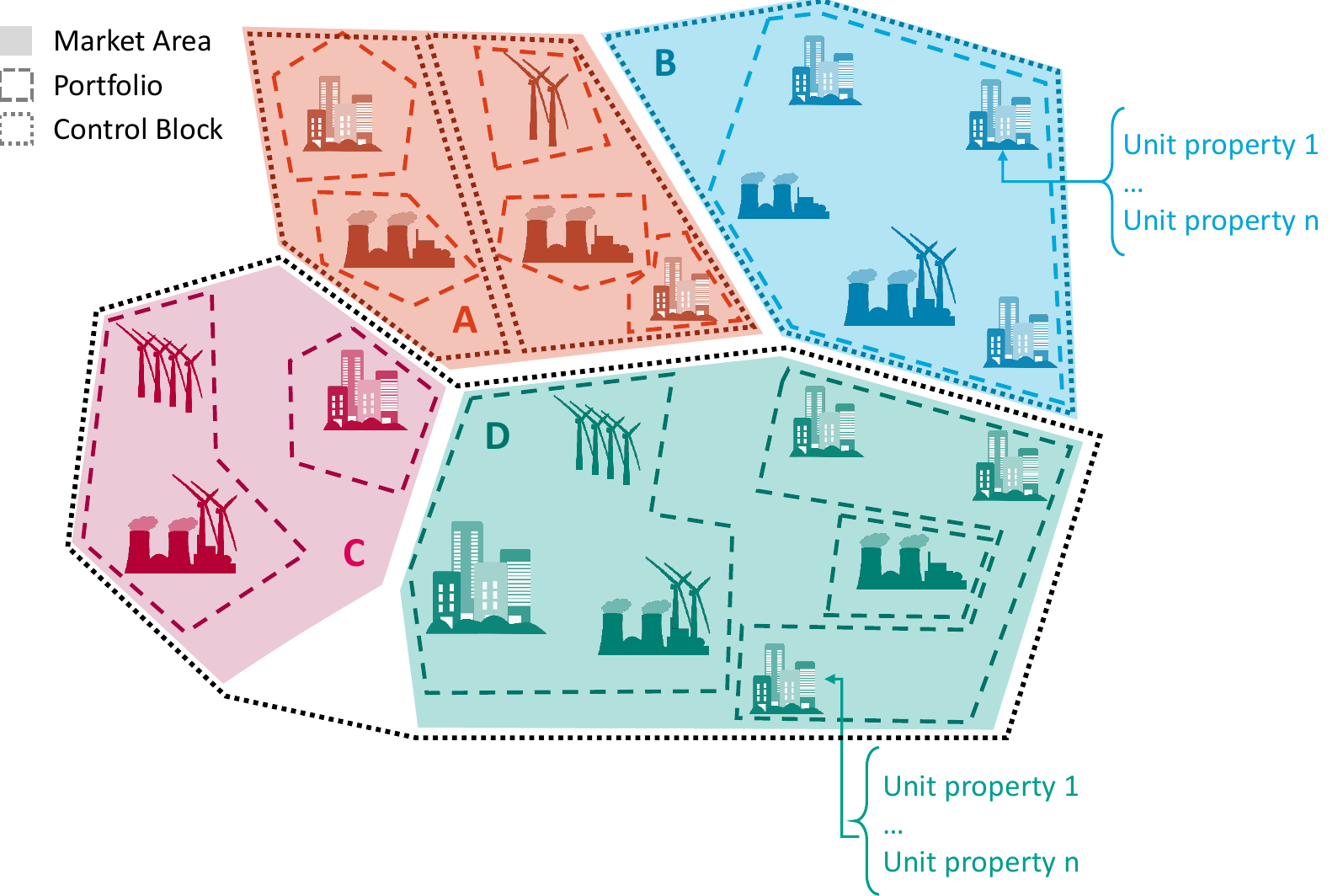}
    \caption{Representation of market actors in the ATLAS model}
    \label{fig:marketActors}
\end{figure}

The ATLAS model consists of several modules coded in python that can be used in a variety of combinations in order to model the chain of electricity markets from day ahead to close-to-real time. These modules consist of a variety of optimization problems or heuristic models to represent the actions of different agents in the electricity market environment. Each module is a set of codes developed in python that rely on the same data structure in order to ensure their ability to remain modular. The full data structure with all properties, classes and links is shown in Figure \ref{fig:ATLASmodel}. The market actors are represented at different levels (unit, portfolio, market area and control block), as shown in Figure \ref{fig:marketActors}. Each unit is represented by a specific class comprising all relevant technical constraints. The current version of ATLAS supports seven different unit types: Thermic, Hydraulic, Storage, Load, Photovoltaic, Wind, and other Non-dispatchable production (this includes for instance, biomass, biogas, hydraulic run-of-river, geothermal and waste). Using these unit types, it is also of course possible to model certain specificities where necessary, and most studies include various representations of each type with unique parameters. For instance, electric vehicles can be modeled as load or storage, depending on the desired vehicle-to-grid behavior. Each unit is part of a \textbf{portfolio}, which in turn is located in a specific \textbf{market area} and controlled by a \textbf{control area}, representing the transmission system operator.  

Using these delimitations, it is possible to model a variety of combinations of shared information. We can model the impacts of different ownership paradigms. For instance, in Figure \ref{fig:marketActors}, Market Area A has several market players, each one owning a single generation unit or demand center, whereas Market Area B has a single monopoly that possesses all units within in the market area. In Market Area C, the portfolio is split into a single generation company and a single demand supplier, while Market Area D has several heterogeneous portfolios. The ATLAS model also allows for different architectures of system operators. A single market area can be represented by multiple system operators, as in Germany and shown in Market Area A in Figure \ref{fig:marketActors}. A system operator can also operate multiple market areas, as in Norway, Sweden and Italy and shown across Market Areas C and D in our reference case. 

\begin{figure}[!ht]
    \centering
    \begin{subfigure}[b]{0.44\textwidth}
        \includegraphics[width=\textwidth]{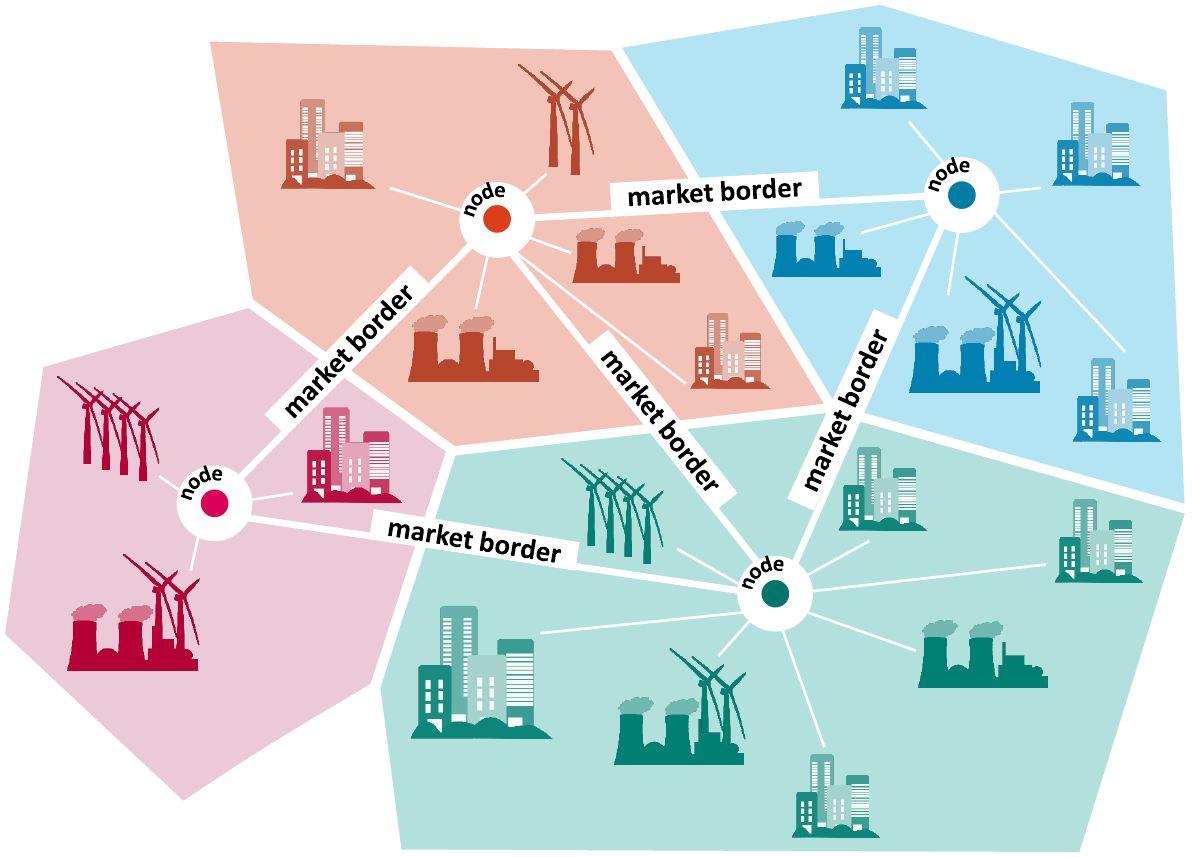}
        \caption{Simple ATC Representation}
        \label{fig:subplot_atlasATC}
    \end{subfigure}
    \begin{subfigure}[b]{0.49\textwidth}
        \centering
        \includegraphics[width=\textwidth]{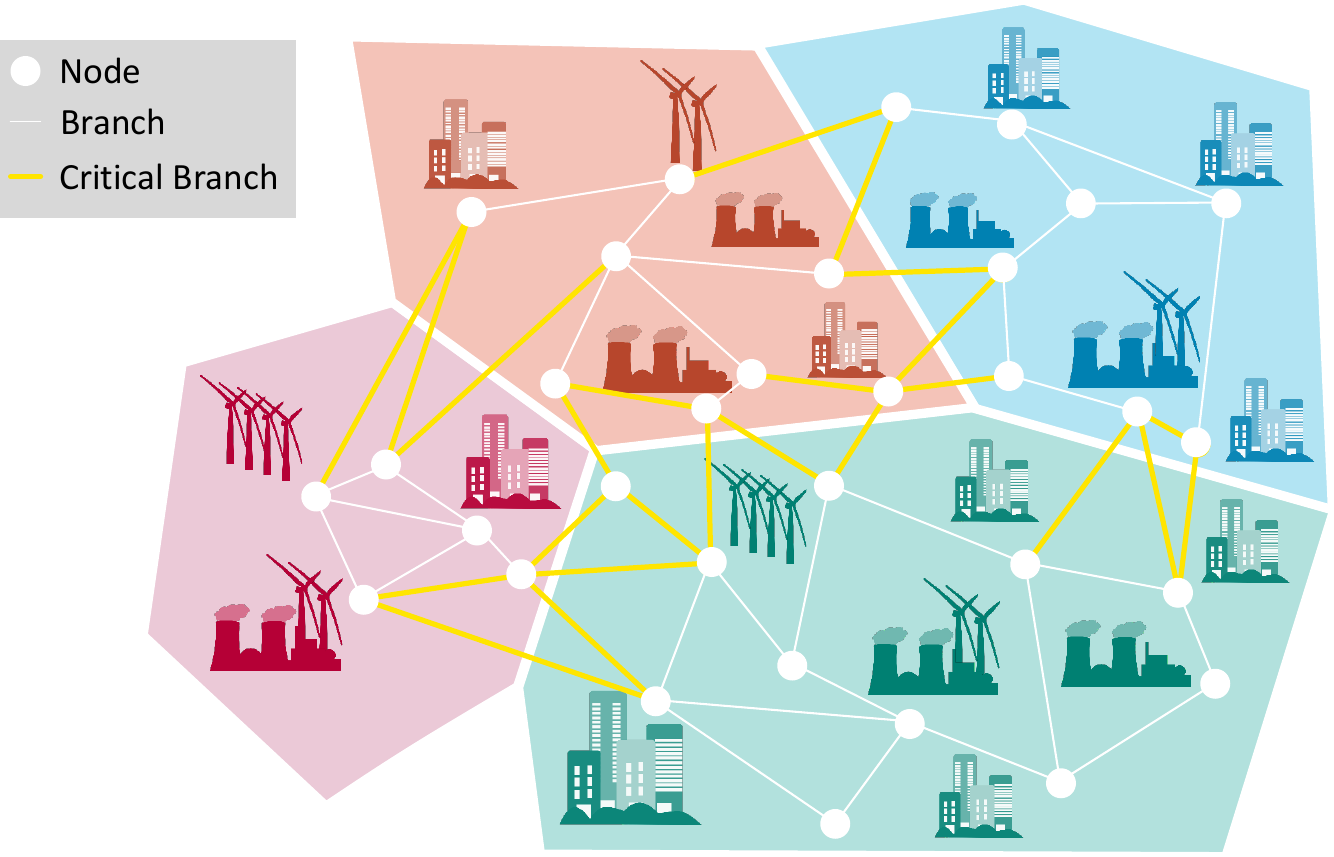}
        \caption{Flow-based Representation}
        \label{fig:subplot_atlasFB}
    \end{subfigure}
    \caption{ATLAS Network Constraint Representations}
    \label{fig:atlasNetwork}
\end{figure}

The ATLAS model also allows for different network representations, shown in Figures \ref{fig:subplot_atlasATC} and \ref{fig:subplot_atlasFB}. It is possible to represent each zone by a single node and connected by a market border. In this case, each unit is connected to this main node, as shown in Figure \ref{fig:subplot_atlasATC}. Figure \ref{fig:subplot_atlasFB} shows the flow-based representation. In general, the model is used for zonal studies, so only the zonal power transfer distribution factor (PTDF) matrix are input. However, nothing constrains the model from being used in nodal studies as well (and in fact it was used in a nodal model during the Osmose project - https://www.osmose-h2020.eu/). 

\begin{figure}[ht!]
    \centering
    \includegraphics[width=0.7\textwidth]{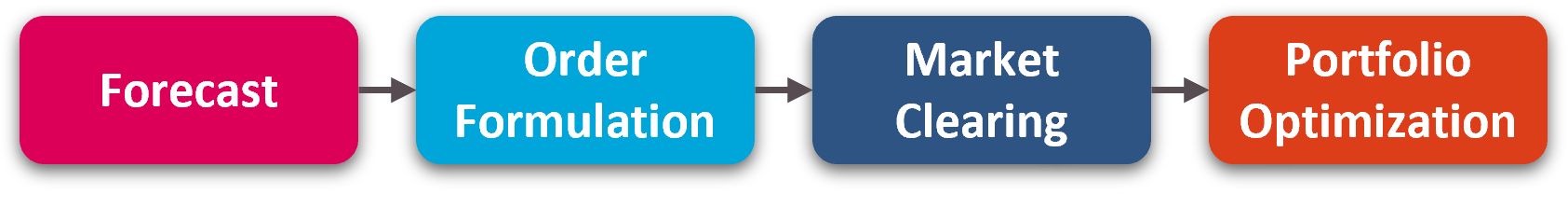}
    \caption{ATLAS Basic Chain of Modules}
    \label{fig:ATLASminichain}
\end{figure}

Each market modeled by ATLAS is generally represented by a chain of four modules, shown in Figure \ref{fig:ATLASminichain}. There are additionally several ghost modules not depicted that will be explained in this work. The inputs to each module depend on several factors, including the outcomes of previous modules, the execution date of the module and the actor involved. The information shared between each actor depends on "published values": market clearing price, accepted offer status, etc. They can also have their own private information that is not shared to the other actors. 

First, a forecast module generates a forecast of renewable production and load as well as a forecast of the market clearing price that the units will use to generate their orders. The order generation modules are separate optimization problems or heuristics resolved from the perspective of each unit. The Market Clearing then takes the orders created in these steps and solves a problem similar to EUPHEMIA\footnote{See Appendix \ref{app_EUPHEMIA} for a brief overview of the EUPHEMIA algorithm.} (the current day ahead market clearing algorithm in Europe) and LIBRA (the RR/mFRR balancing reserve market clearing algorithm) to associate purchase and sale offers. The information passed to the clearing does not include any of the equipment technical constraints, other than that which can be translated into a standard order book. The Portfolio Optimization phase can be run either from a unit-based or portfolio-based perspective. This module takes in the quantities accepted during the prior clearing and assesses whether they are technically feasible for each of the units within its scope. The asset owner can also attempt to find a program with a reduced overall cost for all its units. This will be discussed in more detail in Section \ref{subsec:PortfolioOptim}, but this optimization includes an arbitrage between the cost of producing the accepted quantities and a forecasted imbalance price that would be charged ex-post. 

\begin{figure}[ht!]
    \centering
    \includegraphics[width=\textwidth]{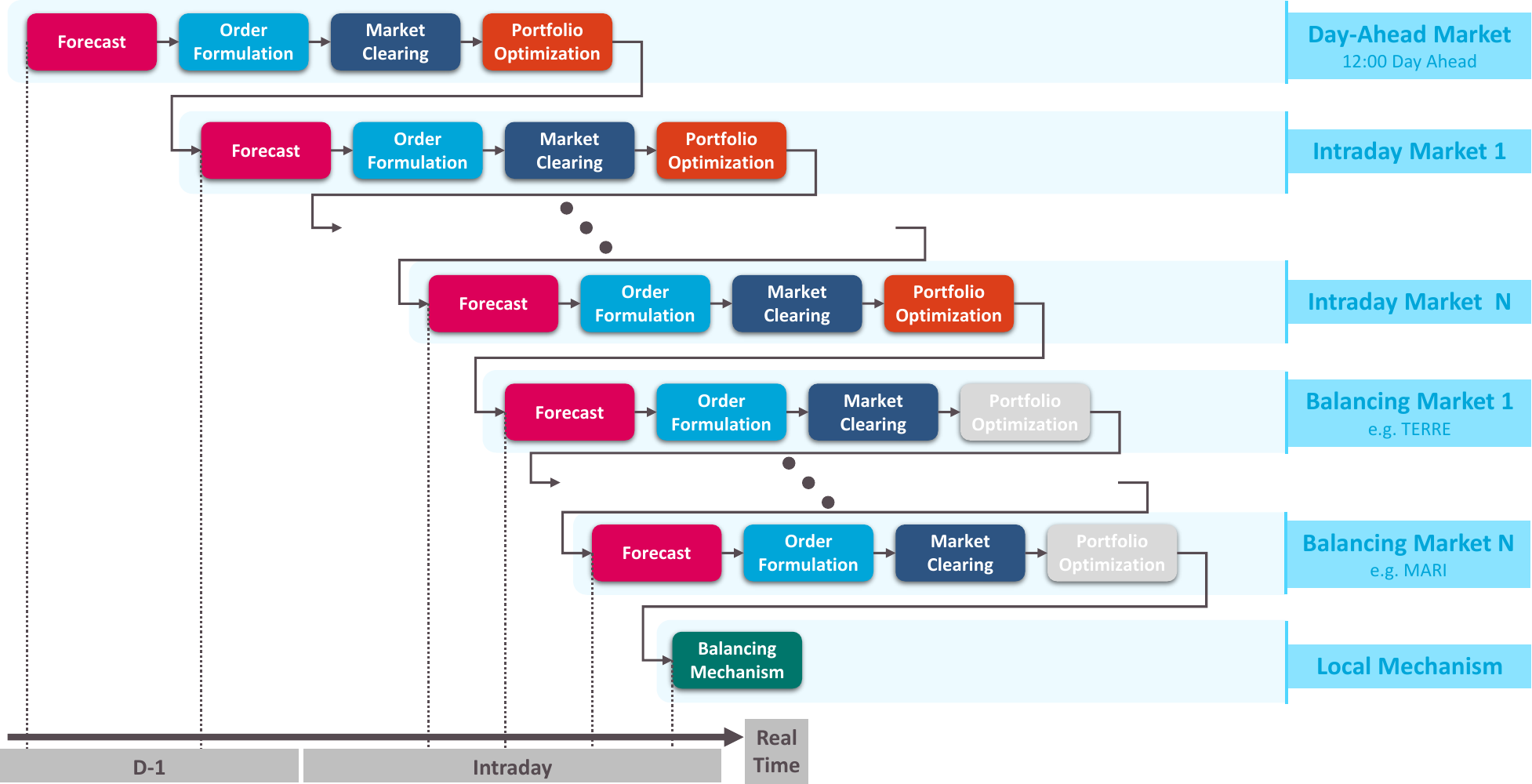}
    \caption{ATLAS Example Chain of Modules}
    \label{fig:ATLASchain}
\end{figure}

Figure \ref{fig:ATLASchain} shows how these modules can be chained together to model multiple processes from day ahead to real time. 

\section{Time Frames}
\begin{figure}[ht!]
    \centering
    \includegraphics[width=\textwidth]{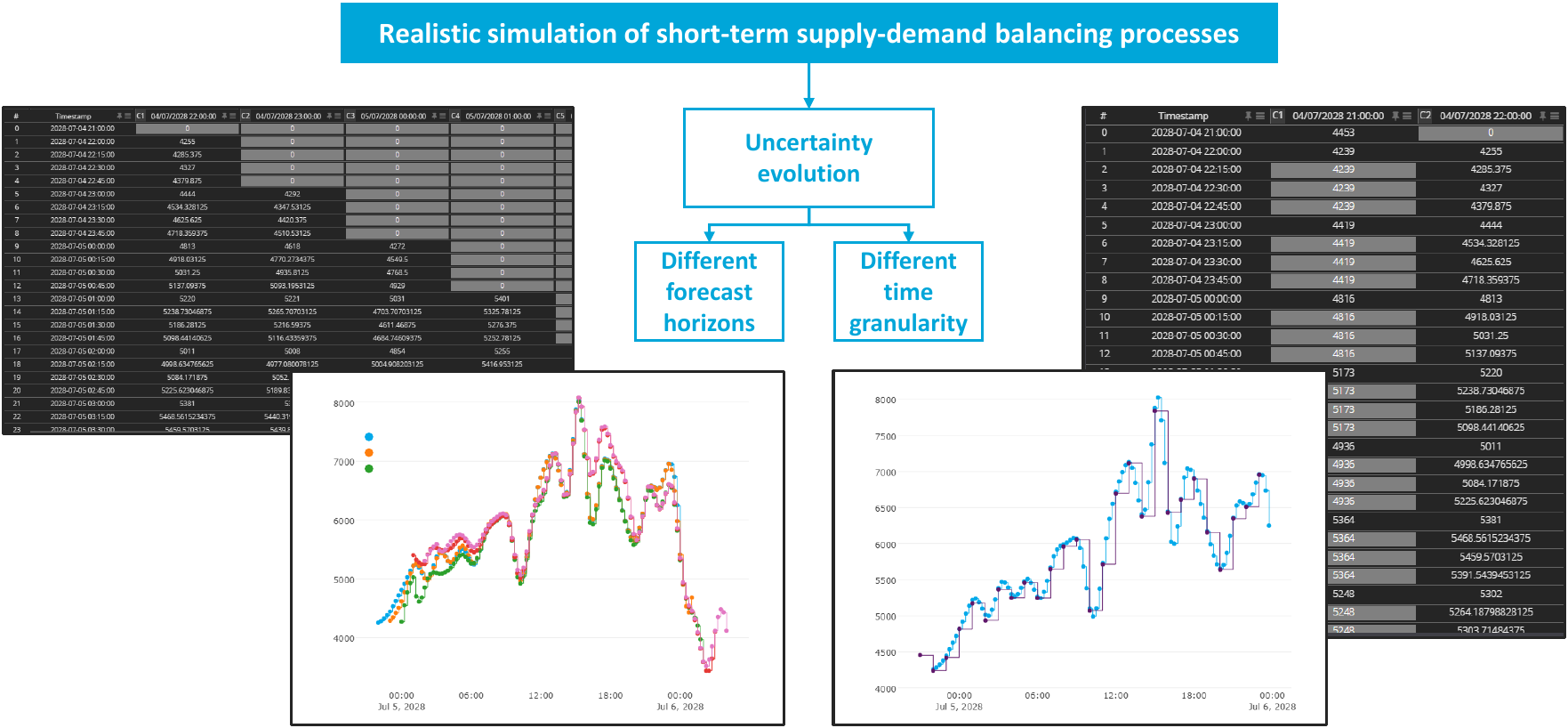}
    \caption{ATLAS model forecast evolution}
    \label{fig:forecastUncertainty}
\end{figure}

The information available to each participant is not only dependent on the structure determined by the delimitations described above, but also on the time frame in question. The ATLAS model models the impacts of uncertainty between different phases. To do this, it relies on the concept of \textit{forecast matrices} to manage uncertainties across different time frames. The concept is relatively straightforward. A single matrix stores the data from "forecasts" performed at different moments in time. Generally, there is an initial forecast for the entire year. Then, new forecasts are generated at different increments for a set period, giving for example the data shown in Figure \ref{fig:forecastUncertainty}. The uncertainties are generated in the Forecast Error Model, shown in Figure \ref{fig:ATLASminichain}. It will be described in more detail in Section \ref{sec:PowerForecast}. ATLAS includes this model to generate forecast errors as well as a model to interpolate between data points in order to simulate the impact of different time granularities. 

The different modules act on different time periods that are handled slightly differently in the ATLAS model than in existing models (say \cite{bacci_new_2023} for instance). The models are centered around what happens in different market phases. As a result, we define the \textit{optimization window} as the period of the market in question. In order for the final production plans to be feasible, we add a period prior to the \textit{optimization window}--referred to as the \textit{traceback period}. These additional periods depend on the technical constraints of the unit in question. This will be explained in more detail in Section \ref{sec:OD_ProblemSetup}.

% The Intraday markets have historically been simulated through two methods. In order to emulate a continuous market, \textcolor{red}{XXXX ADD EXPLANATION}

% \begin{figure}[ht!]
%     \centering
%     \includegraphics[width=\textwidth]{images/timeframes.png}
%     \caption{ATLAS model timeframes}
%     \label{fig:timeframes}
% \end{figure}

\section{Nomenclature} \label{sec:Nomenclature}

Remark: For sets, the notation $A_{b}$ refers to the subset of $A$ linked with variable $b$. For instance, $Z_{ca}$ indicates the subset of market areas belonging to the control area $ca$.

\renewcommand{\arraystretch}{1.3} % Default value: 1
\begin{longtable}[!ht]{!{\color{Grey}\vrule}>{\centering\arraybackslash}p{2.5cm} !{\color{Grey}\vrule} p{10cm}!{\color{Grey}\vrule} p{2cm} !{\color{Grey}\vrule}}
    \arrayrulecolor{Grey} \hline
    \rowcolor{Grey} \multicolumn{3}{|c|}{\large{\textbf{Basic concepts and physical quantities}}} \\ \arrayrulecolor{Grey} \hline
    \rowcolor{Grey} 
    \textbf{Notation} & \textbf{Meaning} & \textbf{Units} \\
    % \multicolumn{3}{|c|}{\textbf{Indices}} \\ \hline
    t & Time & h \\ \hline
    % $e$ & Energy Notation & - \\
    % $r_{u}$ & Up-reserve Notation & - \\
    % $r_{d}$ & Down-reserve Notation & - \\
    $q$ & Quantity, generally of a market offer& MW \\ \hline
    $x$ & Amount Accepted & Frac. of $q$ \\ \hline
    $p$ & Price & \euro/MWh \\ \hline
    $P$ & Power & MW \\ \hline
    $E$ & Energy & MW \\ \hline
    $t$ & Time & h \\ \hline
    $z$ & Zone & -  \\ \hline
    $ca$ & Control Area & - \\ \hline
    $cb$ & Critical Branch & - \\ \hline
    $mb$ & Market Border & - \\ \hline
    $o$ & Offer & - \\     \hline
    $pf$ & Portfolio & - \\ \hline
    $u$ & Unit & - \\ \hline
    $exp/imp$ & Export/Import  & - \\ \hline
    $DS/US$ & Downstream/Upstream  & - \\ \hline
    % $I/D/H$ & Indivisible/Divisible/Hybrid & -\\
    \arrayrulecolor{Grey} \hline 
\end{longtable}

\begin{longtable}[!ht]{!{\color{Grey}\vrule}>{\centering\arraybackslash}p{2.5cm} !{\color{Grey}\vrule} p{12.5cm}!{\color{Grey}\vrule} }
    \arrayrulecolor{Grey} \hline
    \rowcolor{Grey} \multicolumn{2}{|c|}{\large{\textbf{Global parameters and sets}}} \\ \arrayrulecolor{Grey} \hline
    \rowcolor{Grey} 
    \textbf{Notation} & \textbf{Meaning}\\
    $T$ & Discretized time period \\ \hline
    $\Delta t$ & Time step used for discretizing the simulated period \\ \hline
    $Z$ & Set of all zones, $z \in Z$  \\ \hline
    $CA$ & Set of all control areas: $c \in C$ \\ \hline
    $CB$ & Set of all critical branches: $cb_k \in CB$ \\ \hline
    $MB$ & Set of all market borders: $mb_l \in MB$ \\ \hline
    $O$ & Set of all offers: $o \in O$ \\     \hline
    $PF$ & Set of all portfolios: $pf \in PF$\\ \hline
    $U$ & Set of all units: $u \in U$\\ \hline
    $U^{unit\_type}$ & Set of all units of type $unit\_type \in [g, fl, th, h, st, w, pv,ndp,ndl]$. ($g$ = generation, $fl$ = flexible load, $th$ = thermal, $h$ = hydraulic, $st$ = storage, $w$ = wind, $pv$ = photovoltaic, $ndp$ = non-dispatchable production, $ndl$ = non-dispatchable load)\\
    \hline 
    \hline 
\end{longtable}

\begin{longtable}[!ht]{!{\color{Grey}\vrule}>{\centering\arraybackslash}p{2.5cm} !{\color{Grey}\vrule} p{12.5cm}!{\color{Grey}\vrule} }
    \arrayrulecolor{Grey} \hline
    \rowcolor{Grey} \multicolumn{2}{|c|}{\large{\textbf{Temporal parameters}}} \\ \arrayrulecolor{Grey} \hline
    \rowcolor{Grey} 
    \textbf{Notation} & \textbf{Meaning}\\ \hline
    $\Delta t$ & Time step used for discretizing the simulated period \\ \hline
    $t^{ex}$ & Execution date of the simulated phase \\ \hline
    $t^{start}$ & Start date of the simulation \\  \hline
    $t^{end}$ & End date of the simulated phase \\ \hline
    $\Delta t$ & Simulation timestep [hrs] \\  \hline
    $T^{traceback}$ & Number of timesteps to take into account previously-defined programs. The value is defined by the unit's technical constraints. \\ \hline
    $T^{prev}$ & The list of datetimes covering $T^{traceback}$ prior to the optimization window. The optimization is not performed across this timeframe, but the state of the unit is taken into account. \\ \hline
    $T^{sim}$ & The period of time of the simulation, $[t^{start}:t^{end})$ \\ \hline
    $T^{addl}_{tech}$ & The additional timeframe to be considered during the optimization. A value per technology should be provided. This parameter allows for instance for storage units to be optimized across a weekly timescale. \\ \hline
    $T^{opt}$ & The period of time of the optimization to be performed, which depends on the technology. $[t^{start}:t^{end} + T^{addl}_{tech})$. For notational purposes, the $technology$ indice is left off, but it can be assumed to be covered by the section header in each case. \\ \hline
    $T^{ext}$ & The extended time frame from $[t^{start} - T^{traceback}:t^{end} + T^{addl}_{tech} + T^{traceback})$ \\ \hline
    \hline 
\end{longtable}

\begin{longtable}[!ht]{!{\color{Grey}\vrule}>{\centering\arraybackslash}p{2.5cm} !{\color{Grey}\vrule} p{12.5cm}!{\color{Grey}\vrule} }
    \arrayrulecolor{Grey} \hline
    \rowcolor{Grey} \multicolumn{2}{|c|}{\large{\textbf{Zonal characteristics}}} \\ 
    \hline
    \rowcolor{Grey} 
    \textbf{Notation} & \textbf{Meaning}\\ \hline
    $CA_z$ & Control Area $c$ within Zone $z$ \\ \hline
    $z_l^U$, $z_l^D$ & Upstream and downstream zone defined for market border $mb_l$, respectively \\ \hline
    $(\Delta q)^{bal}_{z,t} $ & Supply/demand balance in zone $z$ at time $t$ \\ \hline
    $(\Delta q)^{bal,ref}_{z,t} $ & Supply/demand reference balance in zone $z$ at time $t$ \\ \hline
    $p^{min}_{z,t}$, $p^{max}_{z,t}$ & Maximum and minimum market price allowed for zone $z$ \\
    \hline 
\end{longtable}

\begin{longtable}[!ht]{!{\color{Grey}\vrule}>{\centering\arraybackslash}p{2.5cm} !{\color{Grey}\vrule} p{12.5cm}!{\color{Grey}\vrule} }
    \arrayrulecolor{Grey} \hline
    \rowcolor{Grey} \multicolumn{2}{|c|}{\large{\textbf{Critical branch characteristics (Flow-based Mode)}}} \\ 
    \hline
    \rowcolor{Grey} 
    \textbf{Notation} & \textbf{Meaning}\\ \hline
    $q_{k,t}^{CB}$ & Flow of critical branch $cb_k$ at time $t$ \\ \hline
    $q_{k,t}^{CB,max}$ & Max flow on critical branch $cb_k$ at time $t$ \\ \hline
    $q_{k,t}^{CB,FRM}$ & Flow reliability margin set by TSO on critical branch $cb_k$ at time $t$ \\ \hline
    $q_{k,t}^{CB,ref}$ & Reference flow of critical branch $cb_k$ at time $t$ \\ \hline
    $D_{k,t,z}^{Z}$ & Zonal power transfer distribution factor for zone $z$ on critical branch $cb_k$ at time $t$ \\
    \hline 
\end{longtable}

\begin{longtable}[!ht]{!{\color{Grey}\vrule}>{\centering\arraybackslash}p{2.5cm} !{\color{Grey}\vrule} p{12.5cm}!{\color{Grey}\vrule} }
    \arrayrulecolor{Grey} \hline
    \rowcolor{Grey} \multicolumn{2}{|c|}{\large{\textbf{Market border characteristics}}} \\ 
    \hline
    \rowcolor{Grey} 
    \textbf{Notation} & \textbf{Meaning}\\ \hline
    $(\Delta q)_{l,t}^{MB}$ & Net exchange of traded power going through border $mb_l$ (defined from upstream zone $z_l^{U}$ to downstream zone $z_l^D$) at time $t$ \\ \hline
    $(\Delta q)_{l,t}^{MB,exp}$ & Power exported on border $mb_l$ (defined from  upstream zone $z_l^{U}$ to downstream zone $z_l^D$) at time $t$ \\    \hline
    $(\Delta q)_{l,t}^{MB,imp}$ & Power imported on border $mb_l$ (defined from  upstream zone $z_l^{U}$ to downstream zone $z_l^D$) at time $t$ \\   \hline
    $(\Delta q)_{l,t}^{MB,max}$ & Maximum power allowed to pass on border $mb_l$ from upstream zone $z_l^{U}$ to downstream zone $z_l^D$ at time $t$ \\   \hline 
    $(\Delta q)_{l,t}^{MB,min}$ & Minimum power allowed to pass on border $mb_l$ from upstream zone $z_l^{U}$ to downstream zone $z_l^D$ at time $t$ \\ \hline
    $\sigma_{l,z}$ & $\sigma_{l,z} = 1$ if $z = z_l^{U}$, $\sigma_{l,z} = -1$ else\\ \hline
    $a_l$ & Technical parameter used for modeling losses on a DC border \\ \hline
    $\nu_{l,t}$ & Coefficient used to model losses on DC borders (see Section \ref{ch:MarketClearing}) \\ \hline
    $\xi_{l,t}$ & Helper variable for linearizing constraints on DC borders (see Section \ref{ch:MarketClearing}) \\    
    \hline 
\end{longtable}

\begin{longtable}[!ht]{!{\color{Grey}\vrule}>{\centering\arraybackslash}p{2.5cm} !{\color{Grey}\vrule} p{12.5cm}!{\color{Grey}\vrule} }
    \arrayrulecolor{Grey} \hline
    \rowcolor{Grey} \multicolumn{2}{|c|}{\large{\textbf{Order characteristics}}} \\ 
    \hline
    \rowcolor{Grey} 
    \textbf{Notation} & \textbf{Meaning}\\ \hline
    $q_o$ & Amount of power accepted for order $o$ \\ \hline
    $\delta_o$ & Binary unit commitment variable for order $o$ \\ \hline
    $p_o$ & Price of order $o$ \\ \hline
    $q_o^{min}$ & Minimum quantity of power offered for order $o$ \\ \hline
    $q_o^{max}$ & Maximum quantity of power offered for order $o$ \\ \hline
    $t_o^{start}$ & Start time of order \\ \hline
    $t_o^{end}$ & End time of order \\ \hline
    $d_{o}$ & Duration of order \\ \hline
    $\sigma_{o}$ & Sale/Purchase indicator, $\sigma = 1$ for purchase, -1 for sale \\ 
    \hline 
\end{longtable}

\begin{longtable}[!ht]{!{\color{Grey}\vrule}>{\centering\arraybackslash}p{2.5cm} !{\color{Grey}\vrule} p{12.5cm}!{\color{Grey}\vrule} }
    \arrayrulecolor{Grey} \hline
    \rowcolor{Grey} \multicolumn{2}{|c|}{\large{\textbf{Global unit characteristics}}} \\ 
    \hline
    \rowcolor{Grey} 
    \textbf{Notation} & \textbf{Meaning}\\ \hline
    $P_{u,t,t^{ex}}^{plan}$ & Generation (or consumption) plan of unit $u$ at time $t$, seen from $t_ex$ \\ \hline
    $P_{u,t}^{max}$ & Maximum power output of unit $u$ at time $t$ \\ \hline
    $P_{u,t}^{min}$ & Minimum power output of unit $u$ at time $t$ \\ \hline
    $\Delta P_{u}^{max}$ & Maximum ramping capability of unit $u$ \\ \hline
    $c_u^{var}$ & Variable cost of unit $u$ \\     \hline
    $c_u^{SU}$ & Startup cost of unit $u$ \\ \hline
    $d_u^{notice}$ & Notice delay duration of unit $u$ \\ \hline
    $R^{[ResType,ResDir]}_{u,t,t^{ex}}$ & Procured reserves of $ResTypes \in [FCR, aFRR, mFRR, RR]$ in direction $ResDir \in [up, down]$ on unit $u$ at time $t$ seen from $t^{ex}$ \\ \hline
    \rowcolor{Grey} \multicolumn{2}{|c|}{\large{\textbf{Thermal-specific unit characteristics}}} \\ 
    $d_u^{SU}$ & Startup duration of unit $u \in U_{th}$  \\ \hline
    $d_u^{SD}$ & Shutdown duration of unit $u \in U_{th}$  \\ \hline
    $T_u^{SU}$, $T_u^{SD}$ & Number of timesteps in startup (resp. shutdown) periods of unit $u \in U_{th}$ \\ \hline
    $d_u^{minOn}$ & Minimum time on duration of unit $u \in U_{th}$  \\ \hline
    $T_u^{on}$ & Number of timesteps in \textit{minimum time on} of unit $u \in U_{th}$\\ \hline
    $d_u^{minOff}$ & Minimum time off duration of unit $u \in U_{th}$ \\ \hline
    $d_u^{minStable}$ & Minimum stable power duration of unit $u \in U_{th}$ \\ \hline
    $\Delta P_{u}^{step,up},$ \newline $ \Delta P_{u}^{step,dn}$ & Power variation during startup or shutdown phase of unit $u$ \\ \hline
    \rowcolor{Grey} \multicolumn{2}{|c|}{\large{\textbf{Hydraulic- and Storage-specific unit characteristics}}} \\  \hline
    $E_{u,t,t^{ex}}^{stored}$ & Stored energy in the reservoir of unit $u \in U_H \cup U_{st}$ at time $t$, seen from $t_ex$ \\ \hline
    $E_{u,t}^{max}$ & Maximum storage level in the reservoir of unit $u \in U_H \cup U_{st}$ at time $t$ \\ \hline
    $E_{u,t}^{min}$ & Minimum storage level in the reservoir of unit $u \in U_H \cup U_{st}$ at time $t$  \\ \hline
    $d_{u}^{tran}$ & Transition duration between pumping and turbining of unit $u \in U_{PHS}$ \\ \hline$P^{buy}_{u,t}$, $P^{sell}_{u,t}$ & Power in the purchase and sale direction for storage nit $u$ at time $t$\\ \hline
    $\eta^d_u, \eta^c_u$ & Discharge (resp. Charging) efficiency for the unit $u$ \\ \hline
    $\sigma^{sell}_{u,t}$ & Sale/Purchase indicator for the storage technology $u$ at time $t$ \\ \hline
    $\sigma^{EV}_{u}$ & Indicator used to denote a unit $u$ of type Electric Vehicle ($\sigma^{EV}_{u}\text{ = 1 if EV, } =0 \text{ else}$) \\ \hline
    $E^{disp}_{u, t^{start}-1,t}$ &  Displacement Energy\\ \hline
    \rowcolor{Grey} \multicolumn{2}{|c|}{\large{\textbf{Photovoltaic- and Wind-specific unit characteristics}}} \\  \hline
    $P_{u,t,t^{ex}}^{spill}$ & Power spillage on unit $u \in U_{pv} \cup U_w$ at time $t$, seen from $t_ex$ \\ 
    \hline 
\end{longtable}

\begin{longtable}[!ht]{!{\color{Grey}\vrule}>{\centering\arraybackslash}p{2.5cm} !{\color{Grey}\vrule} p{12.5cm}!{\color{Grey}\vrule} }
    \arrayrulecolor{Grey} \hline
    \rowcolor{Grey} \multicolumn{2}{|c|}{\large{\textbf{Additional parameters}}} \\ 
    \hline
    \rowcolor{Grey} 
    \textbf{Notation} & \textbf{Meaning}\\ \hline
    $\lambda_1$ & Multiplier in the market clearing objective function that allows to increase or decrease the social welfare for computational performance purposes \\ \hline
    $\lambda_2$, $\lambda_3$, $\lambda_4$ & Multipliers in the market clearing objective function that allow for more control on border exchanges \\ \hline
    $\alpha$, $\beta$, $M$ & Multipliers in the first and second pricing objective functions \\ 
    \hline 
\end{longtable}

\begin{figure}[H]
    \centering
    \includegraphics[width=0.95\textheight,angle=90]{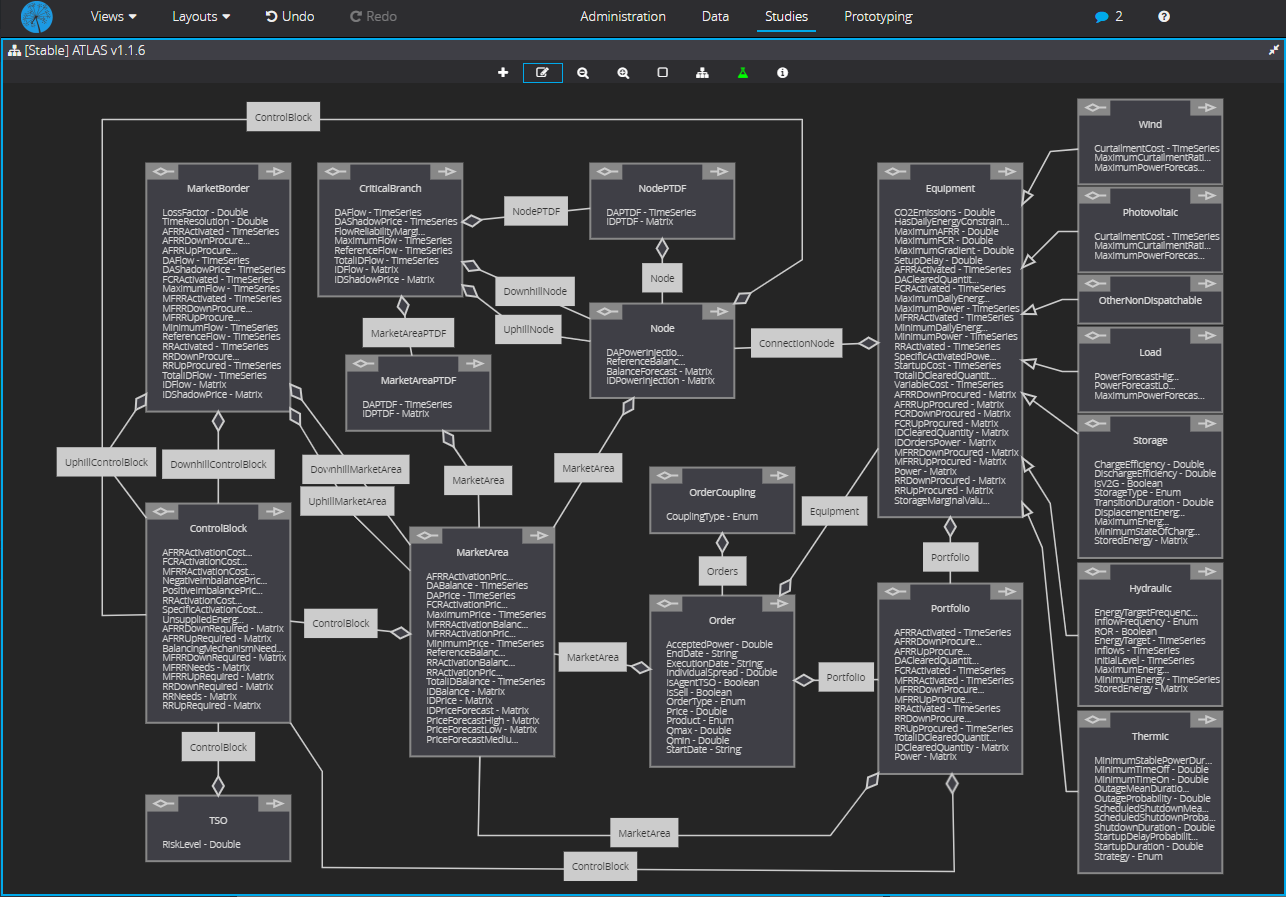}
    \caption{ATLAS Data Format}
    \label{fig:ATLASmodel}
\end{figure}

\renewcommand{\arraystretch}{1.1} % Default value: 1

\chapter{Forecast Models}  \label{ch:ForecastModels}

\section{Power Forecast} \label{sec:PowerForecast}

\subsection{Summary}
  
The PowerForecastSimulation module of the ATLAS model contains all the functions needed to generate forecasts from different time horizons.

The functions allow to learn the statistical properties of a set of forecast archives in order to preserve these properties during a simulation.

Learning is performed using a learning function, while simulations are generated using a simulation function.

\subsection{Methodology}

\subsubsection{Nomenclature}

\begin{longtable}[!ht]{!{\color{Grey}\vrule}>{\centering\arraybackslash}p{2.5cm} !{\color{Grey}\vrule} p{12.5cm}!{\color{Grey}\vrule} }
    \arrayrulecolor{Grey} \hline
    \rowcolor{Grey} \multicolumn{2}{|c|}{\large{\textbf{Unit sets}}} \\ 
    \textbf{Notation} & \textbf{Meaning}\\ \hline
        $P_T$ &  Power output (VRE generation or consumption) observed at time $T$. \\
        \hline
        $\hat{P}_{t,T}$ &  Power output (VRE generation or consumption) forecast at time $T$ seen from time $t$.  \\
        \hline
        $\Delta P_{t,T} = \hat{P}_{t,T}  - \hat{P}_{t-1,T}$ &  Update of the forecast value for time $T$ between times $t$ and $t-1$.  \\
        \hline
        $\epsilon_T = \hat{P}_{T,T} - P_T$ &  Real-time estimation error of the quantity $P$  \\
        \hline
\end{longtable}

We can write the following equality: 

\begin{align} 
    \hat{P}_{T,T} & = \hat{P}_{T-1,T} + (\hat{P}_{T-1,T} - \hat{P}_{T-1,T}) \\
    & = \hat{P}_{T-1,T} + \Delta P_{T,T} \\
    & = \hat{P}_{T-2,T} + (\hat{P}_{T-1,T} - \hat{P}_{T-2,T}) + \Delta P_{T,T} \\
    & = \hat{P}_{T-2,T} + \Delta P_{T-1,T} + \Delta P_{T,T} \\
    &  ... \\
    & = \hat{P}_{T-N,T} + \Delta P_{T-(N-1),T} + ... + \Delta P_{T,T}  \\
    & = \hat{P}_{T-N,T} +  \sum_{N < h\le 0} \Delta P_{T-h,T} \\
    & = \hat{P}_{t,T} + \sum_{t< i\le T} \Delta P_{i,T}\\
\end{align}

Finally:

\begin{align} 
    \hat{P}_{t,T} = P_T  -\sum_{t< i\le T} \Delta P_{i,T}  - \epsilon_T 
\end{align}

The principle of the simulator is therefore to obtain the value of the forecast from the value of the realization and a random generation of forecast updates.

\subsubsection{Analysis of the statistical properties of the dataset}

We analyze the following properties: 
\begin{itemize}
    \item Inter-forecast correlations: $cor(\hat{P}_{t,T}, \hat{P}_{t-1,T})$.

    The correlations between two forecasts targeting the same time ${T}$ but generated at 2 different times ${t}$ vary according to the nature of the input data that generated the update.
    A new weather forecast may constitute a significant update, whereas a reset (new observation of the quantity to be forecast) will have a lesser effect.

    \item Intra-forecast correlations: $cor(\hat{P}_{t,t+h_1},\hat{P}_{t,t+h_2})$.

    The correlation between two forecasts generated at the same time ${t}$ but targeting different times ${T}$ varies according to the duration between these two times. If they are close, the forecasts will be highly correlated. Conversely, if they are spaced out in time, the correlation between these forecasts will be weak.

    \item Forecast performance: $RMSE_h = \sqrt{\frac{1}{N} \sum_t (\hat{P}_{t,t+h} - P_{t+h})^2}$.

    A forecast has larger errors as the forecast horizon increases.
\end{itemize}

\subsection{Hypotheses}

\subsubsection{No self-correlations between updates}

We assume that there is no dependency between two successive updates: 

$cor(\Delta P_{t,T}, \Delta P_{t+i,T}) = 0$.

\subsubsection{Correlation of updates of the same forecast}

We assume that there is a dependency between the forecasts of different horizons within the same update:

$cor(\Delta P_{t,t+h},\Delta P_{t,t+h+i}) > 0$.

\subsubsection{Dependence of discounting on the forecast horizon}

We assume that the forecast error grows non-linearly with the forecast horizon. Consequently, a discount should depend on the forecast horizon $h$.

\subsection{Learning stage} 

\subsubsection{Standardisation of forecast series}
This pre-processing stage consists of normalizing the time series in order to make them as stationary as possible.
This normalization consists of modeling the maximum amplitude of the series as a function of the calendar using quantile regression of the following type 

$Max_t = \sum_{i=1}^{M} \alpha_i sin(\frac{2\pi}{365.25}y(t))+ \beta_i cos(\frac{2\pi}{365.25}y(t))$

where ${1 < y(t) < 366}$ is the position in the year of time $t$.

The time series is then divided by this maximum: 
$p_{t,T} = \frac{P_{t,T}}{Max(T)}$.

\subsubsection{Calculating forecast updates}

An update is defined as follows: 

$\Delta p_{t,T} = \hat{p}_{t,T} - \hat{p}_{t-1,T}$.

where $\hat{p}_{t,T}$ is the forecast of quantity $P$ for time $T$ generated at time $t$.

It is therefore the revision of the forecast compared with the last forecast value for a given date.
It is this value that we are seeking to reproduce.

\subsubsection{Estimation of marginal distributions}

For each forecast horizon and each decile of forecast value, or even each forecast launch time, 
the empirical probability densities of the updates are saved. The updated values are then transposed into their corresponding quantile values.

\subsubsection{Estimation of Copula parameters}
The Spearman intra-prediction correlations of the update values are calculated $cor(\Delta p_{t, t+h_1}, \Delta p_{t, t+h_2})$ in order to apply Gaussian Copulas to the marginal distributions of the dataset and thus avoid learning the joint distributions.

\subsection{Simulation step}

\subsubsection{Normalisation of forecast series}
The normalization model is reused to normalize the observation time series: 
$p_t = \frac{P_t}{Max_t}$.

\subsubsection{Random draw in the copula}
The quantiles of the normalized time series are simulated according to the dependency structure learned during the learning stage.

\subsubsection{Transpose simulated quantiles into updated values}.
The quantiles simulated in this way are transposed into updated values using the probability density learned during the learning stage.

\subsubsection{Calculation of expected values}
By iteratively adding the various simulated discount values to the observed production value 
$hat{p}_{t,T} = p_T -sum_{t< i\le T} \Delta p_{i,T} - \epsilon_T$.

while complying with the constraints on the definition of the quantity to be forecast, we obtain the forecast value.

\subsubsection{Denormalisation of forecast series}
The normalization model is reused to obtain the final forecast value 

$P_{t,T} = p_{t,T} * Max(T)$

\subsection{Data format}

There are two types of data required for the simulator to function correctly:

\subsubsection{Forecasts}
An archive of forecasts is required to learn the simulation model, using rolling forecast horizons corresponding to the desired time step.

\subsubsection{Observations}

Once the simulation model has been learned, a history of observations must be supplied to the simulation function in order to generate the corresponding forecasts. Note that if the observations used for simulation do not correspond to the same perimeter as that used for learning, a transformation must be applied before running the simulation.

\textcolor{red}{TO BE COMPLETED}

\section{Price Forecast} \label{sec:PriceForecast}

\textcolor{red}{TO BE COMPLETED}

\section{Interpolation Method} \label{sec:Interpolation}

ATLAS also includes a method to interpolate between time steps. This allows for studies to be performed at the sub-hourly time step. This method was developed in the European project Osmose (Grant agreement ID: 773406). The documentation of this model will therefore not be included in detail in this document, but can be found in Deliverable 2.3, Section 4.4 and Appendix 7 (\url{https://www.osmose-h2020.eu/wp-content/uploads/2022/02/OSMOSE-D2.3-ModelsForMarketSimulation_Final.pdf}). The algorithm involves an optimization problem with an objective function that differs slightly for solar generation, wind generation, load or price forecasts. 
\chapter{The Optimal Dispatch Problem} \label{ch:OptimalDispatch}

The core optimization problem at the heart of many of the ATLAS modules is the optimal dispatch problem. In this chapter, the definition of this problem in the ATLAS model is described. The model can be modified to ignore certain operating constraints\footnote{as discussed in more detail in the PhD work of Florent Cogen.} and allow for studies focused on assessing specific aspects. However, here we will describe the full problem with all operating constraints. Since this methodology will be used in several of the existing modules (albeit with variations on objective functions, variables and sets of constraints), the problem is described in some detail. 

The objective function varies slightly depending on the module in question. It is used in three particular cases throughout this document: 
\begin{enumerate}
    \item In certain cases, a unit-based optimal dispatch problem is run to generate day ahead market orders. This is the case notably for thermal plants employing an intermediate pricing strategy (see Section \ref{subsec:IntermediateStrategy}) and for storage units. 
    \item Similarly, the optimal dispatch is run to generate intraday orders. These take into account the difference between the program determined post-day ahead phase and the most up-to-date forecast. This use is detailed in Section \ref{ch:IntradayOrders}.
    \item Finally, just following a market clearing, the optimal dispatch problem is run. This phase is generally portfolio-based and performs the following functions: 1) establishes technically feasible production and demand dispatch curves as the quantity accepted in the previous market does not necessarily reflect all technical constraints, 2) allows for arbitrage between forecast imbalance prices and operating costs, 3) allows a portfolio to potentially reduce its costs through a shift of dispatch between accepted quantities in more expensive plants and rejected quantities from cheaper units in case of poorly generated offers. It also then saves this dispatch for each unit (and updates the energy stored in reservoir-based technologies). Details about this mode are included in Section \ref{subsec:PortfolioOptim}.
\end{enumerate}

Section \ref{sec:OD_ProblemSetup}, the problem setup is described, including user-input parameters and the description of the time frame that is common to all use cases. Sections \ref{sec:OD_ThermalPlants} - \ref{sec:OD_AddlConstraints} show the constraints that are also common to each of the use cases. Finally, Section \ref{sec:ObjFxns} explores the specificities of the three use cases.  

\section{Problem Setup} \label{sec:OD_ProblemSetup}

\subsection{User-input Parameters}
We also retrieve the following prototype-specific parameters imputed by the user. These are common to all modules that apply the constraints defined in this chapter: 

\begin{longtable}[!ht]{!{\color{LightGrey}\vrule}>{\centering\arraybackslash}p{2.5cm} !{\color{LightGrey}\vrule} p{12.5cm}!{\color{LightGrey}\vrule} }
    \arrayrulecolor{LightGrey} \hline
    \rowcolor{Grey} \multicolumn{2}{|c|}{\large{\textbf{User-input parameters}}} \\ 
    \hline
    \rowcolor{Grey} 
    \textbf{Notation} & \textbf{Meaning}\\ \hline
    $\Delta t$ & Simulation timestep \\ \hline
    $t^{start}$ & Start date of the simulation \\ \hline
    $t^{end}$ & End optimization date of the simulation \\ \hline
    $\eta^{manu}$ & A penalty for unprovided manual reserves, in Euros \\ \hline
    $\eta^{auto}$ & A penalty for unprovided  automated reserves, in Euros \\ \hline
    $T^{addl}_{tech}$ & The additional timeframe to be considered during the optimization. A value per technology should be provided. This parameter allows for instance for storage units to be optimized across a weekly timescale. \\ \hline
    $\varepsilon$ & A float equal by default to $10^{-3}$, which is a slack parameter for the constraints taking the form of an equality \\ \hline
    \hline
\end{longtable}

\subsection{Timeframes}

\begin{figure}[!ht]
    \centering
    \includegraphics[width=0.8\textwidth]{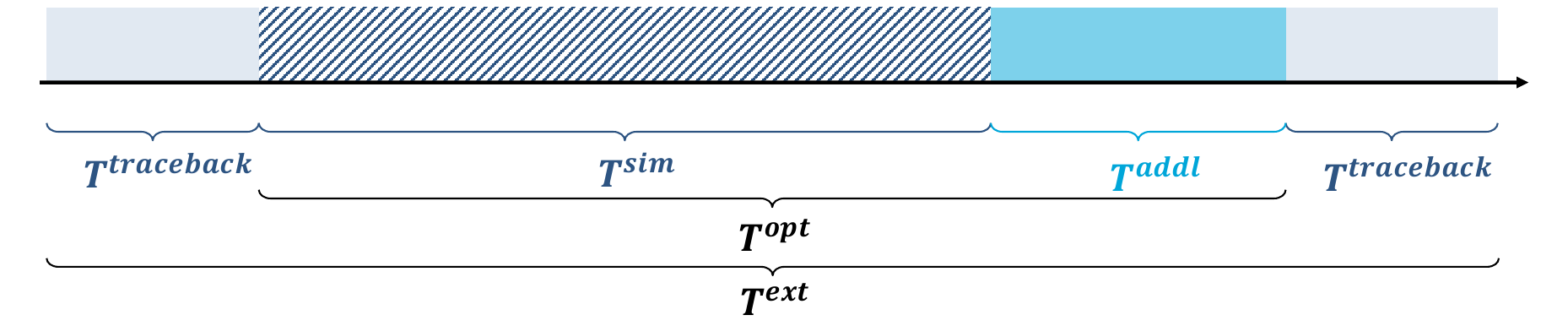}
    \caption{Optimal Dispatch Timeframes}
    \label{fig:timeframes_PO}
\end{figure}

\begin{longtable}[!ht]{!{\color{LightGrey}\vrule}>{\centering\arraybackslash}p{2.5cm} !{\color{LightGrey}\vrule} p{12.5cm}!{\color{LightGrey}\vrule} }
    \arrayrulecolor{LightGrey} \hline
    \rowcolor{Grey} \multicolumn{2}{|c|}{\large{\textbf{Time Frames}}} \\ 
    \hline
    \rowcolor{Grey} 
    \textbf{Notation} & \textbf{Meaning}\\ \hline
    $\Delta t$ & Simulation timestep \\ \hline
    $t^{start}$ & Start date of the simulation \\ \hline
    $t^{end}$ & End optimization date of the simulation \\ \hline
    $T^{traceback}$ & Number of timesteps to take into account previously-defined programs. The value is defined by the unit technical constraints. \\ \hline
    $T^{prev}$ & The list of datetimes covering $T^{traceback}$ prior to the optimization window. The optimization is not performed across this timeframe, but the state of the unit is taken into account. \\ \hline
    $T^{sim}$ & The period of time of the simulation, $[t^{start}:t^{end})$ \\ \hline
    $T^{addl}_{tech}$ & The additional timeframe to be considered during the optimization. A value per technology should be provided. This parameter allows for instance for storage units to be optimized across a weekly timescale. \\ \hline
    $T^{opt}$ & The period of time of the optimization to be performed, which depends on the technology. $[t^{start}:t^{end} + T^{addl}_{tech})$. For notational purposes, the $technology$ indice is left off, but it can be assumed to be covered by the section header in each case. \\ \hline
    $T^{ext}$ & The extended time frame from $[t^{start} - T^{traceback}:t^{end} + T^{addl}_{tech} + T^{traceback})$ \\ \hline
    \hline
\end{longtable}

The time frame over which the current simulation is conducted is denoted $T^{sim}$ and is defined as:
\[
T^{sim} \coloneqq \left\{ t \in \mathbb{N} : \textrm{{\tt StartDate}}+ t\times \Delta t \in [\textrm{{\tt StartDate}}, \textrm{{\tt EndOptimizationDate}}]\right\}
\]
so we have $T^{sim} = \{0,1,\dots, T\}$, with $T = \max T^{sim}$
\footnote{$T^{sim} \neq \emptyset$ because it will at least contain $t = 0$. However, if $T^{sim}$ is a singleton, it means that {\tt EndOptimizationDate} is less than one $\Delta t$ {\it after} {\tt StartDate}. In such a case, an error is returned to the user and the execution of the optimization program is stopped.}. For most technologies, the constraints and objective functions are described across this time period, so: 
\begin{equation}
    T^{opt} = T^{sim} \coloneqq \left\{ t \in \mathbb{N} : \textrm{{\tt StartDate}}+ t\times \Delta t \in [\textrm{{\tt StartDate}}, \textrm{{\tt EndOptimizationDate}}]\right\}
\end{equation}

However, for some technologies, it is important to optimize over a longer time period. For these an additional time period is added on to the optimization period: 
\begin{equation}
    T^{opt} \coloneqq \left\{ t \in \mathbb{N} : \textrm{{\tt StartDate}}+ t\times \Delta t \in [\textrm{{\tt StartDate}}, T^{addl} + \textrm{{\tt EndOptimizationDate}}]\right\}
\end{equation}

These additional hours are set by the user. The default values are as follows: 
\begin{longtable}[!ht]{!{\color{LightGrey}\vrule}>{\centering\arraybackslash}c !{\color{LightGrey}\vrule} c !{\color{LightGrey}\vrule} }
    \caption{Default Values for Additional Hours Parameter} \label{tab:addlHours} \\
    \arrayrulecolor{LightGrey} \hline
    \rowcolor{Grey} 
    \textbf{Technology} &  \\ \hline
    [Storage] Electric Vehicle &  1 day (24 hours) \\ \hline
    [Storage] Pumped Hydraulic  &  6 days (144 hours) \\ \hline
    [Storage] Battery  &  1 day (24 hours) \\ \hline
    Hydraulic  & 12 hours \\ \hline
    Thermal & 0 hours \\ 
    \hline
    \hline
\end{longtable}

The optimization program is solved on $T^{opt}$ but given that many constraints depend on previous time steps, we need to define the optimization problem on a wider, called {\it extended} time frame, defined as follows : 
\[
\begin{aligned}
    T^{ext}  &\coloneqq T^{prev} \cup T^{opt}  \\
    & = \{- T^{traceback}, \dots, -1, 0, 1, \dots, T\}
\end{aligned}
\]
Where $T^{prev} \coloneqq \{-T^{traceback}, \dots , -1\}$ captures the time steps {\it prior} to the optimization time frame, and $T^{traceback}$ is defined as $T^{traceback} \coloneqq \max(T_u^{on} + T_u^{SU},T_u^{off} + T_u^{SD}) + 1 $. We round $T^{traceback}$ by adding one in order to avoid out-of-bounds errors when we are working with the time frame $\{-1\}\cup T^{opt}$. 

Optimization variables, which are defined over $T^{opt}$, will therefore be able to use known values in $T^{prev}$ so that intertemporal constraints will be defined even if we are in the first time steps of the optimization time frame $T^{opt}$ (which translates into the fact that for these $t\in T^{opt}$, $t - T_u^{on} < 0$ for instance). In some cases, the future dispatch is already known. For instance, when the optimal dispatch is used to define intraday market orders, the day-ahead dispatch for the next day may already be known. For this reason, the state of the unit after the optimization period (for a lenght of time equal to $T^{traceback}$) is considered as well, as shown in Figure \ref{fig:timeframes_PO}. In this case, $T^{ext}$ is extended to include these known values: 

\begin{equation}
    T^{ext}  = \{- T^{traceback}, \dots, -1, 0, 1, \dots, T, \dots, T + T^{addl}, \dots, T + T^{addl} + T^{traceback}\}
\end{equation}

\section{Thermal Plants} \label{sec:OD_ThermalPlants}
\begin{figure}[!ht]
\centering
\begin{subfigure}{.32\linewidth}
    \centering
    \includegraphics[width=\linewidth]{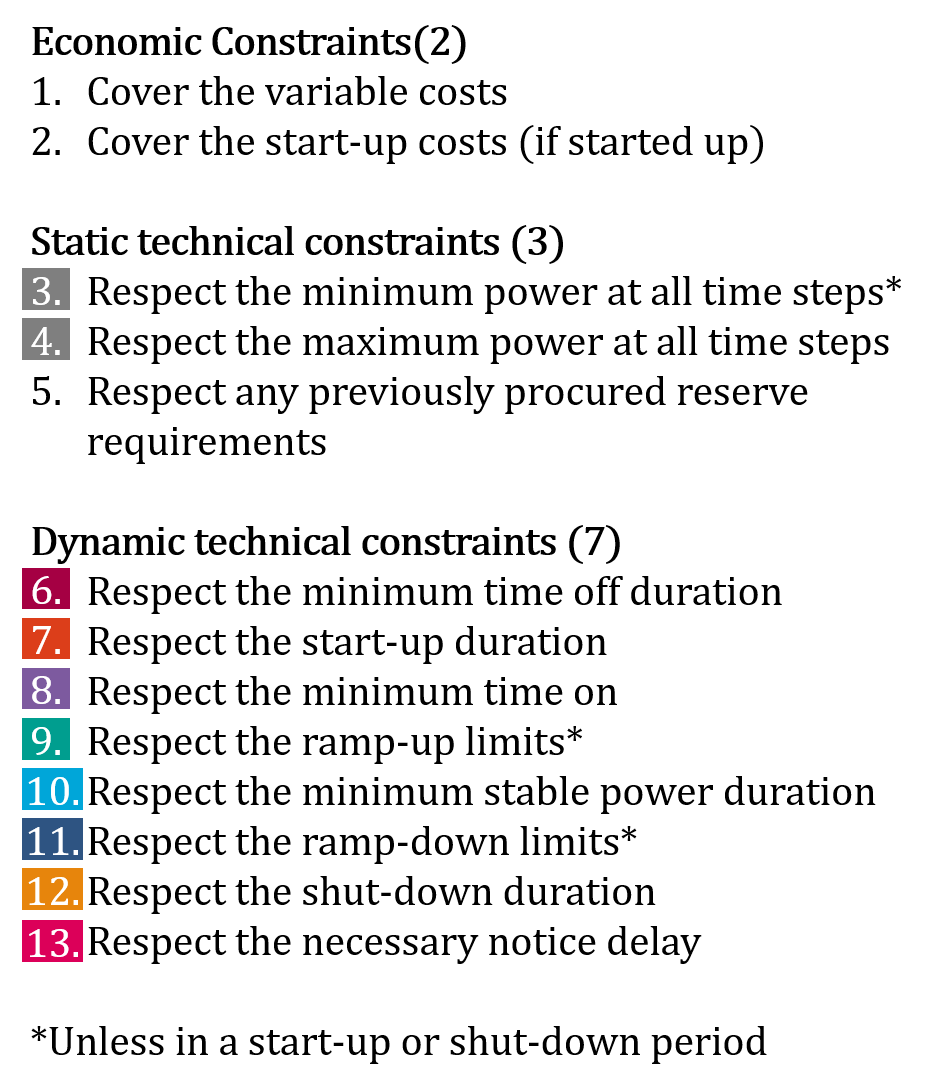}
    % \caption{}\label{fig:thermalConstraints_text}
\end{subfigure}
    % \hfill
\begin{subfigure}{.67\linewidth}
    \centering
    \includegraphics[width=\linewidth]{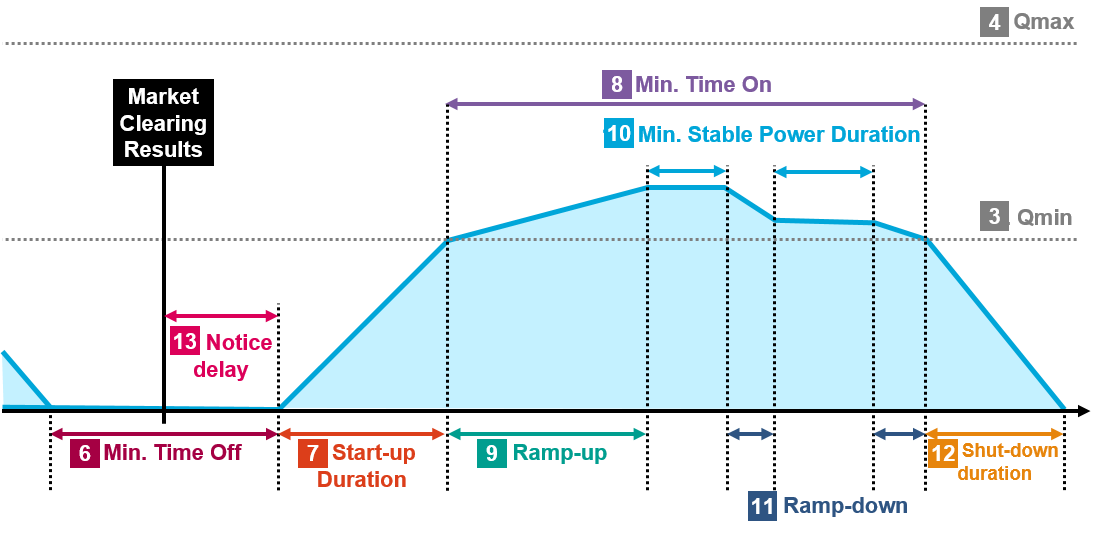}
    % \caption{}\label{fig:thermalConstraints_img}
\end{subfigure}

\RawCaption{\caption{Constraints of a Thermal Power Plant modeled in ATLAS}
\label{fig:thermalConstraints}}
\end{figure}

In the ATLAS model, we consider 13 main constraints for thermal power plants, shown in Figure \ref{fig:thermalConstraints}.

The first two constraints in the list are related to cost-covering and will be discussed in the price setting methods of the order creation modules. \textbf{Constraints 3} and \textbf{4} are the most straightforward and do not merit further discussion. \textbf{Constraint 5} refers to any capacity that has been contractualized in an earlier timeframe. However, the model does take into account the fact that in reality, penalties for unsupplied reserve capacity may not fully incentivize actors to maintain their commitment. For this reason, the difference between previously procured reserves and the provided reserve level is multiplied by a user-defined penalty in the objective function. The unit can provide reserves only when it is online and stable (excluding startup and shutdown ramping periods). In Europe now, there is a common definition of four reserve products\footnote{To be clear, the consensus around the definition-and especially the usage-of these products is still quite heterogenous across Europe.}:

\begin{enumerate}
    \item FCR: Frequency Containment Reserves (\textit{Réserve primaire, RP} in French)
    \item aFRR: Automatic Frequency Restoration Reserves (\textit{Réserve secondaire, RS} in French)
    \item mFRR: Manual Frequency Restoration Reserves (\textit{Réserve rapide, RR} in French)
    \item RR: Restoration Reserves (\textit{Réserve complémentaire, RC} in French)
\end{enumerate}

In these dispatch optimizing phases, in order to simplify slightly the model, the four European reserve products have been grouped into two: automated reserves (FCR and aFRR) and manual reserves (mFRR and RR). 

\textbf{Constraints 6-13} will be detailed in the model description.

\subsection{Inputs}

The optimization program uses the following attributes from the ATLAS model Thermic class : 

\renewcommand{\arraystretch}{1.3} % Default value: 1
\begin{longtable}[!ht]{!{\color{Grey}\vrule}>{\centering\arraybackslash}p{2.5cm} !{\color{Grey}\vrule} p{10cm}!{\color{Grey}\vrule} p{2cm} !{\color{Grey}\vrule}}
    \arrayrulecolor{Grey} \hline
    \rowcolor{Grey} \multicolumn{3}{|c|}{\large{\textbf{Global unit characteristics}}} \\ 
    \hline
    \rowcolor{Grey} 
    \textbf{Notation} & \textbf{Meaning} & \textbf{Units}\\ 
    $d_u^{minStable}$ & {\tt MinimumStablePowerDuration}: a scalar value such that \eqref{eq:consistency_dura} holds & h\\ \hline
    $d_u^{minOn}$ & {\tt MinimumTimeOn}: a scalar value to be such that \eqref{eq:consistency_dura} holds and corresponding to the time the unit has to be on, startup and shutdown periods {\bf excluded} & h\\ \hline
    $d_u^{minOff}$ & {\tt MinimumTimeOff}: a scalar value corresponding to the time the unit has to be off, startups and shutdowns {\bf excluded} & h\\ \hline
    $d_u^{SU}$ & {\tt StartupDuration}: the time it takes for the unit to start up & h\\ \hline
    $d_u^{SD}$ & {\tt ShutdownDuration}: the time it takes for the unit to shut down & h\\ \hline
    $R_{u}^{aFRR,max}$ & {\tt MaximumAFRR}: the maximum aFRR for the unit $u$ & MW\\ \hline
    $R_{u}^{FCR,max}$ & {\tt MaximumFCR}: the maximum FCR for the unit $u$e & MW \\ \hline
    $R_{u,t}^{proc,aFRR,up}$ & {\tt AFRRUpProcured}: the previously procured aFRR for unit $u$ at time $t$ in the upward direction & MW \\ \hline
    $R_{u,t}^{proc,aFRR,dn}$ & {\tt AFRRDownProcured}:  the previously procured aFRR for unit $u$ at time $t$ in the downward direction & MW \\ \hline
    $R_{u,t}^{proc,FCR,up}$ & {\tt FCRUpProcured}:  the previously procured FCR for unit $u$ at time $t$ in the upward direction & MW \\ \hline
    $R_{u,t}^{proc,FCR,dn}$ & {\tt FCRDownProcured}:  the previously procured FCR for unit $u$ at time $t$ in the downward direction & MW \\ \hline
    $R_{u,t}^{proc,mFRR,up}$ & {\tt MFRRUpProcured}:  the previously procured mFRR for unit $u$ at time $t$ in the upward direction & MW \\ \hline
    $R_{u,t}^{proc,mFRR,dn}$ & {\tt MFRRDownProcured}:  the previously procured mFRR for unit $u$ at time $t$ in the downward direction & MW \\ \hline
    $R_{u,t}^{proc,RR,up}$& {\tt RRUpProcured}:  the previously procured RR for unit $u$ at time $t$ in the upward direction & MW \\ \hline
    $R_{u,t}^{proc,RR,dn}$& {\tt RRDownProcured}:  the previously procured RR for unit $u$ at time $t$ in the downward direction & MW \\ \hline
    $c_{u,t}^{var}$ & {\tt VariableCost}: the variable cost of unit $u$ & Euros/MWh \\ \hline
    $c_{u,t}^{SU}$ & {\tt StartupCost}: the startup cost of unit $u$ & Euros \\ \hline
    $P_{u,t}^{min}$ & {\tt MinimumPower}: the minimum power for unit $u$ at time $t$ & MW \\ \hline
    $P_{u,t}^{max}$ & {\tt MaximumPower}: the maximum power for unit $u$ at time $t$ & MW \\ \hline
    $\Delta P_{u}^{max}$ & {\tt MaximumGradient}: a scalar. By convention, if the user inputs {\tt MaximumGradient} = 0, we consider that the unit has no gradient constraint. Initially given in MW/minutes, this value is converted to MW/timestep & MW/timestep\\ \hline
    % & {\tt HasDailyEnergyConstraint} a boolean indicating whether the unit has an energy constraint. This attribute is useful to model load curtailment as thermic units \\ \hline
    $E_{u,t}^{max,daily}$ & {\tt MaximumDailyEnergy} a time series & MW \\
    \hline
\end{longtable}

\subsection{Initial Assumptions, Definitions and Notations} \label{subsec:notations_optidis}
The approach used in the ATLAS model assumes the following:

\begin{enumerate}
    \item {\bf Start-up and shutdown costs}: the startup cost is given and equal to $c_u^{SU}$ and the shutdown cost is null.
    \item {\bf Fixed costs}: the only fixed cost the unit faces is its startup cost.
    \item {\bf Variable cost}: the cost function is linear in the output produced, i.e. $c_{u,t}(P_{u,t}) = c_t\times P_{u,t}$ where $c_t$ is an exogenous (given) and possibly time-varying cost and $P_{u,t}$ is the power output produced at time $t$. In the following, we will refer to $c_t$ as the {\it variable cost}.
    \item {\bf Energy limits}: the unit can be turned on or off as often as necessary but we allow for an upper bounding of the maximum daily energy the unit can provide. 
    \item {\bf Power gradients}: the ramp-up and ramp-down constraints, i.e. the maximum increase or decrease in power output with respect to the previous period are assumed to be symmetric.
    \item {\bf Startup and shutdown ramps}: it is assumed that both startup and shutdown consist in changing deterministically the power output (from $0$ to the minimum power output $P_{u,t}^{min}$ for startup, and conversely from $P_{u,t}^{min}$ to $0$ for shutdown) in a fixed amount of time ($d_u^{SU}$ for startups and $d_u^{SU}$ for shutdowns).
    \item {\bf Reserves}: For this problem, FCR and aFRR are referred to as {\it automated} reserves, mFRR and RR as {\it manual} reserves. Additionally, if the FLAT state is defined, the unit can only provide manual reserves in the FLAT state. The FLAT state is defined when the unit needs to respect a minimum stable power duration and therefore may not be able to respond to fast variations.
\end{enumerate}

Next, we will introduce the key terms and definitions for the thermal optimal dispatch problem. 

\subsubsection{Power gradients}

The power gradients are symmetric, so:
\begin{equation}
    \Delta P^{max,+}_u = -\Delta P^{max,-}_u \equiv \Delta P^{max}_u
\end{equation}
  
and are expressed in terms of MW/time step.\footnote{By convention, if the user imputed {\tt MaximumGradient} = 0, then it means that the unit has no gradient constraint. In this case, we defined an {\it unconstrained} gradient equal to the maximum power output of the unit (with the required conversions) so that in practice, there will not be any gradient constraint for $P_{u,t} \in [P_{u,t}^{min},P_{u,t}^{max}]$. This unconstrained gradient is arbitrarily set as follows:
\[
\Delta P_{unconstrained} \coloneqq P_{u,t}^{max} 
\]
Again, this unconstrained gradient is expressed in terms of MW/ time step. We still implement a gradient constraint (formally, $\Delta P_{unconstrained} < \infty$) for computational reasons. However, in practice, as mentioned above, the gradient will not be constrained with these values.}

\subsubsection{Unit-specific Timeframes and Durations}

We  convert the durations coming from the properties of the input unit in terms of $\Delta t$ and define the following quantities :

\renewcommand{\arraystretch}{2} % Default value: 1

\begin{longtable}[!ht]{!{\color{Grey}\vrule}>{\centering\arraybackslash}p{7cm} !{\color{Grey}\vrule} p{9cm}!{\color{Grey}\vrule} }
    \arrayrulecolor{Grey} \hline
    \rowcolor{Grey} \multicolumn{2}{|c|}{\large{\textbf{Discretized time periods}}} \\ 
    \hline
    \rowcolor{Grey} 
    \textbf{Notation} & \textbf{Meaning}\\ \hline
    $\displaystyle{  T_u^{on} \coloneqq \max\left(1,
    \left\lceil\frac{d_u^{minOn} }{\Delta t} \right\rceil
    \right)}$ & The minimum number of time steps the unit has to be online \\ \hline
    $\displaystyle{ T_u^{off} \coloneqq \max\left(1,
    \left\lceil\frac{d_u^{minOff}}{\Delta t} \right\rceil
    \right)}$ & The minimum number of time steps the unit has to be offline \\ \hline
    $\displaystyle{ T_u^{SU} \coloneqq
    \left\lfloor\frac{d_u^{SU}}{\Delta t} \right\rfloor
    }$ & The number of time steps for startup \\ \hline
    $\displaystyle{ T_u^{SD} \coloneqq
    \left\lfloor\frac{\textrm{$d_u^{SU}$}}{\Delta t} \right\rfloor
    }$ & The number of time steps for shutdown \\ \hline
    %\item $\displaystyle{ T_u^{stable} \coloneqq
    %\left\lceil\frac{d_u^{minStable} \times 60}{\Delta t} \right\rceil\mathds{1}\left(T_u^{stable}\ge 2\right)
    %}$ the minimum number of time steps between two power variations or changes in the gradient expressed in terms of $\Delta t$. 
    
    $\displaystyle{ T_u^{stable} \coloneqq \left\lceil \frac{d_u^{minStable}}{\Delta t} \right\rceil}$ & The minimum number of time steps between two power variations or changes in the gradient expressed in terms of $\Delta t$\footnote{The {\tt MinimumStablePowerDuration} and its associated constraints (see Section \ref{sec:min_stable_time} for more details) are only explicitly taken in consideration if $T_u^{stable}\ge 2$ because if $T_u^{stable} = 1$ we model the fact that the unit remains stable for one time step due to the constant interpolation of the power output. For this reason, if the value calculated is strictly less than 2, $T_u^{stable}$ is set to 0.} \\ \hline
\end{longtable}

We take the ceiling for the minimum durations and the floor for the startup and shutdown times for the following reasons :
\begin{itemize}
    \item If for example the user inputs a duration {\tt MinimumStablePowerDuration} such that
    \[
    k-1 <\displaystyle{\frac{d_u^{minStable}}{\Delta t} }\le k
    \]
    we want the minimum duration to be at least $k$ time steps in order to actually take this minimum duration into account.
    \item On the other hand, for the startup and shutdown durations, if these durations were to be such that either $0\le T_u^{SU}\le 1$ or $0\le T_u^{SD}\le 1$, then we want one time step to "jump" over this startup or shutdown duration. For instance, if the time step is 60 minutes and the startup duration fifteen minutes, then we want the startup duration to be accounted for in one time step from $t$ to $t+1$, which happens in practice when $T_u^{SU} = 0$. 
    \item Finally, we cannot have by assumption $T_u^{on} = 0$ or $T_u^{off} = 0$: the minimum time on or off cannot be shorter than one time step. As such, if the user inputs {\tt MinimumTimeOn} = 0 or {\tt MinimumTimeOff} = 0, then $T_u^{on}$ or $T_u^{off}$ will be set to 1. 
\end{itemize}

\paragraph{Remark} For consistency, we require that the minimum time on for the unit always be greater or equal to the minimum stable power duration:
    \begin{equation}\label{eq:consistency_dura}
            d_u^{minStable} \le d_u^{minOn}
    \end{equation}
If \eqref{eq:consistency_dura} is not satisfied, a warning message is sent to the user and we set  $d_u^{minOn} \leftarrow d_u^{minStable}$. Due to equation \eqref{eq:consistency_dura}, we also have $T_u^{stable} \le T_u^{on}$ in practice.

% As we will see in section \ref{sec:constraints}, the definition of the {\it conditional} constraints will depend on the value of $T_{\bullet}$.

\subsubsection{Unit States}

The idea behind the modeling method is to view the unit as a set of states. Each state will then define its own constraints on the feasible power outputs. In order to improve computational performance, only necessary states are defined. Constraints relating to startups, shutdowns and stable phases are therefore only included when necessary. To begin with, we define the following six basic states for the unit:

\begin{itemize}
    \item $S_{u,t}^{OFF}$ indicating if the unit is off at $t$. The power output $P_{u,t}$ in this state is such that $P_{u,t} = 0$
    \item $S_{u,t}^{START}$ indicating if the unit is in its startup phase at $t$. The power output $P_{u,t}$ in this state is such that $0 < P_{u,t} < P_{u,t}^{min}$.
    \item $S_{u,t}^{STOP}$ indicating if the unit is in its shutdown phase at $t$. The power output $P_{u,t}$ in this state is such that $0 < P_{u,t} < P_{u,t}^{min}$.
    \item $S_{u,t}^{ON\_{UP}}$ indicating when the unit is on at $t$ and that its power output is increased during the time interval $[t,t+1[$ so that $P_{u,t+1} \ge P_{u,t}$. The power output in this state is such that $P_{u,t}^{min} \le P_{u,t} \le P_{u,t}^{max}$.
    \item $S_{u,t}^{ON\_{FLAT}}$ indicating when the unit is on at $t$ and that its power output remains constant during the time interval $[t,t+1[$, so that $P_{u,t+1} = P_{u,t}$. The power output in this state is such that $P_{u,t}^{min} \le P_{u,t} \le P_{u,t}^{max}$.
    \item $S_{u,t}^{ON\_{DOWN}}$ indicating when the unit is on at $t$ and that its power output is decreased during the time interval $[t,t+1[$, so that $P_{u,t+1} \le P_{u,t}$. The power output in this state is such that $P_{u,t}^{min} \le P_{u,t} \le P_{u,t}^{max}$.
\end{itemize}

Among the six states defined above, three will always be defined : OFF, ON\_UP and ON\_DOWN. On the other hand, the ON\_FLAT state can be ignored if $T_u^{stable}  <2$. In this case, it is still possible to have constant power output variations, because the ON\_UP and ON\_DOWN states allow for a power variation comprised between $[0,\Delta P^{max}_{u}]$ and $[- \Delta P^{max}_{u}, 0]$ respectively. But if $T_u^{stable} \ge 2$ we want to {\it require} the unit to be stable for at least one time step. Finally, START and STOP states are only defined if $T_u^{SU} \ge 1$ or $T_u^{SD} \ge 1$ respectively.\footnote{Note that the definition of $T_u^{SU}$ and $T_u^{SD}$ differs from the definition of $T_u^{on}$, $T_u^{off}$ and $T_u^{stable}$ because startup and shutdown ramps are not considered as minimum durations and we want to be able to encompass the startup or the shutdown duration within a time step. This happens if for instance if $\Delta t= 60$ minutes and $d_u^{SU}$ $= 15$ minutes. }

The overall logic behind the definition of the conditional constraints defined from Section \ref{sec:startup_constraint} to Section \ref{sec:minimum_time_off_cons} will always be the same :  if $T_{\bullet} \ge 1$ then we need to forbid additional transitions between states (e.g. force the unit to go through the START state). Moreover, if $T_{\bullet} \ge 2$, then the unit has to remain in this state for more than one time step, so we need to define an additional constraint locking the unit in this state for several time steps. Note that the constraints in this case are cumulative (only the transitions are forbidden if $T_{\bullet} \ge 1$ and if $T_{\bullet} \ge 2$, both the transitions and lock need to be taken into account). As said above, since $T_u^{on} \ge 1$ and $T_u^{off} \ge 1$ by definition, for these states we only define the "lock" constraints if we have $T_u^{on} \ge 2$ or $T_u^{off} \ge 2$.
 
An additional complexity arises when the FLAT state is defined, i.e. if $T_u^{stable}> 0$. In this case, we need to define the state variables $S_t^{ON\_UP}$, $S_t^{ON\_DOWN}$ and $S_t^{ON\_FLAT}$ not only on the time frame $T^{opt}$, but also on the time step $-1$, i.e. the last time step of the previous time frame. We do this because the state variable $S_{-1}^{ON\_\bullet}$ requires the knowledge of the power output $P_{u,t^{start}}$, which is a decision variable, hence $S_{-1}^{ON\_\bullet}$ also needs to be a decision variable. The rest of the program is unchanged, but as a consequence the auxiliary variables associated with these states as well as the gradient auxiliaries are defined on this non standard time frame. See Sections \ref{sec:delta_t_stable}, \ref{sec:delta_t_entered_up} and \ref{sec:delta_t_entered_down} for more details on the auxiliaries and \ref{sec:min_stable_time} for the gradient auxiliaries added to the program when the FLAT state is defined.

\paragraph{Remark} In this optimization program we are required to handle both time steps and time intervals. The convention is the following : {\bf time step} $t$ corresponds to the {\bf time interval} $[t,t+1[$. So for instance, if the unit is ON at $t$ and OFF at $t+1$, it means that it will be ON from date $t$ until date $t+1-\varepsilon$ (for some $\varepsilon >0)$ and that it will be OFF from $t+1$ until (at least) some $t+2-\varepsilon$. We chose this convention because the overall idea is to model the unit as a set of states from which it is possible to switch and that each state constrains the set of possible actions for the "next" time interval. 

\begin{table}[H]
    \centering
    \begin{tabular}{|c|c|c|c|c|c|c|c|c|c|c|c|c|c|c|c|c|c|c|c|c|c|c|c|c|}
    \hline
    $t$     & 1 & 2 & 3 & 4 & 5 & 6 & 7 & 8 & 9 & 10 & 11 & 12 & 13 & 14 & 15 & 16 & 17 & 18 & 19 & 20  \\
    \hline
    \hline
    $S_{u,t}^{OFF}$               & 1 & 1 & 1 & 1 & 0 & 0 & 0 & 0 & 0 & 0 & 0 & 0  & 0 & 0 & 0 & 0 & 0 & 1 & 1 & 1   \\ 
    \hline
    $S_{u,t}^{START}$             & 0 & 0 & 0 & 0 & 1 & 1 & 1 & 0 & 0 & 0 & 0 &  0 & 0 & 0 & 0 & 0 & 0 & 0 & 0 & 0   \\
    \hline
    $S_{u,t}^{STOP}$             & 0 & 0 & 0 & 0 & 0 & 0 & 0 & 0 & 0 & 0 & 0 & 0 &  0 & 0 & 1 & 1 & 1 & 0 & 0 & 0   \\ 
    \hline
        $S_t^{ON\_\bullet}$       & 0 & 0 & 0 & 0 & 0 & 0 & 0 & 1 & 1 & 1 & 1  & 1 & 1 & 1 & 0 & 0 & 0 & 0 & 0 & 0   \\ 
    \hline
    \end{tabular}
    \caption{Example of the transitions of a unit. All ON states are collapsed together.}
    \label{tab:succession_frame_example}
\end{table}

\begin{figure}[H]
    \centering
    \includegraphics[width = 12cm]{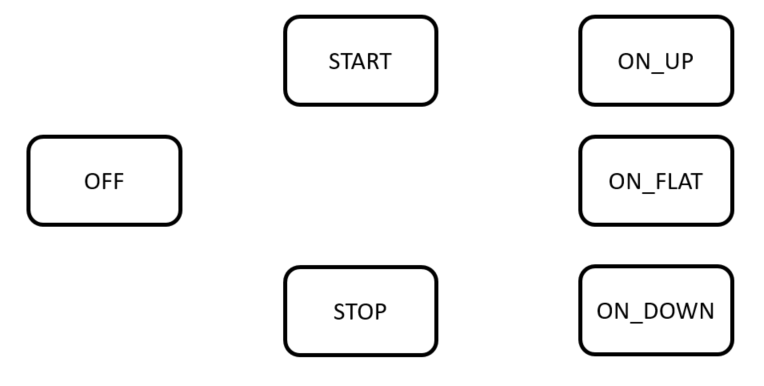}
    \caption{Graph of the possible states of the unit}
    \label{fig:states_graph}
\end{figure}

Table \ref{tab:succession_frame_example} illustrates a startup phase encoded with these binary variables.  Figure  \ref{fig:states_graph} shows the six possible states associated with this case. 

The transition between states is constrained and all states are not always defined, depending on the value of $\Delta t$ and the thermal unit's characteristics. For instance, if $\Delta t > T_u^{SU}$, then the unit can be "instantaneously" started-up, so there is no point in defining an intermediate state START.\footnote{To clarify, in this case "instantaneously" is relative to the value of $\Delta t$. In other words, the duration in question is wholly contained in one time step.}

\paragraph{Definition of the transition states START and STOP}
The states START and STOP are two deterministic states, meaning that if the unit is in this state, then its power output will be gradually increased (resp. decreased) from 0  to $P_{u,t}^{min}$ (resp. from $P_{u,t}^{min}$ to 0) in an amount of time equal to the unit's startup duration (resp. shutdown duration). In order to do so, we define $\Delta P_{u}^{step,up}$ (resp. $\Delta P_{u}^{step,dn}$) such that $\displaystyle{
\Delta P_{u}^{step,up} = \frac{P_{u,t}^{min}}{T_u^{SU} +1}
}$ (resp. $\displaystyle{
\Delta P_{u}^{step,dn} = \frac{P_{u,t}^{min}}{T_u^{SD} +1}
}$). These $\Delta P_{u}^{step,\bullet}$ parameters correspond to the power variation for each timestep during the startup or the shutdown phase.

The idea is that the startup and shutdown phases consist in deterministic increments of the power output. More precisely, during the startup phase, we have:
\[
\forall s \in\{0,\dots,T_u^{SU}-1\},\; P_{u,t+s} = (s+1) \Delta P_{u}^{step,up}
\]

\begin{figure}[H]
    \centering
    \includegraphics[width = 12cm]{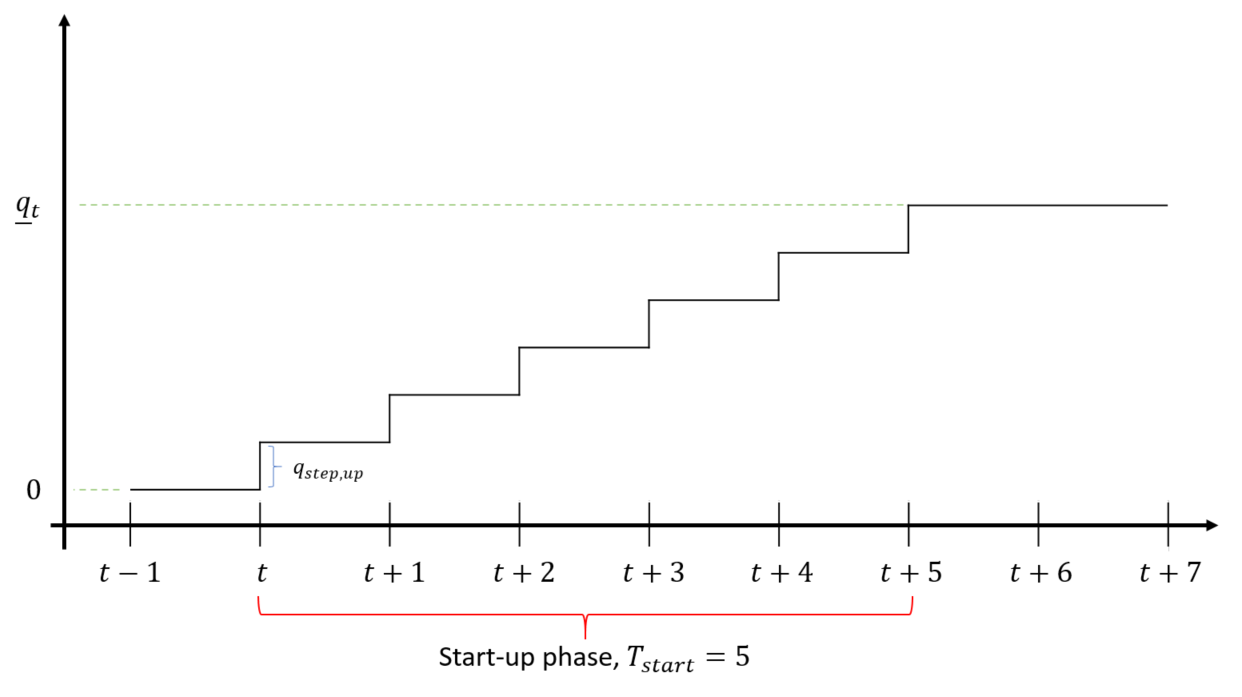}
    \caption{Shape of the startup ramp of a unit}
    \label{fig:startup_ramp}
\end{figure}

So that when the unit enters the startup phase, its power has already been increased and is equal to $\Delta P_{u}^{step,up}$ (as it can be seen in Figure \ref{fig:startup_ramp}) and when the unit leaves the state $S_{u,t}^{START}$ at time step $t + T_u^{SU}$, its power level has been raised to $P_{u,t}^{min}$.

Symmetrically for the shutdown phase, we have :
\[
\forall s \in\{0,\dots,T_u^{SD}-1\},\; P_{u,t+s} = P_{u,t}^{min}-(s+1) \Delta P_{u}^{step,dn}
\]

\paragraph{Modeled Combinations}
In order to keep the number of binary variables and constraints as low as possible, the variables and constraints will vary as all states are not necessarily defined. More precisely, the states OFF, ON\_UP and ON\_DOWN are always defined. Then, depending on the values of $T_u^{SU}$, $T_u^{SD}$ and $T_u^{stable}$ the states START, STOP and ON\_FLAT are defined, as well as their associated variables. 

As a consequence, the implementation is done by reviewing all possible state combinations and by writing the associated constraints (using only the relevant state and auxiliary variables) in those cases. We are left with eight cases:
\begin{enumerate}
    \item $T_u^{SD} = T_u^{stable} = T_u^{SU} = 0$
    \item $T_u^{SD} \ge 1, T_u^{stable} = T_u^{SU} = 0$
    \item $T_u^{SD} = 0, T_u^{stable} \ge 1, T_u^{SU} = 0$
    \item $T_u^{SD} = T_u^{stable} = 0, T_u^{SU} \ge 1$
    \item $T_u^{SD} = T_u^{stable} \ge 1,  T_u^{SU} = 0$
    \item $T_u^{SD} = 0, T_u^{stable} = T_u^{SU} \ge 1$
    \item $T_u^{SD} \ge 1, T_u^{stable} = 0, T_u^{SU} \ge 1$
    \item $T_u^{SD} = T_u^{stable} = T_u^{SU} \ge 1 $
\end{enumerate}
 
The following sections describe the variables and constraints for these different cases. 

\subsection{Variables}\label{sec:decision_variables}

There are three sets of decision variables :
\begin{itemize}
    \item The control variables :
    \begin{itemize}
        \item $P_{u,t}\in[0,P_{u,t}^{max}]$ corresponding to the power output of the unit at time $t$.
        \item Variables corresponding to the reserve level at $t$ : $R_{u,t}^{auto,up}$, $R_{u,t}^{auto,dn}$, $R_{u,t}^{manu,up}$ and $R_{u,t}^{manu,dn}$. All these variables are defined over $[0,P_{u,t}^{max}]$.
        \item {\it Mirror} variables corresponding to the amount of reserve that is {\it not} provided at $t$ :  $R_{u,t}^{unpr,up}$ and $R_{u,t}^{unpr,dn}$. All these variables are defined over $[0,P_{u,t}^{max}]$ 
        \item A {\it relaxed} reserve variable $R^{rel}$ allowing fill-up constraints to be satisfied even if the unit is not online. This variable is always defined and takes values in the $[0,P_{u,t}^{min}]$ range. 
    \end{itemize}
    \item The state variables: $S_{u,t}^{OFF}$, $S_{u,t}^{START}$, $S_{u,t}^{STOP}$, $S_t^{UP}$, $S_t^{DOWN}$, $S_{u,t}^{ON\_{FLAT}}$, $S_{u,t}^{ON\_{DOWN}}$ and $S_{u,t}^{ON\_{UP}}$ that are all binary variables $\in\{0,1\}$ and are indicating in which state the unit is at time $t$. These states are mutually exclusive, according to \eqref{eq:mutual_exclusion}.
    \item The auxiliary variables $\delta_t^{turned\_on}$, $\delta_t^{turned\_off}$, $\delta_t^{stable}$, $\delta_t^{entered\_up}$, $\delta_t^{entered\_down}$, that are binary variables as well\footnote{In practice, these variables will be equivalently defined as non-negative real-valued variables. Their defining constraints will ensure us that they will always be equal to 0 or 1. However, defining real values instead of binary speeds up the computations.}. These variables keep track of precise events occurring during the optimization. More details and specific definitions are provided in \ref{sec:aux_variables}.
    \item A final class of auxiliary variables is also defined, namely the {\it gradients} auxiliaries. These variables allow for a finer control of the gradient either when the STOP state is defined or if the FLAT state is defined. These variables are:
    \begin{itemize}
        \item $U_t$ and $D_t$ if the FLAT state is defined (see Section \ref{sec:min_stable_time}, Equations \eqref{eq:u_t_star} and \eqref{eq:d_t_star}),
        \item $DD_t$ and $\delta_t^{flat,down,stop}$ if additionally to the FLAT state, the STOP state is defined (see Section \ref{sec:shutdown_constraint}, Equations \eqref{eq:flat_down_stop} and \eqref{eq:dd_t_star}),
        \item $\delta_t^{down\_to\_stop}$ if we are in the case where the STOP state is defined but not the FLAT state (see Section \ref{sec:shutdown_constraint}, Equation \eqref{eq:down_to_stop}),
    \end{itemize}

\end{itemize}

\subsection{Constraints}\label{sec:constraints}

Let us now state all the constraints of this problem. We distinguish the constraints on the state variables (see Section \ref{sec:status_constraint}) and the constraints on the control variable (see Section \ref{sec:constraints_control}). Finally, Section \ref{sec:initial_cond} reviews the initial conditions of the problem.

The constraints on the state variables will be broken down into two parts:  {\it permanent} constraints, which represent the fact that some transitions can never occur, no matter the value of $\Delta t$ and {\it conditionnal} constraints, which will be enforced under some condition on $\Delta t$. For instance, if $\Delta t$ is smaller than the startup duration, then we need to implement a constraint regarding the startup phase. 

Before introducing these constraints, we define in Section \ref{sec:aux_variables} auxiliary variables that do not have an immediate interpretation in terms of unit status or power output but are necessary to define some constraints. 

\subsubsection{Auxiliary variables}\label{sec:aux_variables} 

The definition of the constraints on the state variables $S_t^{\bullet}$ and the control variable $P_{u,t}$ requires the introduction of auxiliary binary variables that model the fact that some terms in the constraints will be nonzero if and only if a precise condition is satisfied on the state variables. More precisely, we need up to five of these binary variables : 
\begin{itemize}
    \item A variable that will indicate whether the unit has started up on $t$. We want this variable to be equal to 1 only when the unit begins its startup phase, i.e. at $t$ such that $S_{u,t}^{OFF} = 0$ {\it and} $S_{t-1}^{OFF} = 1$. We will also use this variable to enforce the minimum time on constraint (see Section \ref{sec:minimum_time_on_cons}) once the unit is on. 
    % on fait Ã§a pour Ã©viter de dÃ©finir deux jeux de contraintes .Juste ajouter + Delta start dans la contrainte de minimum time on (et si Delta start = 0, alors tout va bien). Si on ne veut pas dÃ©finir de delta start, mettre des ifs de partout. Un peu fastidieux mais fonctionne quand mÃªme. 
    \item A variable that will indicate whether the unit has been shut down on $t$. We want this variable to be equal to 1 when the unit leaves the $ON\_\bullet$ state, i.e. at $t$ such that either $S_{u,t}^{OFF} = 1$ {\it and} $S_{t-1}^{OFF} = 0$ or $S_{u,t}^{STOP} = 1$ {\it and} $S_{t-1}^{STOP} = 0$ if the STOP state is defined.
    \item A variable that will indicate if the unit enters the stable state at $t$, i.e. equal to 1 when the event $S_{u,t}^{ON\_{FLAT}} = 1$ {\it and} $S_{t-1}^{ON\_FLAT} = 0$ occurs. 
    \item A variable that will indicate if the unit enters in the up state at $t$, i.e. equal to 1 when the event $S_{u,t}^{ON\_{UP}} = 1$ {\it and} $S_{t-1}^{ON\_UP} = 0$ occurs. 
    \item A variable that will indicate if the unit enters in the down state at $t$, i.e. equal to 1 when the event $S_{u,t}^{ON\_{DOWN}} = 1$ {\it and} $S_{t-1}^{ON\_DOWN} = 0$ occurs. 
\end{itemize}

\subsubsection{Indicator that the unit has started on $t$}\label{sec:started_on_t}

This variable indicates when the unit is turned on, i.e. formally either enters the START or a ON state (depending on whether or not the START state is defined, i.e. either $T_u^{SU}>0$ or $T_u^{SU} = 0$). Due to the mutual exclusion constraint (see Equation \ref{eq:mutual_exclusion}) and the transition constraints between states, it is equivalent to define:

\[\left\{\delta_t^{turned\_on}=1\right\}\iff \left\{S_{t-1}
^{START} = 0 \wedge S_{t}
^{START} = 1\right\}\vee\left\{S_{t-1}
^{ON\_\bullet} = 0 \wedge S_{t}
^{ON\_\bullet} = 1\right\}
\]
or:
\[\left\{\delta_t^{turned\_on}=1\right\}\iff \left\{S_{t-1}
^{OFF} = 1 \wedge S_{t}
^{OFF} = 0\right\}
\]

The advantage of the last definition being that the auxiliary variable does not need to change its definition depending on which states are defined. Therefore, we choose the latter expression for $\delta_t^{turned\_on}$, which translates into the set of linear constraints given by equation \ref{eq:started_on_t}:

\begin{equation}\label{eq:started_on_t}
\begin{aligned}
\forall t \in T^{opt},&\;\delta_t^{turned\_on}\le 1 - S_{u,t}^{OFF} \\
&\; \delta_t^{turned\_on}\le  S_{t-1}^{OFF} \\
&\; \delta_t^{turned\_on}\ge  S_{t-1}^{OFF} - S_{u,t}^{OFF} 
\end{aligned}
\end{equation}

\subsubsection{Indicator that the unit is being shut down on $t$}\label{sec:shutdown_on_t}

This variable indicates when the unit is turned off, i.e. formally either enters the STOP or OFF state (depending on whether or not the STOP state is defined, i.e. either $T_u^{SD}>0$ or $T_u^{SD} = 0$). Due to the mutual exclusion constraint (see Equation \eqref{eq:mutual_exclusion}) and the transition constraints between states, it is equivalent to define:
\[\left\{\delta_t^{turned\_off}=1\right\}\iff \left\{S_{t-1}
^{STOP} = 0 \wedge S_{t}
^{STOP} = 1\right\}\vee\left\{S_{t-1}
^{OFF} = 0 \wedge S_{t}
^{OFF} = 1\right\}
\]
Equations \eqref{eq:stopped_on_t_off} and \eqref{eq:stopped_on_t_stop} display the set of constraints that define $\delta_t^{turned\_off}$, depending on whether the STOP state is defined or not.

\begin{equation}\label{eq:stopped_on_t_off}
\begin{aligned}
\forall t \in T^{opt},&\;\delta_t^{turned\_off}\le 1 - S_{t-1}^{OFF} \\
&\; \delta_t^{turned\_off}\le  S_{t}^{OFF} \\
&\; \delta_t^{turned\_off}\ge  S_{t}^{OFF} - S_{t-1}^{OFF} 
\end{aligned}
\end{equation}

\begin{equation}\label{eq:stopped_on_t_stop}
\begin{aligned}
\forall t \in T^{opt},&\;\delta_t^{turned\_off}\le 1 - S_{t-1}^{STOP} \\
&\; \delta_t^{turned\_off}\le  S_{u,t}^{STOP} \\
&\; \delta_t^{turned\_off}\ge  S_{u,t}^{STOP} - S_{t-1}^{STOP} 
\end{aligned}
\end{equation}

\paragraph{Remark} Formally, we implicitly assume with these conventions that the unit is on when it is starting up or running and that it is off when it is being shut down or offline.

\subsubsection{Indicator that the unit is stable at $t$}\label{sec:delta_t_stable}

We denote this variable $\delta_t^{stable}$. This variable indicates whether at $t$, the unit entered the stable state, i.e. that it will provide a constant power output from $t$ until at least the next time step. The logic relationship this variable models is the following : $\delta_t^{stable} = 1 \iff S_{u,t}^{ON\_{FLAT}} = 1 \wedge S_{t-1}^{ON\_FLAT} = 0$. Equation \eqref{eq:delta_stable} defines the variable.

\begin{equation}\label{eq:delta_stable}
\begin{aligned}
\forall t \in \{-1\}\cup T^{opt},&\;\delta_t^{stable}\le 1 - S_{t-1}^{ON\_FLAT} \\
&\; \delta_t^{stable}\le  S_{u,t}^{ON\_{FLAT}} \\
&\; \delta_t^{stable}\ge  S_{u,t}^{ON\_{FLAT}} - S_{t-1}^{ON\_FLAT} 
\end{aligned}
\end{equation}

As mentioned previously, this auxiliary variable involves $S_{t}^{ON\_FLAT}$ and is therefore defined over the time frame $\{-1\}\cup T^{opt}$.

\subsubsection{Indicator that the unit entered the ON\_UP state at $t$}\label{sec:delta_t_entered_up}

We denote this variable $\delta_t^{entered\_up}$. This variable indicates whether at $t$, the unit entered the up state, i.e. that it will increase its power output from $t$ until at least the next time step. The logic relationship this variable models is the following : $\delta_t^{entered\_up} = 1 \iff S_{u,t}^{ON\_{UP}} = 1 \wedge S_{t-1}^{ON\_UP} = 0$. Equation \eqref{eq:delta_entered_up} defines the variable.

\begin{equation}\label{eq:delta_entered_up}
\begin{aligned}
\forall t \in \{-1\}\cup T^{opt},&\;\delta_t^{entered\_up}\le 1 - S_{t-1}^{ON\_UP} \\
&\; \delta_t^{entered\_up}\le  S_{u,t}^{ON\_{UP}} \\
&\; \delta_t^{entered\_up}\ge  S_{u,t}^{ON\_{UP}} - S_{t-1}^{ON\_UP} 
\end{aligned}
\end{equation}

As with $\delta_t^{stable}$, this auxiliary variable involves $S_{t}^{ON\_UP}$ when  $S_{t}^{ON\_FLAT}$ is defined and is therefore defined over the time frame $\{-1\}\cup T^{opt}$.

\subsubsection{Indicator that the unit entered the ON\_DOWN state at $t$}\label{sec:delta_t_entered_down}

We denote this variable $\delta_t^{entered\_down}$. This variable indicates whether at $t$, the unit entered the down state, i.e. that it will decrease its power output from $t$ until at least the next time step. The logic relationship this variable models is the following : $\delta_t^{entered\_down} = 1 \iff S_{u,t}^{ON\_{DOWN}} = 1 \wedge S_{t-1}^{ON\_DOWN} = 0$. Equation \eqref{eq:delta_entered_down} defines the variable.

\begin{equation}\label{eq:delta_entered_down}
\begin{aligned}
\forall t \in \{-1\}\cup T^{opt},&\;\delta_t^{entered\_down}\le 1 - S_{t-1}^{ON\_DOWN} \\
&\; \delta_t^{entered\_down}\le  S_{u,t}^{ON\_{DOWN}} \\
&\; \delta_t^{entered\_down}\ge  S_{u,t}^{ON\_{DOWN}} - S_{t-1}^{ON\_DOWN} 
\end{aligned}
\end{equation}

Again, this auxiliary variable is defined over the time frame $\{-1\}\cup T^{opt}$.

\subsubsection{Constraints between transitions}\label{sec:status_constraint}

The transitions between states are constrained. Whether these constraints are enforced or not depends on the value of the time step parameter $\Delta t$ imputed by the user. In the following sections, we review all constraints and their associated conditions if relevant. 

\begin{figure}[H]

\begin{subfigure}{0.5\textwidth}
\includegraphics[width=0.8\linewidth]{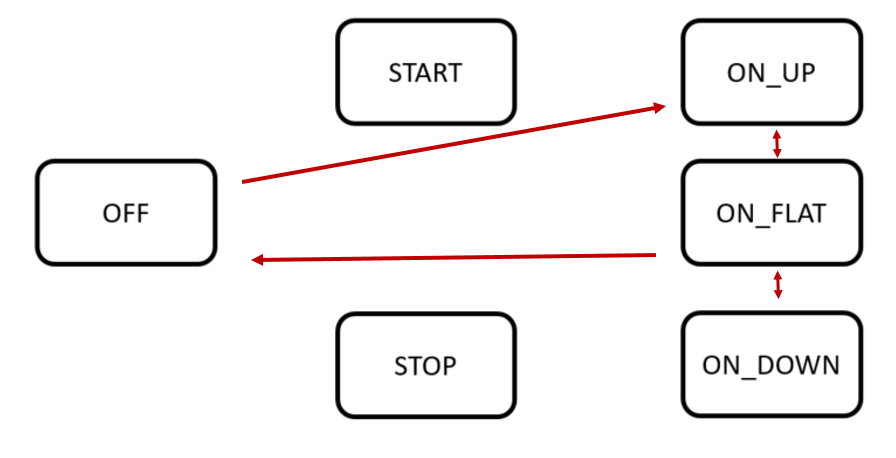} 
\caption{Less constrained case}
\label{fig:not_so_cons}
\end{subfigure}
\begin{subfigure}{0.5\textwidth}
\includegraphics[width=0.8\linewidth]{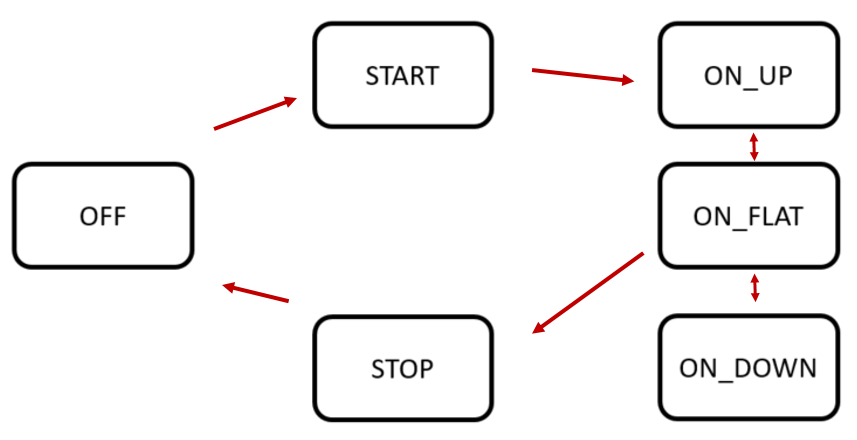}
\caption{Most constrained case}
\label{fig:very_cons}
\end{subfigure}

\caption{Two corner cases in terms of constraints}
\label{fig:different_cases}
\end{figure}

Figure \ref{fig:different_cases} illustrates the possible transitions depending on the level of constraints on these transitions. Figure \ref{fig:very_cons} illustrates the most constraining case where the time step is such that the unit will have to go through all intermediary states whereas Figure \ref{fig:not_so_cons} illustrates a situation where constraints \ref{sec:startup_constraint} and \ref{sec:shutdown_constraint} are not defined, so it is possible to "jump" from OFF to ON and vice versa. 

\subsubsection{Permanent constraints}\label{sec:perm_constraint}

\begin{figure}[H]
    \centering
    \includegraphics[width = 9cm]{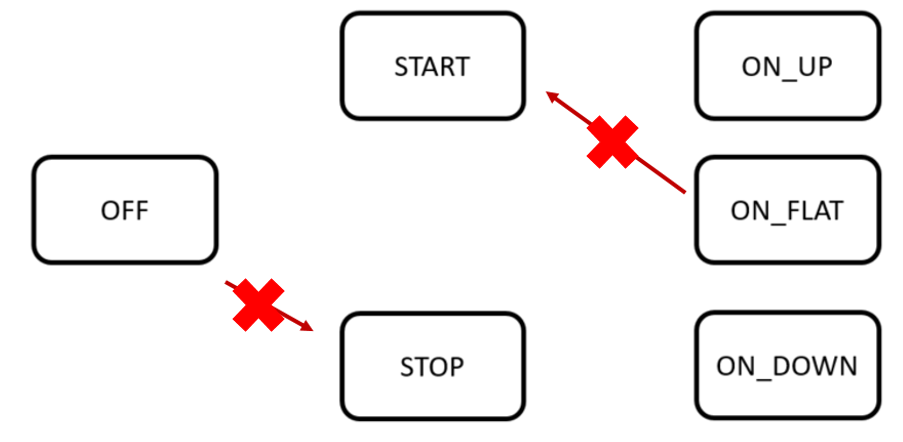}
    \caption{Illustration of two forbidden cases}
    \label{fig:forbidden_case}
\end{figure}

The constraints listed hereafter are always defined - provided that the states they rely on are defined. One such constraint is displayed in Figure \ref{fig:forbidden_case} as an example: we do not want the unit to go back from OFF to STOP or from ON to START.

\subsubsection{Mutual exclusion of all the states} The unit can only be in one state at a time. This corresponds to the implication $S_t^i = 1 \Rightarrow S_t^{-i} = 0$ for $i \in I \coloneqq \{START, STOP, OFF, ON\_UP, ON\_DOWN, ON\_FLAT\}$ and $-i\coloneqq I\backslash\{i\}$. Using the binary variables defined in Section \ref{subsec:notations_optidis}, the constraint is as follows:
\begin{equation}\label{eq:mutual_exclusion}
\forall t \in T^{opt},\; S_{u,t}^{OFF}+S_{u,t}^{START}+S_{u,t}^{STOP}+S_{u,t}^{ON\_{FLAT}}+S_{u,t}^{ON\_{DOWN}}+S_{u,t}^{ON\_{UP}} = 1
\end{equation}

\subsubsection{Transitions between ON\_UP, ON\_DOWN, ON\_FLAT and START} When the unit is online, it is never possible to switch back to the START state. This means that $ON\_FLAT_{t-1} \vee ON\_UP_{t-1} \vee ON\_FLAT_{t-1} = 1 \Rightarrow START_t = 0$. This is accounted for in Equation \eqref{eq:on_start}.

\begin{equation}\label{eq:on_start}
\begin{aligned}
\forall t \in T^{opt},\, & S_{t-1}^{ON\_FLAT} + S_t^{START } \le 1\\
& S_{t-1}^{ON\_UP} + S_t^{START } \le 1\\
& S_{t-1}^{ON\_DOWN} + S_t^{START } \le 1\\
\end{aligned}
\end{equation}

% Cette contrainte n'as plus de raison  d'être, lorsqu'on est en START on peut aller où on veut.
% \paragraph{Transitions between START and ON\_FLAT or ON\_DOWN} By convention, when the unit is started-up, the "ON" state it falls into is ON\_UP. Therefore, we do not want the unit to transition from START to ON\_FLAT or ON\_DOWN (equation \eqref{eq:start_on}). In addition, we do not want the unit to go back from one of the ON states to START  (equation \eqref{eq:on_start}). 

% Equation \eqref{eq:start_on} translates the following implications : $START_{t-1} = 1 \Rightarrow ON\_FLAT_t = 0$ and  $START_{t-1} = 1 \Rightarrow ON\_DOWN_t = 0$ while equation \eqref{eq:on_start} translates the following implications : $ON\_\bullet_{t-1} = 1 \Rightarrow START_t = 0$

% \begin{equation}\label{eq:start_on}
% \begin{aligned}
% \forall t \in T^{opt},&\; S_{t-1}^{START}+S_{u,t}^{ON\_{FLAT}}\le 1\\
% &\; S_{t-1}^{START}+S_{u,t}^{ON\_{DOWN}}\le 1
%\end{aligned}
%\end{equation}
%\begin{equation}\label{eq:on_start}
%\begin{aligned}
%\forall t \in T^{opt},&\; S_{t-1}^{ON\_UP}+S_{u,t}^{START}\le 1\\
%&\; S_{t-1}^{ON\_DOWN}+S_{u,t}^{START}\le 1\\
%&\; S_{t-1}^{ON\_FLAT}+S_{u,t}^{START}\le 1
%\end{aligned}
%\end{equation}

\subsubsection{Transitions between OFF and START and OFF and STOP} Equation \eqref{eq:start_off} translates the implication $START_{t-1} = 1 \Rightarrow OFF_t = 0$ and Equation \eqref{eq:off_stop} the implication $OFF_{t-1} = 1 \Rightarrow STOP_t = 0$
\begin{equation}\label{eq:start_off}
    \forall t \in T^{opt},\; S_{t-1}^{START} + S_{u,t}^{OFF} \le 1
\end{equation}

\begin{equation}\label{eq:off_stop}
    \forall t \in T^{opt}, \; S_{t-1}^{OFF} + S_{u,t}^{STOP} \le 1
\end{equation}

\subsubsection{Transitions between STOP and ON\_UP, ON\_DOWN and ON\_FLAT} With Equation \eqref{eq:stop_on}, we prevent the unit from going back from STOP to ON, which corresponds to the implication $STOP_{t-1} = 1 \Rightarrow ON\_\bullet_t = 0$
\begin{equation}\label{eq:stop_on}
\begin{aligned}
\forall t \in T^{opt},&\; S_{t-1}^{STOP}+S_{u,t}^{ON\_{DOWN}}\le 1\\
&\; S_{t-1}^{STOP}+S_{u,t}^{ON\_{UP}}\le 1\\
&\; S_{t-1}^{STOP}+S_{u,t}^{ON\_{FLAT}}\le 1
\end{aligned}
\end{equation}

%\paragraph{Transitions between OFF and ON\_FLAT or OFF and ON\_DOWN} Equation \eqref{eq:off_on} is analoguous to the one defined by equation \eqref{eq:start_on}. It translates the implication $OFF_{t-1}=1 \Rightarrow ON\_FLAT_t = 0$ and $OFF_{t-1}=1 \Rightarrow ON\_DOWN_t = 0$

%\begin{equation}\label{eq:off_on}
%\begin{aligned}
%\forall t \in T^{opt},&\; S_{t-1}^{OFF}+S_{u,t}^{ON\_{FLAT}}\le 1\\
%&\; S_{t-1}^{OFF}+S_{u,t}^{ON\_{DOWN}}\le 1
%\end{aligned}
%\end{equation}

\subsubsection{Transitions between START and STOP} Finally with Equation \eqref{eq:start_stop} we prevent the unit from jumping from its startup to its shut-down phase. This corresponds to the implications $START_{t-1} = 1 \Rightarrow STOP_t = 0$ and $STOP_{t-1} = 1 \Rightarrow START_t = 0$

\begin{equation}\label{eq:start_stop}
\begin{aligned}
\forall t \in T^{opt},&\; S_{t-1}^{START}+S_{u,t}^{STOP}\le 1\\
&\; S_{t-1}^{STOP}+S_{u,t}^{START}\le 1
\end{aligned}
\end{equation}

\subsubsection{Constraints enforced if and only if $T_u^{SU} > 0$}\label{sec:startup_constraint}

%% A REPRENDRE ET VÉRIVIER DANS CETTE SECTION 
%% 1 . LE LOCK SUR LE START
%% 2. POUR TOUTES LES CONTRAINTES, LES INDICES QUI FONT N'IMPORTE QUOI, VÉRIFIER QUE C'EST COHÉRENT ET NOTAMMENT QUE LE \delta NE "BOUGE PAS" QUAND T BOUGE

% question : pour start et stop, est-ce qu'il ne faut pas, en plus d'interdire la transition, bloquer les états (ce qui se fait dans ce cas si T_start > 1 et T_stop > 1). En fait le seule chose qui change par rapport aux autres T, c'est que dans ce cas il y a possibilité qu'ils soient égaux à 0 et donc non définis. 

% bilan des trucs à reprendre dans les contraintes sur les états :

% lier start et stop (?)
% vérifier les idices de partout et notamment que la dummy qui vaut 1 à un moment fixé vaut toujours un (c'est une sorte de pivot)

% ensuite faire les modifications dans les gradients (notamment défini en t+1 et t et jusqu'en T+1
% vérifier et faire tous les cas de bord

\paragraph{If  $T_u^{SU} \ge 1$} (or equivalently $\Delta t \le d_u^{SU}\times 60$) The unit has to go through the startup procedure, meaning that a new implication is to be taken into account, namely $OFF_{t-1} = 1 \Rightarrow ON\_\bullet_t = 0$. Equation \eqref{eq:off_on_cond} enforces this constraint.

\begin{equation}\label{eq:off_on_cond}
\begin{aligned}
    \textrm{If $T_u^{SU} >0$, then }\forall t \in T^{opt},\; &S_{t-1}^{OFF} + S_{u,t}^{ON\_{UP}} \le 1 \\
    & S_{t-1}^{OFF} + S_{u,t}^{ON\_{DOWN}} \le 1 \\
    & S_{t-1}^{OFF} + S_{u,t}^{ON\_{FLAT}} \le 1 
\end{aligned}
\end{equation}

Moreover, we want to impose the unit to leave the START state after $T_u^{SU}$ time steps. This so-called "eviction" constraint is necessary because if the startup gradient is greater than the regular gradient, then the unit might want to stay in the START state for an additional time step in order to benefit from this greater gradient. In order to guarantee that the unit can be started-up after a short period of time, we define this constraint using the auxiliary variable indicating that the unit was turned-on (because there will always be at least $T_u^{SU}$ time steps between two startups). The constraint is then given by Equation \eqref{eq:eviction_start}.
\begin{equation}\label{eq:eviction_start}
    \forall t\in T^{opt}\; \delta_{t-T_u^{SU}}^{turned\_on} + S_{t}^{START} \le 1
\end{equation}

\paragraph{If $T_u^{SU} \ge 2$} We must additionally lock the unit for several time steps in the START state. The constraint can expressed as the following implication :

\begin{equation}\label{eq:implication_start}
\forall t \in T^{opt}, \; \delta_t^{turned\_on} = 1 \Rightarrow\;\forall s \in \{1,\dots, T_u^{SU}-1\},\;
S_{t+s}^{START} = 1\tag{Locking in START}
\end{equation}

Where $\delta_t^{turned\_on}$ is the auxiliary variable defined in Section \ref{sec:started_on_t}. Equation \eqref{eq:implication_start} translates in the linear constraint defined in Equation \eqref{eq:minium_time_start} : 

\begin{equation}\label{eq:minium_time_start}
\begin{array}{l}
\textrm{If $T_u^{SU}\ge2$, then :}\\
\forall t \in T^{opt},\;\forall s \in \{1,\dots, T_u^{SU}-1\}\;
    \delta_{t-s}^{turned\_on} \le  S_{t}^{START} 
\end{array}
\end{equation}

\paragraph{Remark} In practice, equations are implemented "backwards" as in Equation \eqref{eq:minium_time_start} in order to avoid out-of-bounds errors on the boundary of the time frame. 

\subsubsection{Constraints enforced if and only if $T_u^{SD} > 0$}\label{sec:shutdown_constraint}

\paragraph{If $T_u^{SD} \ge 1$} We force the unit to go through the shutdown procedure when it is turned off. It means that we now have the implication $ON\_UP_{t-1} \vee ON\_FLAT_{t-1} \vee ON\_DOWN_{t-1} = 1 \Rightarrow OFF_t = 0$. Equation \eqref{eq:on_off_cond} enforces this constraint.

\begin{equation}\label{eq:on_off_cond}
\begin{aligned}
\textrm{If $T_u^{SD} > 0$, then }\forall t \in T^{opt},&\; S_{t-1}^{ON\_UP}+S_{u,t}^{OFF}\le 1\\
&\; S_{t-1}^{ON\_DOWN}+S_{u,t}^{OFF}\le 1\\
&\; S_{t-1}^{ON\_FLAT}+S_{u,t}^{OFF}\le 1
\end{aligned}
\end{equation}

Moreover, we want to impose the unit to leave the STOP state after $T_u^{SD}$ time steps. This so-called "eviction" constraint is also necessary here because the unit may remain in the STOP state for more than two time steps again in order to take advantage of a possible greater shutdown gradient. The equation implementing this constraint is very similar to Equation \eqref{eq:eviction_start} and is given by  \eqref{eq:eviction_stop}.
\begin{equation}\label{eq:eviction_stop}
    \forall t\in T^{opt}\; \delta_{t-T_u^{SD}}^{turned\_off} + S_{t}^{STOP} \le 1
\end{equation}

\paragraph{STOP-related gradient auxiliaries} These variables are needed in order to force the unit to lower its power output to $P_{u,t}^{min}$ before entering the shutdown phase. We need to distinguish two cases, depending on whether or not the FLAT state is also defined. The overall purpose of these variables is to adjust the gradient when the unit transitions to the STOP state. They are therefore added to the gradient constraints \eqref{eq:upward_grad_cons}-\eqref{eq:upward_grad_uncons} and \eqref{eq:downward_grad_cons}-\eqref{eq:downward_grad_uncons} when relevant. 

\begin{enumerate}
    \item[(i)] If the FLAT state is not defined, then we need one gradient auxiliary, namely $\delta_t^{down\_to\_stop}$, which will adjust the gradient when the unit transitions from the DOWN to the STOP state.
    \item[(ii)] If the FLAT state is defined, we need two auxiliaries, $DD_t$ if the unit's gradient was bounded by $D_t$ before entering the STOP state and $\delta_t^{flat,down,stop}$ if the unit was in the DOWN state one time step before entering the STOP state. 
\end{enumerate}

In case (i) we define the gradient auxiliary $\delta_t^{down\_to\_stop}$, which is simply the indicator that the unit is in STOP at $t$ and was in DOWN at $t-1$, i.e. 
\begin{equation}\label{eq:down_to_stop}
    \delta_t^{down\_to\_stop} = 1 \iff S_{u,t}^{STOP} = 1\wedge S_{t-1}^{ON\_DOWN} = 1 \tag{Definition of $\delta_t^{down\_to\_stop}$}
\end{equation}

This requires adding the three following constraints to the problem :

\begin{equation}\label{eq:down_to_stop_implementation}
\begin{aligned}
\forall t \in T^{opt},&\;\delta_{t}^{down\_to\_stop}\le S_{u,t}^{STOP} \\
&\; \delta_{t}^{down\_to\_stop}\le S_{t-1}^{ON\_DOWN} \\
&\; \delta_{t}^{down\_to\_stop}\ge S_{u,t}^{STOP} + S_{t-1}^{ON\_DOWN} - 1 
\end{aligned}
\end{equation}

In case (ii), we need to adjust the gradient when the unit transitions from the ON state to the STOP state. As the FLAT state is defined, we need to take into account three cases: either the unit was in the FLAT state (no adjustment to the gradient is then required) or the unit was in the DOWN state for one or two time steps (then the gradients need to be corrected). Note that the unit can never transition from the UP to the STOP state when the FLAT state is defined, but in order to make sure that this transition is impossible, we pass the transition constraint given by \eqref{eq:up_to_down_constraint}\footnote{Since this constraint can be assimilated as a gradient constraint, we do not need to define it on the time frame $\{-1\}\cup T^{opt}$} to the program.

\begin{equation}\label{eq:up_to_down_constraint}
\begin{aligned}
\textrm{If  $T_u^{stable}, T_u^{SD} \ge 1 $, then }\forall t \in T^{opt},&\; S_{t-1}^{ON\_UP}+S_{u,t}^{STOP}\le 1\\
\end{aligned}
\end{equation}

In the case where the unit transitions from the DOWN state to the STOP state, if the unit was in the DOWN state for only one time step, we define $\delta_t^{flat,down,stop}$, given by \eqref{eq:flat_down_stop}:

\begin{equation}\label{eq:flat_down_stop}
    \forall t \in T^{opt},\;\delta_t^{flat,down,stop} \& \coloneqq S_{t}^{STOP}\times S_{t-1}^{ON\_DOWN}\times S_{t-2}^{ON\_FLAT} 
    \tag{Definition of $\delta_t^{flat,down,stop}$}
\end{equation}

This auxiliary variable is implemented in practice by the set of equations given by Equation \eqref{eq:delta_fds_linear} :

\begin{equation}\label{eq:delta_fds_linear}
\begin{aligned}
\forall t \in T^{opt},&\;\delta_{t}^{flat,down,stop}\le S_{u,t}^{STOP} \\
&\; \delta_{t}^{down\_to\_stop}\le S_{t-1}^{DOWN} \\
&\; \delta_{t}^{down\_to\_stop}\le S_{t-2}^{FLAT} \\
&\; \delta_{t}^{down\_to\_stop}\ge S_{u,t}^{STOP} + S_{t-1}^{DOWN} + S_{t-2}^{ON\_FLAT} - 2 
\end{aligned}
\end{equation}

The correction of the gradient when the unit was in the DOWN state for more than one time step is implemented by the variable $D_t^*$ formally defined by the expression given in equation \eqref{eq:dd_t_star} :

\begin{equation}\label{eq:dd_t_star}
    \forall t \in \textrm{\texttt{gradientsTimeFrame}},\; DD_t^* \coloneqq S_{t+1}^{STOP}\times D_t^* = S_{t+1}^{STOP} S_{u,t}^{ON\_{DOWN}} S_{t-1}^{ON\_DOWN}\left(P_{u,t}-P_{u,t-1}\right) \tag{Definition of $DD_t^*$}
\end{equation}

Where the variable $D_t^*$ is defined in Section \ref{sec:min_stable_time}. This variable does not require initial conditions and is implemented in practice by the set of constraints given in Equation \eqref{eq:dd_t_linear}. Denoting $\underline{m}$ and $\overline{m}$ constants\footnote{in practice, we set $\underline{m}= -P_{u,t}^{max}$ and $\overline{m} = P_{u,t}^{max}$} such that $\underline{m}\le P_{u,t}-P_{u,t-1}\le \overline{m}$, we have:

\begin{equation}\label{eq:dd_t_linear}
    \begin{aligned}
    \forall t \in T^{grad},\;
    & DD_t\le \overline{m} S_{t+1}^{STOP}\\
    & DD_t\ge \underline{m} S_{t+1}^{STOP}\\
    & DD_t\le D_t -\underline{m} \left(1-S_{t+1}^{STOP}\right) \\
    & DD_t\ge D_t -\overline{m} \left(1-S_{t+1}^{STOP}\right) \\
    \end{aligned}
\end{equation}

Where the variable $D_t$ is defined in Section \ref{sec:min_stable_time}.

\paragraph{If $T_u^{SD} \ge 2$} We must additionally lock the unit for several time steps in the STOP state. 

\begin{equation}\label{eq:implication_stop}
\forall t \in T^{opt}, \; \delta_t^{turned\_off} = 1 \Rightarrow\;\forall s \in \{1,\dots, T_u^{SD}-1\},\;
S_{t+s}^{STOP} = 1\tag{Locking in STOP}
\end{equation}

Where $\delta_t^{turned\_off}$ is the auxiliary variable defined in Section \ref{sec:shutdown_on_t}. Equation \eqref{eq:implication_stop} translates in the linear constraint defined in Equation \eqref{eq:minium_time_stop} : 

\begin{equation}\label{eq:minium_time_stop}
\begin{array}{l}
\textrm{If $T_u^{SD}\ge2$, then :}\\
\forall t \in T^{opt},\;\forall s \in \{1,\dots, T_u^{SD}-1\}\;
    \delta_{t-s}^{turned\_off} \le  S_{t}^{STOP} 
\end{array}
\end{equation}

\subsubsection{Constraints enforced if and only if $T_u^{stable} > 0$}\label{sec:min_stable_time}

If we have $T_u^{stable} > 0$ (or $\Delta t < d_u^{minStable}\times 60$), we have two constraints to take into account :

\begin{itemize}
    \item Forbid the direct transitions between DOWN and UP and conversely. Equation \eqref{eq:on_up_cond} takes this constraint into account.
    \item Prevent the unit from changing its gradient from one time step to another, for example if the unit is increasing the power two time steps in a row, then the amount of the increase should remain constant. This is taken into account by defining the gradient auxiliary variables $U_t^*$ and $D_t^*$, introduced in \eqref{eq:u_t_star} and \eqref{eq:d_t_star}
\end{itemize}

\paragraph{Remark 1.} Since the unit can be stable if it doesn't adjust its power output from one time step to another, we only define this state if $T_u^{stable}\ge2$. More precisely, if {\tt MinimumStablePowerDuration} is such that $T_u^{stable} =0$ or $T_u^{stable} =1$, then the ON\_FLAT state won't be defined. 

\paragraph{Remark 2.} For technical reasons, the state variables associated with the ON states ($S_{u,t}^{ON\_{FLAT}}$, $S_{u,t}^{ON\_{UP}}$ and $S_{u,t}^{ON\_{DOWN}}$ are defined on the time frame $\{-1\}\cup T^{opt}$. As a consequence, unless explicitly specified otherwise, {\bf all} constraints involving these variables ("locking" constraints, mutual exclusion constraint, transition constraints) at time $t$ are defined on this time frame as well.

\paragraph{Forbidden transitions}  The implications of the case $\Delta t < d_u^{minStable}\times 60$ are  $ON\_DOWN_{t-1} = 1 \Rightarrow ON\_UP_{t} = 0$ and  $ON\_UP_{t-1} = 1 \Rightarrow ON\_DOWN_{t} = 0$. Equation \eqref{eq:on_up_cond} enforces these constraints. 

\begin{equation}\label{eq:on_up_cond}
\begin{aligned}
\textrm{If  $T_u^{stable} \ge 1 $, then }\forall t \in \{-1\}\cup T^{opt},&\; S_{t-1}^{ON\_UP}+S_{u,t}^{ON\_{DOWN}}\le 1\\
&\; S_{t-1}^{ON\_DOWN}+S_{u,t}^{ON\_{UP}}\le 1\\
\end{aligned}
\end{equation}

Moreover, since $T_u^{stable}\ge 2$, we want to lock the unit in this state. The implication we want to model is the following : $ON\_FLAT_t = 1\Rightarrow ON\_FLAT_{t+1} = 1 \wedge \dots \wedge ON\_FLAT_{t+T_u^{stable}-2} = 1$, which is equivalently rewritten as :
\[
ON\_FLAT_t = 1\Rightarrow \forall s\in\{1,\dots,\max(1,T_u^{stable}-2)\},\,ON\_FLAT_{t+s} = 1
\]
We can see that with this formulation, the unit remains stable at time steps $t$ and for the $T_u^{stable} - 2 - 1 + 1  = T_u^{stable} - 2 $ of the interval $\{1,\dots,\max(1,T_u^{stable}-2)\}$ so overall, the unit remains stable for $1 + T_u^{stable} - 2  = T_u^{stable} - 1$  time steps. Equation \eqref{eq:min_stable_time} translates this constraint in terms of linear conditions.

\begin{equation}\label{eq:min_stable_time}
\forall t\in T^{opt},\, \forall s\in\{1,\dots, \max(1,T_u^{stable}-2)\},\, \delta_{t-s}^{stable} \le S_{t}^{ON\_FLAT}
\end{equation}

\begin{figure}[H]
    \centering
    \includegraphics[width = 12 cm]{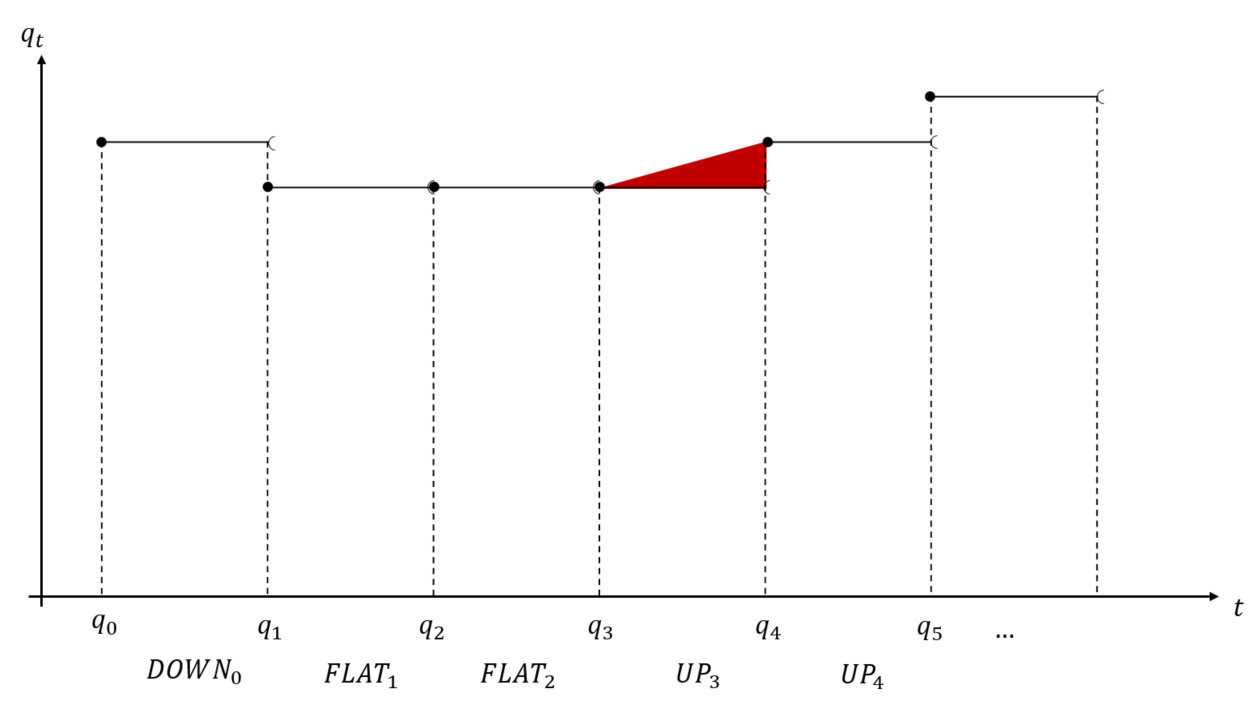}
    \caption{Excess time steps with the FLAT state}
    \label{fig:excess_flat}
\end{figure}

We choose to lock the unit for $T_u^{stable}-1$ time steps because due to the definition of the states, locking the unit in the FLAT for $T_u^{stable}$ time steps leads to locking the power output at a given level for one time step in excess, as it can be seen on Figure \ref{fig:excess_flat}, which illustrates a transition from FLAT to UP. 

In this example, we assume that $T_u^{stable} = 3$ so at $t=1$, the unit enters the FLAT state and has to remain in this state until at least $t=3$.However, at $t=2$, the definition of the FLAT state implies that $P_{u,t_{3}} = P_{u,t_{2}}$. So if we force the unit to remain in the FLAT state until $t=3$ then $P_{u,t_{4}}$ will be equal to $P_{u,t_{3}}$ and it will be "as if" the unit had been stable for 4 time steps instead of 3. However, if we only impose the constraint over $T_u^{stable}-1$ time steps, the unit can change its state and increase its power output as early as when $t=4$ as depicted in the figure. The red triangle represents, when the unit is in the UP state, the range of the power output for $t=4$ when $t=3$.

\begin{figure}[H]
    \centering
\begin{subfigure}{0.49\textwidth}
\includegraphics[width=\linewidth]{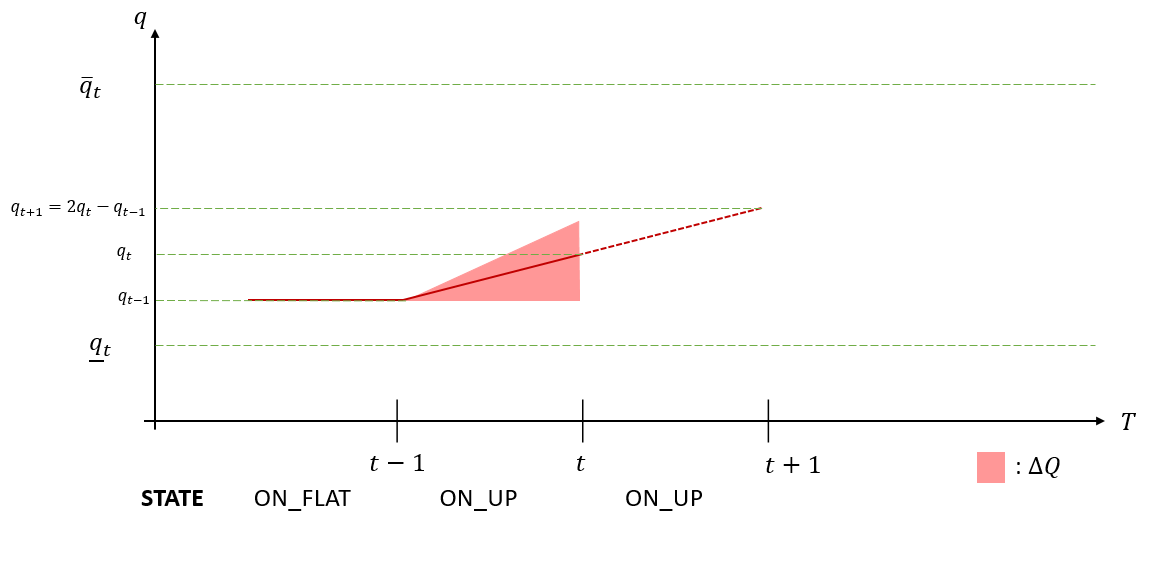}
\caption{Example of constrained gradient for the second time step in the ON\_UP state}
\label{fig:locking_grad}
\end{subfigure}
\begin{subfigure}{0.49\textwidth}
\includegraphics[width=\linewidth]{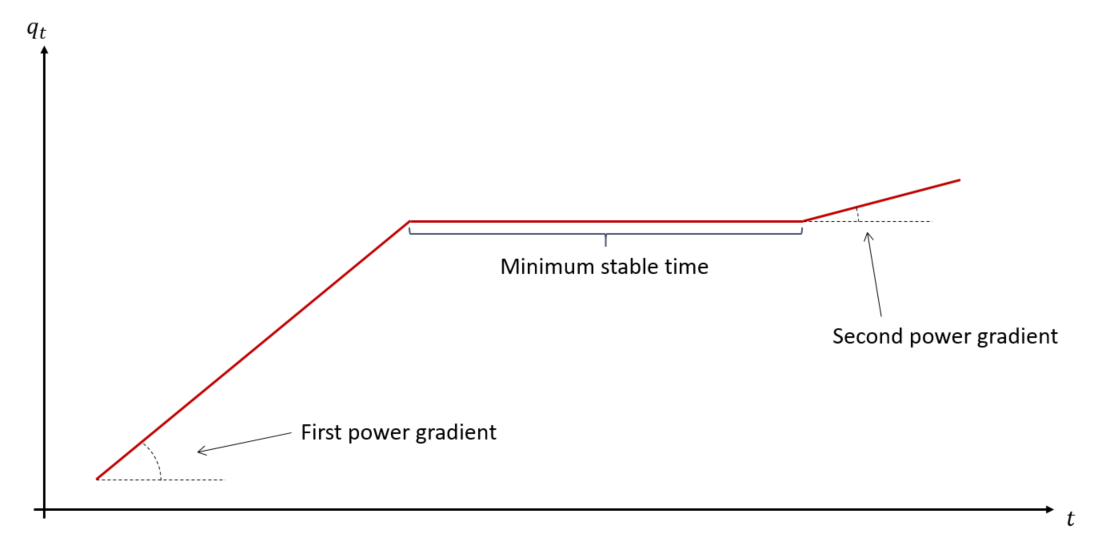}
\caption{Mandatory transition by the FLAT state}
\label{fig:broken_grad}
\end{subfigure}
    \caption{Constraints induced when taking into account the stable state}
    \label{fig:on_flat_illustration}
\end{figure}

\paragraph{Constraints on the gradient} Another constraint we want to model is the fact that if the unit has a minimum stable time constraint, it also means that its power gradient cannot vary from one time step to another. For example, if at $t=1$ the power increase is 75MW, then it cannot vary unless the unit enters the stable state, remains in this state for $T_u^{stable}$ time steps and further increases its output. A visual illustration of this constraint is provided in Figure \ref{fig:on_flat_illustration}: Figure \ref{fig:locking_grad} displays the constraint on the gradient starting at the second time step of increase and Figure \ref{fig:broken_grad} for the mandatory transition by the FLAT state if the rate of increase is to be changed.

This constraint is taken into account by introducing the gradient auxiliaries $D_t^*$ and $U_t^*$ that will be included in the gradient constraints \eqref{eq:upward_grad_cons}-\eqref{eq:upward_grad_uncons} and \eqref{eq:downward_grad_cons}-\eqref{eq:downward_grad_uncons}. 

We define 
\begin{equation}\label{eq:u_t_star}
    \forall t\in T^{opt},\; U_t^* = S_{t}^{ON\_UP}S_{t-1}^{ON\_UP}(P_{u,t}-P_{u,t-1})
    \tag{Definition of $U_t^*$}
\end{equation}
and

\begin{equation}\label{eq:d_t_star}
    \forall t\in T^{opt},\; D_t^* = S_{t}^{ON\_DOWN}S_{t-1}^{ON\_DOWN}(P_{u,t}-P_{u,t-1})
    \tag{Definition of $D_t^*$}
\end{equation}

The idea behind these variables is that if for instance $S_{t-1}^{ON\_UP}=1$ and $S_{t}^{ON\_UP}=1$, then the power gradient is bounded upward and downward by $P_{u,t}-P_{u,t-1}$, which implies  $P_{u,t+1} -P_{u,t} = P_{u,t} - P_{u,t-1} $, i.e. a conservation of the slope of the upward gradient, and symmetrically for the downward gradient. 

It is not possible to directly plug $D_t^*$ and $U_t^*$ in the gradient because these variables involve computing the product of three variables at once. In order to circumvent this issue, we introduce $\tilde{U}_t$, $\tilde{D}_t$ defined as:

\[
\begin{aligned}
\forall t \in T^{opt}, \; & \tilde{U}_t = S_{t-1}^{ON\_UP}(P_{u,t}-P_{u,t-1})\\
& \tilde{D}_t = S_{t-1}^{ON\_DOWN}(P_{u,t}-P_{u,t-1})\\
\end{aligned}
\]
We then define $U_t$ and $D_t$ as follows : 
\[
\forall t \in T^{opt}, \;
U_t = 
\tilde{U}_t S_{t}^{ON\_UP}
\]

\[
\forall t \in T^{opt}, \;
D_t = 
\tilde{D}_t S_{t}^{ON\_DOWN}
\]
So we have that $\forall t \in T^{opt}, \; U^*_t = U_t\textrm{ and }D_t^* = D_t$ but $D_t$ and $U_t$ are computationnally tractable. Indeed, the constraints \eqref{eq:contraint_u} and \eqref{eq:contraint_d} are needed in order to define $U_t$ and $D_t$ using a set of linear constraints. Denoting $\underline{m}$ and $\overline{m}$ as in Section \ref{sec:shutdown_constraint}, we have:

\begin{equation}\label{eq:contraint_u}
\begin{aligned}
\forall t \in T^{opt},\;
& U_t\le \overline{m} S_{u,t}^{ON\_{UP}}\\
& U_t\ge \underline{m} S_{u,t}^{ON\_{UP}}\\
& U_t\le \tilde{U}_t -\underline{m} \left(1-S_{u,t}^{ON\_{UP}}\right) \\
& U_t\ge \tilde{U}_t -\overline{m} \left(1-S_{u,t}^{ON\_{UP}}\right) \\
\end{aligned}
\end{equation}

\begin{equation}\label{eq:contraint_d}
    \begin{aligned}
\forall t \in T^{opt}\;
& D_t\le \overline{m} S_{u,t}^{ON\_{DOWN}}\\
& D_t\ge \underline{m} S_{u,t}^{ON\_{DOWN}}\\
& D_t\le \tilde{D}_t -\underline{m} \left(1-S_{u,t}^{ON\_{DOWN}}\right) \\
& D_t\ge \tilde{D}_t -\overline{m} \left(1-S_{u,t}^{ON\_{DOWN}}\right) \\
\end{aligned}
\end{equation}

As for $\tilde{U}_t$ and $\tilde{D}_t$, the constraints \eqref{eq:constraint_tilde_u} and \eqref{eq:constraint_tilde_d} are needed. Again, denoting $\underline{m}$ and $\overline{m}$ as in \eqref{eq:contraint_u} and \eqref{eq:contraint_d}, we have:

\begin{equation}\label{eq:constraint_tilde_u}
    \begin{aligned}
\forall t \in T^{opt},\;
& \tilde{U_t}\le \overline{m} S_{t-1}^{ON\_UP}\\
& \tilde{U_t}\ge \underline{m} S_{t-1}^{ON\_UP}\\
& \tilde{U_t}\le P_{u,t} - P_{u,t-1} -\underline{m} \left(1-S_{t-1}^{ON\_UP}\right) \\
& \tilde{U_t}\ge P_{u,t} - P_{u,t-1} -\overline{m} \left(1-S_{t-1}^{ON\_UP}\right) \\
\end{aligned}
\end{equation}

\par Moreover, if the STOP state is also defined, the gradients are also driven by the gradient auxiliaries $DD_t$ and $\delta_t^{flat,down,stop}$.

\begin{equation}\label{eq:constraint_tilde_d}
        \begin{aligned}
\forall t \in T^{opt},\;
& \tilde{D_t}\le \overline{m} S_{t-1}^{ON\_DOWN}\\
& \tilde{D_t}\ge \underline{m} S_{t-1}^{ON\_DOWN}\\
& \tilde{D_t}\le P_{u,t} - P_{u,t-1} -\underline{m} \left(1-S_{t-1}^{ON\_DOWN}\right) \\
& \tilde{D_t}\ge P_{u,t} - P_{u,t-1} -\overline{m} \left(1-S_{t-1}^{ON\_DOWN}\right) \\
\end{aligned}
\end{equation}

%\paragraph{Remark} The gradient auxiliaries involve the state variables $S_{u,t}^{ON\_{UP}}$ and $S_{u,t}^{ON\_{DOWN}}$. However, these variables are defined over the regular time frame only (since their goal is to bound the control variable $P_{u,t}$). Moreover, we need an initial condition $U_{-1}$  and $D_{-1}$ (so that the whole T^{grad} is covered). These initial conditions, as they involve a control variables, are not scalars like the other initial conditions but a constraint taking the form of an equality.

\subsubsection{Constraints enforced if and only if $T_u^{on} > 1$}\label{sec:minimum_time_on_cons}

 If $T_u^{on} > 1$, we want the unit to remain online for at least $T_u^{on}$ time steps. The implication can be written as follows, using the auxiliary variable $\delta_t^{turned\_on}$  : 
\begin{equation}\label{eq:linear_cond_525}
\forall t \in T^{opt}, \; \delta_t^{turned\_on} = 1 \Rightarrow\;\forall s \in \{1,\dots, T_u^{on}-1\},\;
S_{t+T_u^{SU}+s}^{ON\_UP}\vee S_{t+T_u^{SU}+s}^{ON\_FLAT}\vee S_{t+T_u^{SU}+s}^{ON\_DOWN} = 1\tag{Minimum time ON}
\end{equation}

We can make three remarks : 
\begin{itemize}
    \item First, we skip all time steps between $t$ and $t+T_u^{SU}$. We do this because $\delta_t^{turned\_on}$ indicates when the unit leaves the OFF state and we assume that the {\tt MinmiumTimeOn} duration excludes both the startup and the shutdown durations. Doing this avoids defining an additional variable indicating when the unit is in the ON status (i.e. has left the startup phase). 
    \item Second, we consider the interval up until $t+T_u^{on}-1$ because due to our convention, the unit will be on from $t$ up until $t+T_u^{on}-1$, covering a period of $t+T_u^{on}-1-t+1 =T_u^{on} $ time steps.
    %\item This implication can be equivalently rewritten using a sum, as in the following expression : 
%\[
%ON\_\bullet_t - ON\_\bullet_{t-1} = 1 \Rightarrow 
%\sum_{s=1}^{T_u^{on}-1}(ON\_UP_{t+s}+ON\_DOWN_{t+s}+ON\_FLAT_{t+s})\ge T_u^{on}
%\]
\end{itemize}
 The constraint \eqref{eq:linear_cond_525} is accounted for in Equation \eqref{eq:minimum_time_on_cond}.
\begin{equation}\label{eq:minimum_time_on_cond}
\begin{array}{l}
\textrm{If $T_u^{on}\ge2$, then :}\\
\forall t \in T^{opt},\;\forall s \in \{1,\dots, T_u^{on}-1\}\;
    \delta_{t-s-T_u^{SU}}^{turned\_on} \le  S_{t}^{ON\_FLAT} + S_{t}^{ON\_DOWN} + S_{t}^{ON\_UP}    
\end{array}
\end{equation}

\paragraph{Remark} Between expressions \eqref{eq:linear_cond_525} and \eqref{eq:minimum_time_on_cond} we shifted the time indices. The resulting constraint is strictly equivalent but shifting the time indices prevents from defining time steps {\it after} the last time step $T$ whereas all necessary time steps before $t=0$ exist by construction of $T^{prev}$.

\subsubsection{Constraints enforced if and only if $T_u^{off} > 1$}\label{sec:minimum_time_off_cons}

Finally if the time step is smaller than the minimum time offline, we need to take into account the implication $S_{u,t}^{OFF} - S_{t-1}^{OFF} = 1 \Rightarrow \forall s\in\{1,\dots,T_u^{off}-1\},\, S_{t+s}^{OFF} = 1$, which can be written using the auxiliary variable $\delta_t^{turned\_off}$ : 
\begin{equation}\label{eq:minimum_time_off_implication}
\forall t \in T^{opt}, \; \forall s\in\{1,\dots, T_u^{off}-1\}\;
\delta_t^{turned\_off} = 1 \Rightarrow S_{t+s}^{OFF} = 1\tag{Minimum time OFF}
\end{equation}

The implication \eqref{eq:minimum_time_off_implication} is enforced by Equation \eqref{eq:minimum_time_off_cond}.

\begin{equation}\label{eq:minimum_time_off_cond}
\begin{array}{l}
\textrm{If $T_u^{off}\ge2$, then }\\
\forall t \in T^{opt},\; \forall s\in\{1,\dots, T_u^{off}-1\}\;
\delta_{t-s-T_u^{SD}}^{turned\_off} \le S_{t}^{OFF}   
\end{array}
\end{equation}

\subsubsection{Constraints on the control variable} \label{sec:constraints_control}

The constraints reviewed in Section \ref{sec:status_constraint} allow us to control the state of the unit and its transitions from one state to another, no matter the value of $\Delta t$. Yet, each state comes with its precise constraints for the control variable $P_{u,t}$:
\begin{itemize}
    \item If the unit is on at $t$, we want the power output $P_{u,t}$ to lie within $[P_{u,t}^{min},P_{u,t}^{max}]$ and the power variation $P_{u,t+1} - P_{u,t}$ between two time steps to be upper bounded by the upward gradient $\Delta P^{max}_{u}$ and lower bounded by the downward gradient $-\Delta P^{max}_{u}$ (at most, and provided that there is a gradient constraint, otherwise unconstrained bounds are implemented.
    \item If the unit is off, at $t$, then the power output and variation should both be equal to 0.
    \item If the unit is in its startup or shutdown phase, then we want the power output to be deterministically increased or decreased, according to the example of Figure \ref{fig:startup_ramp}.
\end{itemize}

Since we model the constraints on the transitions through the state variables, there are only 2 sets of constraints on the power output: the first set deals with upper and lower bounds of the power output and is clarified in Section \ref{sec:power_output}. The second set corresponds to the upward and downward gradients, given in Section \ref{sec:power_grad}. We then review the constraints on the reserve requirements (Section \ref{sec:reserve_constraints}) and on the energy limits (\ref{sec:maximum_energy}).

\subsubsection{Constraints on the power output}\label{sec:power_output}

\paragraph{Lower bound}\label{sec:upper_bond}
\begin{equation}\label{eq:upper_bound}
\begin{aligned}
    \forall t \in T^{opt},\; P_{u,t} &\ge\displaystyle{ P_{u,t}^{min}\left(S_{u,t}^{ON\_{FLAT}}+S_{u,t}^{ON\_{UP}}+S_{u,t}^{ON\_{DOWN}}\right)} \\
    & \ge \displaystyle{+ \delta_t^{turned\_off}(P_{u,t}^{min} - \Delta P_{u}^{step,dn})}
\end{aligned}
\end{equation}

\paragraph{Upper bound}\label{sec:lower_bond}
\begin{equation}\label{eq:lower_bound}
\begin{aligned}
        \forall t \in T^{opt},\; P_{u,t} &\le \displaystyle{P_{u,t}^{max}\left(S_{u,t}^{ON\_{FLAT}}+S_{u,t}^{ON\_{UP}}+S_{u,t}^{ON\_{DOWN}}\right) }\\
    &\displaystyle{+S_{u,t}^{STOP}P_{u,t}^{min} - \delta_t^{turned\_off}\Delta P_{u}^{step,dn}
    } \\
    &\displaystyle{+ S_{u,t}^{START} P_{u,t}^{min}}
\end{aligned}
\end{equation}

The bounds given by \eqref{eq:upper_bound} and \eqref{eq:lower_bound} for the startups and the shutdowns are 0 and $P_{u,t}^{min}$ respectively. These are loose bounds, yet the unit cannot freely adjust its power output within this bandwidth because the gradient constraints are such that the power variations are deterministic, thus recreating the piece-wise constant function depicted in Figure \ref{fig:startup_ramp}.

$\delta_t^{turned\_off}(P_{u,t}^{min} - \Delta P_{u}^{step,dn})$ is added to the lower bound and $-\delta_t^{turned\_off}$ to the upper bound only if the STOP state is defined, in order to control the power output of the unit when the latter is turned off. This is intended to avoid situations where, due to the fact that the shutdown ramp is incomplete the unit will abide by its gradient but provide an incorrect power output, thus skipping a time step at $P_{u,t}^{min}$ and being turned off a time step too early.

\subsubsection{Constraints on the power gradient}\label{sec:power_grad}

\paragraph{Overall principles}

The constraints on the gradient bound the variation of the power output within a time interval $[t,t+1[$. The main ideas are the following : 
\begin{itemize}
\item When the unit is in the ON\_UP state, the variation  must lie in the $[0,\Delta P^{max}_{u}]$ range, if the unit remains in the ON\_UP state, then the power increase is constant,
\item When the unit is in the ON\_DOWN state, the variation must lie in the $[-\Delta P^{max}_{u},0]$ range, if the unit remains in the ON\_DOWN state, then the power decrease is constant ,
\item When the unit is in the ON\_FLAT state, the power variation is 0
\item When the unit is in the START stage, the variation is deterministic: $\Delta P_{u}^{step,up} \le P_{u,t} \le \Delta P_{u}^{step,up}$
\item When the unit is in the STOP stage, the variation is deterministic: $-\Delta P_{u}^{step,dn} \le P_{u,t} \le -\Delta P_{u}^{step,dn}$
\item When the unit is OFF, the variation is 0
\end{itemize}

With the auxiliary variables $\delta_t^{turned\_on}$ and $\delta_t^{turned\_off}$ we allow for a relaxation of the gradient when the unit is started up or shut down (this allows the unit to vary from 0 to $P_{u,t}^{min}$  or to reach 0 when the output is initially in the range $[P_{u,t}^{min},P_{u,t}^{max}]$ in the absence of startup or shut down ramp. 

We also accommodate for the situation where the power output is not constrained (which is done in practice by implementing a relaxed gradient. 

Finally, the gradients are defined over the $T^{grad}$, which is simply the time frame $T^{opt}$ shifted of one time index, i.e. $T^{grad}$ $ \coloneqq \{-1\}\cup T^{opt}\backslash \{T\}$

\paragraph{Corner cases} Let us illustrate one of the problems that may arise for the gradient when the unit is transitioning from two states. 

Suppose that the unit has no stable state and only features a shutdown ramp. In this case, the naive way of writing the gradient is :
\[
\begin{aligned}
- \delta_{t+1}^{turned\_off} \Delta P_{u}^{step,dn} - S_{u,t}^{STOP} \Delta P_{u}^{step,dn} - \Delta P^{max}_u S_{u,t}^{ON\_{DOWN}} 
\le P_{u,t+1}-P_{u,t} \le& 
\Delta P^{max}_u S_{u,t}^{ON\_{UP}} \\
&- \delta_{t+1}^{turned\_off} \Delta P_{u}^{step,dn}\\
& - S_{u,t}^{STOP} \Delta P_{u}^{step,dn}
\end{aligned}
\]
But now if we assume that the unit is in the DOWN state at $t$ and is to be turned off at $t+1$ (i.e. the unit will be in the STOP state at $t+1$), the latter expression boils down to :
\[
- \Delta P_{u}^{step,dn} - \Delta P^{max}_u 
\le P_{u,t+1}-P_{u,t} \le 
- \Delta P_{u}^{step,dn} 
\]
And we can see that on the left-hand side of this expression, since $\Delta P^{max}_u \ge 0$, we allow for a greater gradient than $\Delta P_{u}^{step,dn}$ for the first step of the shutdown procedure. In practice, amounts to violating the gradient when the unit is online. In order to accommodate for this situation, the gradient should therefore be written as follows: 

\begin{multline}\label{eq:corrected_grad}
- \delta_{t+1}^{turned\_off} \Delta P_{u}^{step,dn} + \delta_{t+1}^{down\_to\_stop} \Delta P^{max}_u \\  - S_{u,t}^{STOP} \Delta P_{u}^{step,dn} - \Delta P^{max}_u S_{u,t}^{ON\_{DOWN}} 
\le P_{u,t+1}-P_{u,t} \le
\Delta P^{max}_u S_{u,t}^{ON\_{UP}} \\
- \delta_{t+1}^{turned\_off} \Delta P_{u}^{step,dn}
 - S_{u,t}^{STOP} \Delta P_{u}^{step,dn}
\tag{Corrected Gradient}
\end{multline}

And in this situation if we are in the DOWN-STOP transition, the gradient boils down to :
\[
-  \Delta P_{u}^{step,dn}  \le P_{u,t+1}-P_{u,t} \le  -  \Delta P_{u}^{step,dn} 
\]
and the gradient constraint when the unit is online is no longer violated. This is why we've introduced in Section \ref{sec:shutdown_constraint} the gradient auxiliary $\delta_t^{down\_to\_stop}$ (as well as the auxiliaries $\delta_t^{flat,down,stop}$ and $DD_t^*$).

\paragraph{Upward gradient}\label{sec:upward_grad}

% Dans les gradients, 
The gradient should be accounted for if and only if $\Delta P^{max}_{u}>0$, so that the  upward gradient constraint is either given by :

\begin{equation}\label{eq:upward_grad_cons}
\begin{aligned}
\forall t \in T^{grad}, \; P_{u,t+1} - P_{u,t} \le&  
\displaystyle{\Delta P^{max}_u \delta_t^{entered\_up}}+ U_{t} + D_{t} - DD_t \\
&\displaystyle{+ \delta_{t+1}^{turned\_on}\Delta P_{u}^{step,up}+S_{t}^{START} \Delta P_{u}^{step,up}} \\
&\displaystyle{-\delta_{t+1}^{turned\_off}\Delta P_{u}^{step,dn} - S_{t}^{STOP} \Delta P_{u}^{step,dn}}\\
&\displaystyle{+\delta_{t+1}^{turned\_on}\Delta P_{unconstrained}}
\end{aligned}
\end{equation}
If $\Delta P^{max}_u > 0$. The last line, $\displaystyle{\delta_{t+1}^{turned\_on} \Delta P_{unconstrained}}$ replaces the second one if and only if the START state is not defined. Moreover, if the FLAT state is not defined, then the first line simply boils down to $\Delta P^{max}_u S_{u,t}^{ON\_{UP}}$. Otherwise, if $\Delta P^{max}_u = 0$ then the constraint is given by : 
\begin{equation}\label{eq:upward_grad_uncons}
\begin{aligned}
\forall t \in T^{grad}, \; P_{u,t+1} - P_{u,t} \le&  
\displaystyle{\Delta P_{unconstrained} \delta_t^{entered\_up}}+ U_{t} + D_{t} - DD_t \\
&\displaystyle{+ \delta_{t+1}^{turned\_on}\Delta P_{u}^{step,up}+S_{t}^{START} \Delta P_{u}^{step,up}} \\
&\displaystyle{-\delta_{t+1}^{turned\_off}\Delta P_{u}^{step,dn} - S_{t}^{STOP} \Delta P_{u}^{step,dn}}\\
&\displaystyle{+\delta_{t+1}^{turned\_on}\Delta P_{unconstrained}}
\end{aligned}
\end{equation}

\paragraph{Downward gradient}\label{sec:downward_grad}
The constraint is the symmetric of Equations \eqref{eq:upward_grad_cons} and  \eqref{eq:upward_grad_uncons}, depending on whether or not $\Delta P^{max}_u > 0$ :

\begin{equation}\label{eq:downward_grad_cons}
\begin{aligned}
    \forall t \in T^{grad}, \; P_{u,t+1} - P_{u,t} \ge &\displaystyle{- \Delta P^{max}_u \delta_t^{entered\_down}+U_{t}-D_{t} - DD_t + \left(\delta_{t+1}^{flat,down,stop} + \delta_{t+1}^{down\_to\_stop}\right) \Delta P^{max}_u }\\
&\displaystyle{+ \delta_{t+1}^{turned\_on}\Delta P_{u}^{step,up}+S_{t}^{START} \Delta P_{u}^{step,up}} \\
&\displaystyle{-\delta_{t+1}^{turned\_off}\Delta P_{u}^{step,dn}  + \delta_{t+1}^{down\_to\_stop}\Delta P^{max}_u- S_{t}^{STOP} \Delta P_{u}^{step,dn}}\\
&\displaystyle{-\delta_{t+1}^{turned\_off} \Delta P_{unconstrained}}
\end{aligned}
\end{equation}
Where the last line replaces the third if and only if the STOP state is not defined. Moreover, if the FLAT state is not defined, then the first line simply boils down to $-\Delta P^{max}_u S_{u,t}^{ON\_{DOWN}}$. Otherwise, if $\Delta P^{max}_u = 0$, then the unconstrained condition given by \eqref{eq:downward_grad_uncons} is enforced : 
\begin{equation}\label{eq:downward_grad_uncons}
\begin{aligned}
    \forall t \in T^{grad}, \; P_{u,t+1} - P_{u,t} \ge &\displaystyle{- \Delta P_{unconstrained} \delta_t^{entered\_down}+U_{t}-D_{t} - DD_t} \\
    & \displaystyle{+ \left(\delta_{t+1}^{flat,down,stop} + \delta_{t+1}^{down\_to\_stop}\right) \Delta P_{unconstrained} }\\
&\displaystyle{+ \delta_{t+1}^{turned\_on}\Delta P_{u}^{step,up}+S_{t}^{START} \Delta P_{u}^{step,up}} \\
&\displaystyle{-\delta_{t+1}^{turned\_off}\Delta P_{u}^{step,dn}  + \delta_{t+1}^{down\_to\_stop}\Delta P^{max}_{u}- S_{t}^{STOP} \Delta P_{u}^{step,dn}}\\
&\displaystyle{-\delta_{t+1}^{turned\_off} \Delta P_{unconstrained}}
\end{aligned}
\end{equation}
\paragraph{Remarks :} \begin{itemize}
    \item In \eqref{eq:downward_grad_cons} and \eqref{eq:downward_grad_uncons}, we add the term $\displaystyle{-\delta_{t+1}^{turned\_off} \Delta P_{unconstrained}}$ in order to relax the downward gradient when the unit is shut down. If the unit has no shutdown ramp, then in the worst case it could be at $P_{u,t}=P_{u,t}^{max}$ at $t$ and shut down at $t+1$ (i.e. $P_{u,t+1} = 0$). In order to allow for such a jump, we need to relax the downward gradient when the unit is to be shut down. The unconstrained gradient is large enough to accommodate for such a case. We do not need to include this term in \eqref{eq:upward_grad_cons} and \eqref{eq:upward_grad_uncons} because obviously, when the unit is to be shut down, no relaxation on the upward gradient is necessary. 
    The term $\delta_{t+1}^{turned\_on} \Delta P_{unconstrained}$ is added in \eqref{eq:upward_grad_cons} and \eqref{eq:upward_grad_uncons} for analogous reasons. These terms however are only included if $T_u^{SD} =0$ or $T_u^{SU} = 0$.
\item If the unit has a shutdown ramp, we require that on the last time step before entering the shutdown phase the unit's power output is equal to $P_{u,t}^{min}$. When no FLAT state is defined, the gradient when the unit is to enter the STOP state is therefore given by $\Delta P_{u}^{step,dn}-\Delta P^{max}_{u}$. The term $\Delta P^{max}_{u}$ is added because if the unit is to be turned off at $t+1$, it means that it is in the DOWN state at $t$ and therefore the gradient is given by $\Delta P^{max}_{u}$. As such, we ensure that the unit will lower its power output to $P_{u,t}^{min}$ {\it before} entering the shutdown phase. For similar reasons, the corrections $DD_t$ and $\delta_t^{flat,down,stop}$ are introduced.
\item Finally, the terms $U_{t}$ and $D_{t}$ bound the gradient if the unit is to be in the ON\_UP (resp. ON\_DOWN) state at the next time step. These variables have been introduced and defined in Section \ref{sec:min_stable_time}. If no ON\_FLAT state is defined, the first line of the gradient reads $ - S_{u,t}^{ON\_{DOWN}} \Delta P^{max}_u \le P_{u,t+1}-P_{u,t} \le S_{u,t}^{ON\_{UP}} \Delta P^{max}_u $ instead of $-  \delta_t^{turned\_off} \Delta P^{max}_u \le P_{u,t+1}-P_{u,t} \le \delta_t^{turned\_on} \Delta P^{max}_u $
\end{itemize}

\subsubsection{Constraints on the maximum energy}\label{sec:maximum_energy}

The last constraint we want to take into account is the maximum energy level. This constraint is useful when thermal units are used for modeling load curtailments. Each value of time series $E_{u}^{max,daily}$ corresponds to the maximum daily energy for a given day. The idea is to retrieve the maximum daily energy for all the days over which the optimization program is computed and for each day, to derive an energy upper bound simply by scaling the value {\tt MaximumDailyEnergy}$(d)$ of day $d$ by the number of hours the optimization program "spends" in this day $d$". Put otherwise, if {\tt MaximumDailyEnergy}$(d) = 1000$ and the optimization program spans over 12 hours of day $d$, then the energy limit will be 500MW. 

We denote the number of days over which the optimization program spans $D$ and define the set of days $\Delta D \coloneqq \{1,.\dots,D\}$. We then split the interval $T^{opt}$ over these days : $T^{opt} = \{0,\dots, T_1, \dots T_d,\dots T_{D-1},\dots T\}$ and denote $T^{opt}_d = \{T_{d-1},\dots,T_d\}$ for $d\in \Delta D$ with $T_0 = 0$ and $T_{D} = T$ 

For each day of the optimization program, the power output is bounded as in Equation \eqref{eq:max_daily_energy}:

\begin{equation}\label{eq:max_daily_energy}
    \begin{array}{l}
    \textrm{If {\tt HasDailyEnergyConstraint},\, then $\forall d\in\{1,\dots D\}$} \\
    \displaystyle{\sum_{t\in T^{opt}_d}P_{u,t} \le \sum_{d=1}^{D}\textrm{{\tt MaximumDailyEnergy}}(d)\frac{\Delta t}{1440}}\vert T^{opt}_d\vert
    \end{array}
\end{equation}
Where $ \vert T^{opt}_d\vert$ denotes the cardinality of the set $T^{opt}_d$, $\displaystyle{\frac{\Delta t}{1440}\vert T^{opt}_d\vert}$ is a converting factor that expresses the maximum amount of energy of day $d$ {\tt MaximumDailyEnergy}$(d)$ for the corresponding period expressed in time steps $\Delta t$

\subsubsection{Initial conditions}\label{sec:initial_cond}
% Algorithmes d'initialisation à préciser.

Initial conditions have to be defined on the control, state and auxiliary variables. The gradient auxiliaries $DD_t$, $D_t$ and $U_t$  have their own initialization. To do so, we retrieve the latest known values of the {\tt Power} time series, which contains the realized power outcomes of the unit for the previous periods. 

Retrieving the values of $P_{u,t}$ from the {\tt Power} time series will allow us to deduce the values of the state and auxiliary variables. This is however only required if the program is not solved for the first time. 

In the following, we consider two main cases: either the program that is about to be solved is the first one (i.e. there is no initial value to define, see Section \ref{sec:init_prog}) or the program comes after another optimization program and in this case we will need to define non-arbitrary initial values (Section \ref{sec:not_init})

\paragraph{Initial conditions if the program is "day zero"}\label{sec:init_prog}

This situation happens when the program is run for the first time or if the last time index of the {\tt Power} time series is earlier than {\tt StartDate - AddHours($T^{traceback}$)} where {\tt AddHours} is a C\# method for handling dates.

In this case, we initialize the variables as follows\footnote{Recall that if the FLAT state is defined, state variables $S_{u,t}^{ON\_{UP}}$, $S_{u,t}^{ON\_{DOWN}}$ and $S_{u,t}^{ON\_{FLAT}}$ are defined over $\{-1\}\cup T^{opt}$, as well as the auxiliary variables associated to these variables. This amounts to excluding the last time step of the time frame $T^{prev}$ when relevant.} : 

\begin{equation}\label{eq:init_cond_init}
\begin{aligned}
\forall t \in T^{prev},\; & P_{u,t} = 0 \\
& S_{u,t}^{OFF} = 1\\
& S_{u,t}^{ON\_{UP}} = S_{u,t}^{ON\_{DOWN}} = S_{u,t}^{ON\_{FLAT}} = S_{u,t}^{STOP} = S_{u,t}^{START} = 0\\
& \delta_t^{turned\_on} = \delta_t^{turned\_off} = 0\\ 
& \delta_t^{entered\_up} =  \delta_t^{entered\_down} = \delta_t^{stable} = 0\\
& \delta_t^{flat,down,stop} = \delta_t^{down\_to\_stop} = 0
\end{aligned}
\end{equation}
Due to the definition of $T^{prev}$, the unit can be started as early as $t=0$ if necessary. 

\paragraph{Remark} Due to the fact that for the initialization of the optimization program all units will have to be started up, a warning message will be sent to the user in order to remind him that the optimization program was initialized.

\paragraph{Initial conditions if the program is not "day zero"}\label{sec:not_init}

This situation occurs whenever the last known {\tt Power} value has a date between {\tt StartDate - AddHours($T^{traceback}$)} and {\tt StartDate}. In terms of time stamps, it means that there exists a $k \in T^{prev}$ such that $P_{u,t_{k}}, P_{u,t_{k-1}}, \dots$ are known. We define $T^{opt}_{prev,front} = \{k,\dots,0\}$ and $T^{opt}_{prev,back} = \{-T^{traceback},\dots,k-1\}$ the sub-periods of $T_{init}$ {\it after} and {\it before} the latest known value of the power output $P_{u,t_k}$. Two cases are possible.

\paragraph{Inconsistent program} If $\vert T^{opt}_{prev,front}\vert \ge 2$, it means that the latest value of $P_{u,t_k}$ lies somewhere in $T^{opt}_{init}\backslash\{0\}$. In this case, a warning message is sent to the user informing him that there is an inconsistency between the current optimization time frame and the previous one and the initial conditions are set as in \eqref{eq:init_cond_init}, i.e. as a "day zero" program.

\paragraph{Consistent program} If $T^{opt}_{prev,front} = \{0\}$, that is to say the latest known value is just before {\tt StartDate}, then the program will be considered to be consistent and the initial conditions are successively defined. 
\begin{itemize}
    \item First, the lastest known values of $P_{u,t}$ are retrieved for $t\in T^{prev}$
    \item Based on these values, the value of the state variables are retrieved
    \item Based on the values of the state variables, the auxiliary variables are initialized.
\end{itemize} 

The pseudo-code of the initialization algorithm is given in algorithms \ref{alg:init_q_t_state_var} and \ref{alg:init_aux_variables}. Note that in these pseudo-codes we neglect the boundary constraints on the various time frames handled during the initialization.

\begin{algorithm}[H]
\begin{algorithmic}

\For{$t\in\{-1,\dots,-T^{traceback}\}$}\Comment{Retrieve the values of $P_{u,t}$}
\State $P_{u,t}\leftarrow\textrm{{\tt Power.GetValue}}(t)$
\EndFor

\For{$t\in\{-1,\dots,-T^{traceback}\}$}\Comment{Initial conditions on the state variables}

\If{$P_{u,t} > \textrm{{\tt MinimumPower.GetValue}}(t)$}
    \State $S_{u,t}^{OFF} \leftarrow 0$
    \State $S_{u,t}^{STOP} \leftarrow 0$
    \State $S_{u,t}^{START} \leftarrow 0$
    \If{$P_{u,t} < P_{u,t-1}$}\Comment{Exact initialization required only if the FLAT state is defined.}
        \State $S_{t-1}^{ON\_UP} \leftarrow 0$
        \State $S_{t-1}^{ON\_DOWN} \leftarrow 1$
        \State $S_{t-1}^{ON\_FLAT} \leftarrow 0$
    \ElsIf{$P_{u,t} > P_{u,t-1}$}
        \State $S_{t-1}^{ON\_UP} \leftarrow 1$
        \State $S_{t-1}^{ON\_DOWN} \leftarrow 0$
        \State $S_{t-1}^{ON\_FLAT} \leftarrow 0$
    \ElsIf{$P_{u,t} = P_{u,t-1}$}
        \State $S_{t-1}^{ON\_UP} \leftarrow 0$
        \State $S_{t-1}^{ON\_DOWN} \leftarrow 0$
        \State $S_{t-1}^{ON\_FLAT} \leftarrow 1$
    \EndIf
\ElsIf{$P_{u,t} > 0$}\Comment{Reconstruct the startups and shutdowns}
    \If{$P_{u,t} < P_{u,t-1}$}
        \State $S_{t}^{STOP} \leftarrow 1$
        \State $S_{t}^{START} \leftarrow 0$
    \Else
        \State $S_{t}^{STOP} \leftarrow 0$  
        \State $S_{t}^{START} \leftarrow 1$
    \EndIf
\Else\Comment{Final possibility: the unit is OFF.}
    \State $S_{u,t}^{OFF} \leftarrow 1$
    \State $S_{u,t}^{STOP} \leftarrow 0$
    \State $S_{u,t}^{START} \leftarrow 0$
    \State $S_{t-1}^{ON\_UP} \leftarrow 0$
    \State $S_{t-1}^{ON\_DOWN} \leftarrow 0$
    \State $S_{t-1}^{ON\_FLAT} \leftarrow 0$
\EndIf
\EndFor

\end{algorithmic}
\caption{Initialization of the control and state variables}\label{alg:init_q_t_state_var}
\end{algorithm}

\begin{algorithm}[H]
\begin{algorithmic}
\For{$t\in\{-1,\dots,-T^{traceback}\}$}\Comment{Initialization of $\delta_t^{turned\_on}$ and $\delta_t^{turned\_off}$}
\State $\delta_t^{turned\_on}\leftarrow 0$ 
\State $\delta_t^{turned\_off}\leftarrow 0$ 
\If{$S_{u,t}^{STOP} - S_{t-1}^{STOP} = 1$}\Comment{Replace by $S_{u,t}^{OFF} - S_{t-1}^{OFF} = 1$ if STOP is not defined.}
    \State $\delta_t^{turned\_off}\leftarrow 1$
\ElsIf{$S_{u,t}^{START} - S_{t-1}^{START} = 1$}\Comment{Replace by $S_{u,t}^{OFF} - S_{t-1}^{OFF} = -1$ if START is not defined.}
    \State $\delta_t^{turned\_on}\leftarrow 1$
\EndIf

\EndFor
\For{$t\in\{-1,\dots,-T^{traceback}\}$}\Comment{Initialization of $\delta_t^{stable}$, $\delta_t^{entered\_up}$ and $\delta_t^{entered\_down}$}
\State $\delta_t^{stable}\leftarrow 0$ 
\State $\delta_t^{entered\_up}\leftarrow 0$ 
\State $\delta_t^{entered\_down}\leftarrow 0$ 
\If{$S_{u,t}^{STOP} - S_{t-1}^{STOP} = 1$}
    \State $\delta_t^{stable}\leftarrow 1$
\ElsIf{$S_{u,t}^{ON\_{UP}} - S_{t-1}^{ON\_UP} = 1$}
    \State $\delta_t^{entered\_up}\leftarrow 1$
\ElsIf{$S_{u,t}^{ON\_{UP}} - S_{t-1}^{ON\_UP} = 1$}
    \State $\delta_t^{entered\_down}\leftarrow 1$
\EndIf
\EndFor

\For{$t\in\{-1,\dots,-T^{traceback}\}$}\Comment{Initialization of $\delta_t^{stable}$, $\delta_t^{entered\_up}$ and $\delta_t^{entered\_down}$}
\State $\delta_t^{stable}\leftarrow 0$ 
\State $\delta_t^{entered\_up}\leftarrow 0$ 
\State $\delta_t^{entered\_down}\leftarrow 0$ 
\If{$S_{u,t}^{ON\_{FLAT}} - S_{t-1}^{ON\_FLAT} = 1$}
    \State $\delta_t^{stable}\leftarrow 1$
\ElsIf{$S_{u,t}^{ON\_{UP}} - S_{t-1}^{ON\_UP} = 1$}
    \State $\delta_t^{entered\_up}\leftarrow 1$
\ElsIf{$S_{u,t}^{ON\_{DOWN}} - S_{t-1}^{ON\_DOWN} = 1$}
    \State $\delta_t^{entered\_down}\leftarrow 1$
\EndIf
\EndFor

\For{$t\in\{-1,\dots,-T^{traceback}\}$}\Comment{Initialization of $\delta_t^{flat,down,stop}$ or $\delta_t^{down\_to\_stop}$}
\State $\delta_t^{flat,down,stop}\leftarrow \displaystyle{\left\lfloor \frac{S_{u,t}^{STOP} + S_{t-1}^{ON\_DOWN} + S_{t-2}^{ON\_FLAT}}{3}\right\rfloor}$ 
\State $\delta_t^{down\_to\_stop}\leftarrow 0$ 
\If{$S_{u,t}^{STOP} - S_{t-1}^{ON\_DOWN} = 0$}
    \State $\delta_t^{down\_to\_stop}\leftarrow 1$ 
\EndIf
\EndFor

\State $U_{-1} = S_{-1}^{ON\_UP}\times S_{-2}^{ON\_UP}\times\left(P_{u,t_{-1}} - P_{u,t_{-2}}\right)$\Comment{Initial condition on $U_t$}
\State $D_{-1} = S_{-1}^{ON\_DOWN}\times S_{-2}^{ON\_DOWN}\times\left(P_{u,t_{-1}} - P_{u,t_{-2}}\right)$\Comment{Initial condition on $D_t$}

\end{algorithmic}
\caption{Initialization of the auxiliary variables}\label{alg:init_aux_variables}
\end{algorithm}

\section{Wind and Photovoltaic}

\begin{longtable}[H]{!{\color{Grey}\vrule}>{\centering\arraybackslash}p{3cm} !{\color{Grey}\vrule} p{10.5cm}!{\color{Grey}\vrule} >{\centering\arraybackslash} p{2cm}!{\color{Grey}\vrule}}
    \arrayrulecolor{Grey} \hline
    \rowcolor{Grey} \multicolumn{3}{|c|}{\large{\textbf{Load-, Photovoltaic- and Wind-specific unit characteristics}}} \label{tab:nomenclature_pv_wind}\\
    \hline
    \rowcolor{Grey} 
    \textbf{Notation} & \textbf{Description} & \textbf{Units}\\
    \hline
    $P_{u,t,t^{ex}}^{max}$ & Forecast of the maximum power output of unit $u \in \{U^{w}, U^{pv}\}$ at time $t \in T^{sim}$, forecast from $t^{ex} $ & MW\\
    \hline
    $Curt_{u}$ & Percentage of curtailed power allowed on unit $u \in \{U^{w}, U^{pv}\}$ & - \\
    \hline
\end{longtable}

Wind and photovoltaic units can be curtailed only up to a percentage $Curt_{u}$ of their maximum power forecast $P^{max}_{u,t,t^{ex}}$. The constraints are quite simple for these units. First, we define their maximum and minimum power limits. The maximum power is defined as the maximum power based on the most recent forecast, and minimum power is defined by this value and the curtailment ratio of the unit : 

\qquad \qquad \qquad \qquad
$\forall u\in U^{wind}\cup U^{pv}:$
\begin{align}    
    &P_{u,t} \leq P_{u,t,t^{ex}}^{max} \label{eq:optDisp_windPV1} \\
    Curt_{u}*P_{u,t,t^{ex}}^{max} \leq &P_{u,t} \label{eq:optDisp_windPV2}
\end{align}

\section{Non-dispatchable Production or Load}
Non-dispatchable production represents any must-run units in the system, run-of-river hydraulic units for instance. We also can model inflexible load in the same way. The constraints for these are the same as for wind and photovoltaic above, but with $Curt_{u} = 1$:

\qquad \qquad \qquad \qquad
$\forall u\in U^{ndp}\cup U^{ndl}:$
\begin{equation}    
    P_{u,t} = P_{u,t,t^{ex}}^{max} \label{eq:optDisp_ndp1}
\end{equation}

As for the other equality constraints in the problem, for better resolution this constraint is coded with a small tolerance. 

\section{Flexible Load}
Currently, the only flexible load (that is not covered by the storage modeling) taken into account by ATLAS are power-to-gas units. 

Power-to-gas units are modeled in the following way in the ATLAS model. The main constraints are simple: 

\qquad \qquad \qquad \qquad
$\forall u\in U^{p2g}:$
\begin{align}    
    &P_{u,t} \leq P_{u,t,t^{ex}}^{max} \label{eq:optDisp_flexLoad1}  \\
    P_{u,t,t^{ex}}^{min} \leq &P_{u,t}\label{eq:optDisp_flexLoad2}
\end{align}

% \subsection{Objective Functions}

% max_power = get_time_series_value(opt_Wind.MaximumPowerForecast.GetForecast(GP.ActionHour, ti, ti), ti)
% min_power = (1 - get_time_series_value(opt_Wind.MaximumCurtailmentRatio, ti)) * max_power
% # Maximum and Minimum Power
% constraintList.Add(equipment_wind_pvj.PowerLevel[ti] <= maxPowerti)
% constraintList.Add(equipment_wind_pvj.PowerLevel[ti] >= minPowerti)

% #relaxedReserve disabling condition (eq. (43))
% #constraintList.Add(equipment_wind_pvj.relaxedReserves[ti] <= minPowerti)

% # impossible commitment and stable reserves constraints (eq. (44))
% constraintList.Add(equipment_wind_pvj.automatedReservesUp[ti] <= equipment_wind_pvj.maximumAutomated )
% constraintList.Add(equipment_wind_pvj.automatedReservesDown[ti] <= equipment_wind_pvj.maximumAutomated )
% constraintList.Add(equipment_wind_pvj.reservesUp[ti] <= maxPowerti )
% constraintList.Add(equipment_wind_pvj.reservesDown[ti] <= maxPowerti ) 

% sum_power_level.Add(equipment_wind_pvj.PowerLevel[ti])

\section{Hydraulic Units} 
This section refers only to hydraulic units that have a reservoir of energy and can only produce (meaning that they can’t consume energy, which excludes pumped hydraulic storage that are treated as storage units). In the ATLAS model, these hydraulic units are separated into fragments, each with a price associated. The prices are water values calculated prior to the ATLAS model. The method for this calculation is based on Bellman’s algorithm [\cite{bellman1966dynamic}]. The concept of separating the units into fragments is to represent the spread of historical hydraulic water values that can be explained by differences in the constraints of the various units (aggregated by area in our simulations), and in particular by different energy/power ratios. The water values are calculated by aggregating flexible hydraulic units in each area. In order to represent this dispersion in our simulations, we take this aggregated water value and multiply it by the fragment values that come from the spread. 

The constraints are quite similar to the other units and remain rather simple for the moment: 

\qquad \qquad \qquad \qquad
$\forall u\in U^{H}:$
\begin{align}    
    &P_{u,t} \leq P_{u,t,t^{ex}}^{max} \label{eq:optDisp_hydro1}\\
    P_{u,t,t^{ex}}^{min} \leq &P_{u,t} \label{eq:optDisp_hydro2}
\end{align}

\section{Storage Units}

\begin{longtable}[!ht]{!{\color{Grey}\vrule}>{\centering\arraybackslash}p{2.5cm} !{\color{Grey}\vrule} p{12.5cm}!{\color{Grey}\vrule} }
    \arrayrulecolor{Grey} \hline
    \rowcolor{Grey} \multicolumn{2}{|c|}{\large{\textbf{Storage unit characteristics}}} \\ \arrayrulecolor{Grey} \hline
    \rowcolor{Grey} 
    \textbf{Notation} & \textbf{Meaning}\\
    $P^{buy}_{u,t}, P^{sell}_{u,t}$ & Power in the purchase and sale direction for storage unit $u$ at time $t$\\ \hline
    $\eta^d_u, \eta^c_u$ & Discharge (resp. Charging) efficiency for the unit $u$ \\ \hline
    $\sigma^{sell}_{u,t}$ & Sale/Purchase indicator for the storage technology $u$ at time $t$ \\ \hline
    $\sigma^{EV}_{u}$ & Indicator used to denote a unit $u$ of type Electric Vehicle ($\sigma^{EV}_{u}\text{ = 1 if EV, } =0 \text{ else}$) \\ \hline
    $E^{stored}_{u,t}$ & Storage level of unit $u$ at $t$  \\ \hline
    $E^{stored}_{u,t^{start}}$ & The initial storage level of unit $u$ at $t^{start}$ \\     \hline
    $E^{max}_{u,t}, E^{min}_{u,t}$ & Maximum (resp. Minimum) storage level of unit $u$ at time $t$ \\ \hline
    $E^{disp}_{u, t^{start}-1,t}$ &  Displacement energy of EV unit $u$, corresponding to the energy consumed by its displacement between times $t^{start}-1$ and $t$ \\ \hline
    \hline 
    \hline 
\end{longtable}

Storage units regroup units that have a reservoir and can both produce energy or consume it to store it. Three main types of storage are modeled in ATLAS:
\begin{enumerate}
    \item Batteries
    \item Pumped Hydraulic Storage
    \item Electric Vehicles
\end{enumerate}

For these kinds of units, the optimization problem is solved over the simulation period plus an additional period to avoid having an outcome generation plan heavily influenced by having a short-term vision of a single day. The default values of these for the three different storage types are shown in Table \ref{tab:addlHours}. 

Simply put, based on a price forecast, each storage unit will try to dispatch its buying and selling times by maximizing selling in times when the price is forecasted to be high, and buying when it is low. 

For storage units, we separate the power level into a positive and a negative component with the following definitions: 
\begin{gather}
    P_{u,t} = P^{buy}_{u,t} + P^{sell}_{u,t} \label{eq:optDisp_storage_init1}\\
    0 \leq P^{sell}_{u,t} \leq P_{u,t}^{max} \label{eq:optDisp_storage_init2}\\
    P_{u,t}^{min} \leq P^{buy}_{u,t} \leq 0 \label{eq:optDisp_storage_init3}
\end{gather}
where $P_{u,t}^{min}$ is a negative value (load is modeled as negative in the optimal dispatch problem). 

First off, at the end of the complete optimization period, the unit needs to have the same volume of energy stored as it had at the beginning (Equation \ref{eq:optDisp_storage_c1}). 
\begin{equation}
    \sum_{t} \frac{P^{sell}_{u,t}}{\eta^d_u} = \sum_{t} P^{buy}_{u,t} * \eta^c_u 
\label{eq:optDisp_storage_c1}
\end{equation}
In addition, it has to respect its maximum (Equation \ref{eq:optDisp_storage_c2}) and minimum (Equation \ref{eq:optDisp_storage_c3}) power output constraints, as well as its minimum (Equation \ref{eq:optDisp_storage_c4}) and maximum (Equation \ref{eq:optDisp_storage_c5}) reservoir capacity constraints. 
\begin{gather}
    P^{sell}_{u,t} \leq \sigma^{sell}_{u,t} * P^{max}_{u,t} * \Delta t * \eta^d_u \label{eq:optDisp_storage_c2} \\
    P^{buy}_{u,t} * \eta^c_u  \leq (1 -\sigma^{sell}_{u,t}) * P^{max}_{u,t} * \Delta t \label{eq:optDisp_storage_c3} \\
    E^{stored}_{u,t} \geq E^{min}_{u,t} \label{eq:optDisp_storage_c4} \\
    E^{stored}_{u,t} \leq E^{max}_{u,t} \label{eq:optDisp_storage_c5}
\end{gather}
The energy stored is tracked by Equations \ref{eq:optDisp_storage_c6} and \ref{eq:optDisp_storage_c7}, \ref{eq:optDisp_storage_c6} representing the initial condition constraint.
\begin{gather}
    E^{stored}_{u,t^{start}} = E^{init}_{u} + P^{buy}_{u,t^{start}} * \eta^c_u - \frac{P^{sell}_{u,t^{start}}}{\eta^d_u} - \delta_{u}^{EV} * E^{disp}_{u, t^{start}-1,t} \label{eq:optDisp_storage_c6} \\
    \forall t > t^{start},  \quad E^{stored}_{u,t} = E^{stored}_{u,t-1} +P^{buy}_{u,t} * \eta^c_u - \frac{P^{sell}_{u,t}}{\eta^d_u} - \delta_{u}^{EV} * E^{disp}_{u, t^{start}-1,t} \label{eq:optDisp_storage_c7}
\end{gather}
where $t^{start}$ is the starting date of the simulation, prior to which no market has been simulated. 

In ATLAS, electric vehicles can be modeled either as non-dispatchable load or as storage units, depending on how controllable they are. In general, EVs that can control their recharge, or even produce energy (i.e. vehicle-to-grid units) are modeled as storage. For these units, the $E^{min}_{u,t}$ in Equation \ref{eq:optDisp_storage_c4} is fixed at 30\% of $E^{max}_{u,t}$.

% \subsection{Objective Functions}

\section{Additional Constraints} \label{sec:OD_AddlConstraints}
All units--regardless of their type--must also consider the impacts of previously procured reserves. 
\subsection{Constraints on the reserve requirements}\label{sec:reserve_constraints}

\begin{wrapfigure}{r}{7cm}
    \centering
	\includegraphics[width = 6.5cm]{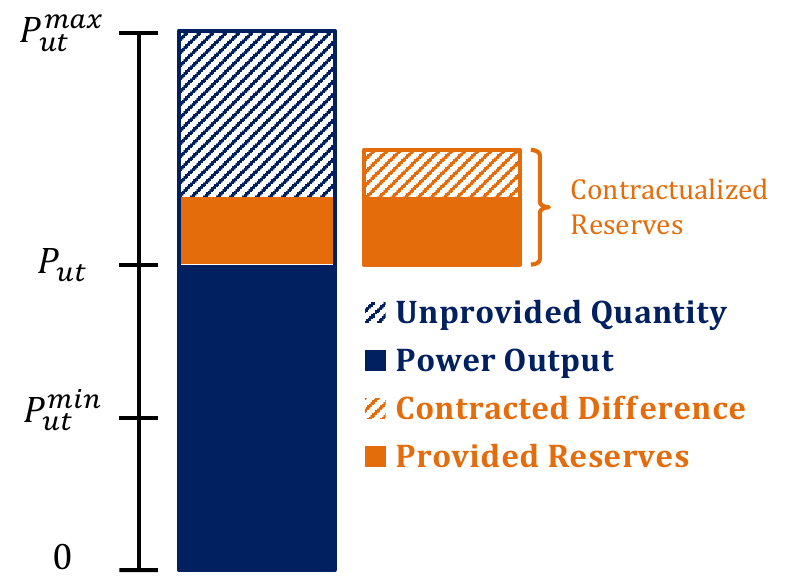}
	\caption[Reserve Constraint Quantities]{Reserve Constraint Quantities} %\footnotemark{}}
	% \vspace{-30pt}
	\label{fig:optDisp_fillupConstraint}
\end{wrapfigure}

Reserve requirements are taken into account in the objective function as a penalized difference between the contracted reserves and the reserves provided by the unit. This difference is the area shown in orange stripes in Figure \ref{fig:optDisp_fillupConstraint} and is referred to as the {\it Contracted Difference}. The penalty coefficients, $\eta^{auto}$ ({\tt automaticUnsuppliedReservePenalty}) and $\eta^{manu}$ ({\ manualUnsuppliedReservePenalty}) are user parameters that should be set high enough to incentivize the unit to minimize the difference between the contracted reserves and the reserve output actually provided. There are two kinds of reserves treated: {\it automated} reserves and {\it manual} reserves. {\it Automated} reserves regroup Frequency Containment Reserves (FCR) and Automatic Frequency Restoration Reserves (aFRR), while {\it manual} reserves cover Manual Frequency Restoration Reserves (mFRR) and Replacement Reserves (RR) (see Section \ref{sec:OD_ThermalPlants}). Each is associated with a penalty which weights their impact in the objective function. FCR and aFRR are referred to as {\it automated} reserves and their corresponding penalty is $\eta^{auto}$ and mFRR and RR as  (and are penalized by an amount equal to $\eta^{manu}$). 

The penalties are expressed in euros/MW per hour, meaning that the baseline bandwidth duration is one hour. However, if the time step is 30 minutes, then the penalty over one time step will be twice as low. Conversely, if the time step is two hours, the penalty for one time step will be twice as large.  

As described in the initial assumptions for thermal units, if the FLAT state is defined, manual reserves can only be provided in the FLAT state. Otherwise, if the FLAT state is not defined, all reserves can be provided as long as the unit is online (i.e. in the ON\_UP or ON\_DOWN states). The unit cannot provide reserves during startups, shutdowns or if the unit is offline. 

Finally, automated reserves are bounded by the maximum capacities (resp. $R^{aFRR,max}_u$ and $R^{FCR,max}_u$). If the procurement is greater than the capacity, the remaining part cannot be provided and will therefore be counted in the penalty. This part is referred to as the {\it infeasible} reserves, as opposed to the {\it feasible} reserves that serve as the baseline amount to be taken into account by the unit. In a nutshell, reserves are taken into account in the following way :

\begin{itemize} 
    \item FCR and aFRR are referred to as {\it automated} reserves, mFRR and RR as {\it manual} reserves, each kind of reserve has its penalty. 
    \item {\it automated} reserves are broken down into two parts namely the {\it feasible} and {\it infeasible} parts, the latter corresponding to the procured amount exceeding the capacity of the unit. The infeasible automated reserves are collapsed together in a {\tt automatedUnsuppliedReserves} scalar that is added as a penalty in the objective function.
    \item If the FLAT state is defined, the unit can only provide manual reserves in the FLAT state. 
\end{itemize}

Let us now introduce the decision variables associated with the reserves requirements. These variables are defined for all $t\in T^{opt}$. There are three kinds of variables : 
\begin{itemize}
    \item Variables corresponding to the provided reserve level : $R_{u,t}^{auto,up}$, $R_{u,t}^{auto,dn}$, $R_{u,t}^{manu,up}$ and $R_{u,t}^{manu,dn}$ 
    \item {\it Mirror} variables corresponding to the quantity that is {\it not} provided : $R_{u,t}^{unpr,up}$ and $R_{u,t}^{unpr,dn}$. These variables represent the blue-striped area in Figure \ref{fig:optDisp_fillupConstraint}.
    \item {\it Contracted Difference} variables that are multiplied by the unsupplied reserve penalties and represent the part of the previously procured reserves that the unit either cannot or will not provide following the cost optimization
    \item A {\it relaxed} reserve variable {\tt relaxedReserves} allowing fill-up constraints to be satisfied even if the unit is not online. This variable is always defined and takes values in the $[0,P_{u,t}^{min}]$ range. 
\end{itemize}

These decision variables but {\tt relaxedReserves} enter the definition of the $\textrm{{\tt contractedDifference}}^{\bullet,\bullet}_t$ variables that are included in the objective function.

\paragraph{Contracted Difference variables} These variables are included in the objective function and the objective for the unit is to minimize their value. As mentioned above, we make a distinction between automated procurement and manual procurement when the FLAT state is defined. 

If the FLAT state is defined, the {\tt contractedDifference} control variables are defined as in Equation \eqref{eq:definition_of_contrated_differnece_flat}. We denote $R^{proc,manu,up}_{u,t} \coloneqq R^{proc,mFRR,up}_{u,t} + R^{proc,RR,up}_{u,t}$ and $R^{proc,manu,dn}_{u,t} \coloneqq R^{proc,mFRR,dn}_{u,t} + R^{proc,RR,dn}_{u,t}$. Additionally, we denote $(x)_+ = \max(0,x)$.
\begin{equation}\label{eq:definition_of_contrated_differnece_flat}
\begin{aligned}
    \forall t \in T^{opt},&\; \Delta R_{u,t}^{contr,auto,up} = \left(R^{proc,auto,up}_{u,t}-R_{u,t}^{auto,up}\right)_+\\
    \forall t \in T^{opt},&\; \Delta R_{u,t}^{contr,auto,dn} = (R^{proc,auto,dn}_{u,t} -R_{u,t}^{auto,dn})_+\\
    \forall t \in T^{opt},&\; \Delta R_{u,t}^{contr,manu,up} = \left(R^{proc,manu,up}_{u,t}-R_{u,t}^{manu,up}\right)_+\\
    \forall t \in T^{opt},&\; \Delta R_{u,t}^{contr,manu,dn} = \left(R^{proc,manu,dn}_{u,t}-R_{u,t}^{manu,dn}\right)_+\\
\end{aligned}
\end{equation}

The variables $\textrm{{\tt contractedDifference}}^{\bullet,\bullet}_t$ include a nonlinearity which is taken into account by adding constraint \eqref{eq:constraint_contract_auto} if relevant and \eqref{eq:constraint_contract_res} to the problem :

\begin{equation}\label{eq:constraint_contract_auto}
\begin{aligned}
    \forall t \in T^{opt},\; &
    \Delta R_{u,t}^{contr,auto,up} \ge R^{proc,auto,up}_{u,t}-R_{u,t}^{auto,up} \\
    & \Delta R_{u,t}^{contr,auto,dn} \ge R^{proc,auto,dn}_{u,t} -R_{u,t}^{auto,dn} \\
\end{aligned}
\end{equation}

\begin{equation}\label{eq:constraint_contract_res}
\begin{aligned}
    \forall t \in T^{opt},\; &
    \Delta R_{u,t}^{contr,manu,up} \ge R^{proc,manu,up}_{u,t} -R_{u,t}^{manu,up} \\
    & \Delta R_{u,t}^{contr,manu,dn} \ge R^{proc,manu,dn}_{u,t} -R_{u,t}^{manu,dn} \\
\end{aligned}
\end{equation}

\paragraph{Remark} We don't need to add a positivity constraint that lower bounds the variable since the {\tt contractedDifference} variables are passed as non-negative variables to the solver.\\

Reserves also require to include "fill up" constraints to the problem that make sure that for each time step $t$, the sum of the provided power, provided and unprovided reserves is equal to the unit's maximum power and symmetrically that the power output minus the provided and unprovided reserves is equal to the unit's minimum power. Constraints \eqref{eq:fill_upward_flat} and \eqref{eq:fill_downward_flat} translate this requirement when the FLAT state is defined (so we distinguish the automated reserves and the reserves) and Equations \eqref{eq:fill_upward_noflat} and \eqref{eq:fill_downward_noflat} translate the same constraint in the case where the FLAT state is not defined. 

\paragraph{Upward "fill up" constraint:}

\begin{equation}\label{eq:fill_upward_flat}
\begin{array}{lr}
    \forall t\in T^{opt},\;
    P_{u,t} + R_{u,t}^{auto,up} +
    R_{u,t}^{manu,up} + R_{u,t}^{unpr,up} = P_{u,t}^{max}
\end{array}
\end{equation}
and
\begin{equation}\label{eq:fill_upward_noflat}
\begin{array}{lr}
    \forall t\in T^{opt},\;
    P_{u,t} +
    R_{u,t}^{manu,up} + R_{u,t}^{unpr,up} = P_{u,t}^{max}
\end{array}
\end{equation}

\paragraph{Downward "fill up" constraint:}

\begin{equation}\label{eq:fill_downward_flat}
\begin{array}{lr}
    \forall t\in T^{opt},\;
    P_{u,t} - R_{u,t}^{auto,dn} + \textrm{{\tt relaxedReserves}}_t -
    R_{u,t}^{manu,dn} - R_{u,t}^{unpr,dn} = P_{u,t}^{min}
\end{array}
\end{equation}

\begin{equation}\label{eq:fill_downward_noflat}
\begin{array}{lr}
    \forall t\in T^{opt},\;
    P_{u,t} + \textrm{{\tt relaxedReserves}}_t -
    R_{u,t}^{manu,dn} - R_{u,t}^{unpr,dn} = P_{u,t}^{min}
\end{array}
\end{equation}

Moreover, we need to add a disabling condition on {\tt relaxedReserves} in order to ensure that this variable will only used when the unit is not in one of the ON states. This disabling constraint is given by Equation \eqref{eq:diabling_cons} (the FLAT state is added only if relevant).

\begin{equation}\label{eq:diabling_cons}
    \forall t\in T^{opt},\; \textrm{{\tt relaxedReserves}}_t \le P_{u,t}^{min}\left(1 - S_{u,t}^{ON\_{UP}} - S_{u,t}^{ON\_{FLAT}} - S_{u,t}^{ON\_{DOWN}} \right)
\end{equation}

We also want to prevent the unit from fulfilling its upward reserve commitment when it is offline or starting-up. This corner case can occur since upward reserve commitments can be understood as lowering the unit's maximum power, and if the power output is 0 then the condition that the power output is lower than the adjusted minimum power is satisfied. In order to prevent this, we include the "unsatisfied commitment" constraint defined in Equation \eqref{eq:impossible_commitment}. In practice, this equation is merged with Equation \eqref{eq:flat_reserve_constraint} when the FLAT state is defined in order to avoid duplicates in the constraints. 

\begin{equation}\label{eq:impossible_commitment}
\begin{aligned}
    \forall t\in T^{opt},\;&  R_{u,t}^{manu,up} \le P_{u,t}^{max} \left(1 - S_{t}^{OFF} - S_{t}^{START} - S_{t}^{STOP}\right)\\
    & R_{u,t}^{auto,up} \le R^{auto,max}_{u} \left(1 - S_{t}^{OFF} - S_{t}^{START} - S_{t}^{STOP}\right)\\
    \forall t\in T^{opt},\;&  R_{u,t}^{manu,dn} \le P_{u,t}^{max} \left(1 - S_{t}^{OFF} - S_{t}^{START} - S_{t}^{STOP}\right)\\
    & R_{u,t}^{auto,dn} \le R^{auto,max}_{u} \left(1 - S_{t}^{OFF} - S_{t}^{START} - S_{t}^{STOP}\right)\\
\end{aligned}
\end{equation}

Finally, if the FLAT state is defined we forbid the unit to provide reserves if the unit is not in the FLAT state. This is taken into account by the Equation \eqref{eq:flat_reserve_constraint}:

\begin{equation}\label{eq:flat_reserve_constraint}
\begin{aligned}
    \forall t\in T^{opt},\;&  R_{u,t}^{manu,up} \le P_{u,t}^{max} \left(1 - S_{t}^{ON\_UP}- S_{t}^{ON\_DOWN}\right)\\
    & R_{u,t}^{manu,dn} \le P_{u,t}^{max} \left(1 - S_{t}^{ON\_UP}- S_{t}^{ON\_DOWN}\right)\\
\end{aligned}
\end{equation}
\subsection{Imbalance Constraints}\label{sec:imbalance_constraints}
For certain of the optimal dispatch use cases, we consider the imbalance respective to a previous dispatch, either from a market clearing or from a previous program (taking into account forecast errors from a different execution date). For these, certain constraints are defined. The imbalance is split into a large and a small imbalance in order to apply differentiated penalties for imbalances that cause larger problems for the system. They are also each separated into positive and negative terms in order to penalize the absolute value of the imbalance. This gives us the following terms, defined for each portfolio $pf$ and each time $t$ :
\begin{gather}
     I^{tot}_{pf} = P_{pf,t} - P^{target}_{pf,t} \label{eq:optDisp_imbalance1} \\
     \text{with } P_{pf,t} = \sum_{u \in U_{pf}} P_{u,t} \text{ and } P^{target}_{pf,t} = \sum_{u \in U_{pf}} P^{target}_{u,t} \label{eq:optDisp_imbalance2} \\
     I^{tot}_{pf} = I^{large,up} + I^{small,up} - I^{large,down} - I^{small,down}  \label{eq:optDisp_imbalance3} \\
     0\leq I^{large,up}, I^{large,down} \leq I^{max,total} \label{eq:optDisp_imbalance4} \\     
     0\leq I^{small,up}, I^{small,down} \leq I^{max,small} \label{eq:optDisp_imbalance5} 
\end{gather}

The value of $P^{target}_{u,t}$ comes from the previous program described above. We also define four corresponding imbalance prices: $p_t^{smallImb,up}$, $p_t^{largeImb,up}$, $p_t^{smallImb,down}$, and $p_t^{largeImb,down}$ that will be used in the objective function. The prices can be calculated in different manners, depending on the user inputs. The calculation most often used is a simple affine function of the forecast market price: 
\begin{align}
    p_t^{smallImb,up} &= p_t^{smallImb,down} = \alpha^{smallImb} p^{forecast}_{t,t^{ex}} + \beta^{smallImb}\\
    p_t^{largeImb,up} &= p_t^{largeImb,down} = \alpha^{largeImb} p^{forecast}_{t,t^{ex}} + \beta^{largeImb}
\end{align}
where $\alpha^{smallImb}$, $\alpha^{largeImb}$, $\beta^{smallImb}$, and $\beta^{largeImb}$ are user-defined parameters.

\section{Objective Functions} \label{sec:ObjFxns}
The optimal dispatch problem is used in several different situations. These situations each have their own objective functions and grouping of the constraints described above. It can be run unit-based, where each unit runs its own optimal dispatch problem, or in portfolio mode, where a portfolio owns a group of assets that it optimizes together. 
The optimization is therefore performed knowing the cost and technical feasibility of multiple units, combined in different manners in order to represent the different market participants (as shown in Figure \ref{fig:marketActors}. There are four main uses of the optimal dispatch problem:

\begin{enumerate}
    \item Day Ahead Market Order Creation
    \item Portfolio Optimization (Post-Market Clearing)
    \item Intraday Market Order Creation
    \item Balancing mechanism, which is entirely described in \cite{cogen_atlas_2024}.
\end{enumerate}

We can define four basic cost functions that make up the different objective functions for each of these use cases: 
\begin{enumerate}
    \item Operation Cost 
    \item Potential Profit
    \item Imbalance Cost
    \item Cost of Unprovided Reserves
\end{enumerate}

These will be defined for each of the use cases in the following sections.
 
\subsection{Day Ahead Market Order Creation} \label{subsec:OptDisp_DAO}
This phase is generally run in unit-based mode, and only for certain units in the system. Many of the units do not perform an optimal dispatch problem for their order creation. These other methods are described in detail in Chapter \ref{ch:OrderCreation}. Certain units, however, base their offer creation on determining their optimal dispatch given a forecast market price (or multiple price scenarios), then generating order books that respect these optimal production plans. 

For each thermal unit with an intermediate strategy, the following optimization problem is solved: 

\qquad \qquad \qquad \qquad \qquad \qquad $\forall u \in U^{th}:$
\begin{minie}|s|
    {}{\pi_{th}^{DAO}  \label{eq:thermalDAO_objfxn}}{\label{eq:thermalDAO_prob}}{}
    \addConstraint{\text{Equations \ref{eq:started_on_t} - \ref{eq:init_cond_init}}}{}{\label{eq:thermalDAO_constraints}}{}
    \addConstraint{\text{Equations \ref{eq:constraint_contract_auto} - \ref{eq:flat_reserve_constraint}}}{}{\label{eq:thermalDAO_resConstraints}}{}
\end{minie}

where 

\begin{equation}
    \pi_{th}^{DAO} = \pi_{th}^{OpCost} - \pi_{th}^{PotProf} + \pi_{th}^{UnprRes}
\end{equation}

with 
\begin{align}
    \pi_{th}^{OpCost} &= \sum_{t\in T^{opt}}( c^{var}_{u,t}P_{u,t} \Delta t - \delta_t^{turned\_on} c_{u,t}^{SU} ) \label{eq:dao_objfxn_th_OpCost} \\
    \pi_{th}^{PotProf} &= \sum_{t\in T^{opt}}( p^{forecast}_{t,t^{ex}}P_{u,t} \Delta t ) \label{eq:dao_objfxn_th_PotProf}\\
    \pi_{th}^{UnprRes} &= \sum_{t \in T^{opt}} \left( \eta^{manu} * (\Delta R_{u,t}^{manu,up} - \Delta R_{u,t}^{manu,dn}) + \eta^{auto} * (\Delta R_{u,t}^{auto,up} - \Delta R_{u,t}^{auto,dn}) \right) \label{eq:dao_objfxn_th_UnprRes}
\end{align}

Where $p^{forecast}_{t,t^{ex}}$ is an exogenous price forecast and  $\delta_t^{turned\_on}$ is a binary variable indicating whether the unit has been started up on $t$. It is by definition equal to 1 at only one time step and will be defined by the set of constraints detailed in Section \ref{sec:started_on_t}, Equation \eqref{eq:started_on_t}.

The parameters $\eta^{manu}$, $\eta^{auto}$, $\eta^{res}$ are set by the user and are used to incentivize the unit to provide a reserve level as close as possible to its procured level. The terms $\Delta R_{ut}^{contr,manu,up}$, $\Delta R_{ut}^{contr,manu,dn}$, $\Delta R_{ut}^{contr,auto,up}$ and $\Delta R_{ut}^{contr,auto,dn}$ correspond to the difference between the procured amount of reserves (either upwards, downwards and potentially automated) and the provided amount of reserves. See Section \ref{sec:reserve_constraints} and Equation \eqref{eq:definition_of_contrated_differnece_flat} for more details. 

Each storage unit solves the following optimization problem: 

\qquad \qquad \qquad \qquad \qquad \qquad $\forall u \in U^{st}:$
\begin{minie}|s|
    {}{\pi_{st}^{DAO}  \label{eq:storageDAO_objfxn}}{\label{eq:storageDAO_prob}}{}
    \addConstraint{\text{Equations \ref{eq:optDisp_storage_init1} - \ref{eq:optDisp_storage_c7}}}{}{\label{eq:storageDAO_constraints}}{}
    \addConstraint{\text{Equations \ref{eq:constraint_contract_auto} - \ref{eq:flat_reserve_constraint}}}{}{\label{eq:storageDAO_resConstraints}}{}
\end{minie}

where 

\begin{equation}
    \pi_{st}^{DAO} =  \pi_{st}^{UnprRes} - \pi_{st}^{PotProf}
\end{equation}

The storage objective function considers only two factors, the potential profit of the storage unit and the cost of unprovided previously procured reserves. As discussed in previous sections, storage units are optimized over longer periods of time. Following the resolution of the problem, only the values corresponding to the simulation period are retained.

\begin{align}
    \pi_{st}^{PotProf} &= \sum_{t\in T^{ext}} \left( p^{forecast}_{t,t^{ex}}*(P^{sell}_{u,t} - P^{buy}_{u,t}) \right) \label{eq:dao_objfxn_storage_PotProf} \\
    \pi_{st}^{UnprRes} &= \sum_{t \in T^{opt}} ( \eta^{manu} * (\Delta R_{u,t}^{manu,up} - \Delta R_{u,t}^{manu,dn}) + \eta^{auto} * (\Delta R_{u,t}^{auto,up} - \Delta R_{u,t}^{auto,dn})  \label{eq:dao_objfxn_storage_UnprRes}
\end{align}

\subsection{Portfolio Optimization} \label{subsec:PortfolioOptim}
This phase is run directly following a market clearing. That means a set of orders for each unit has been accepted. Order couplings created in the prior modules may not be enough to represent all technical constraints. Portfolio optimization then allows for these equipments to verify that the accepted offers are viable given the full inclusion of their technical constraints. If there is a difference, the actor can then look at markets closer to real-time (e.g. intraday markets or the new balancing platforms, TERRE, MARI, etc.) to reduce their potential imbalance.
The goal for this phase of post-market clearing is therefore two-fold. First, to verify that all units respect all their technical constraints. And second, to ensure that the production plan determined by the market for each unit is optimal for this specific market actor. For this second piece, the portfolio manager can arbitrate between the operating cost and a forecast imbalance price. The penalty for not providing previously procured reserves is also assessed in this arbitrage.

Then the constraints of each unit are iteratively added into the problem, giving the following generalized formulation: 

\qquad \qquad \qquad $\forall pf \in PF :$
\begin{minie}|s|
    {}{\pi_{th}^{PO} + \pi_{h}^{PO} + \pi_{st}^{PO} + \pi_{fl}^{PO} + \pi_{pf}^{PO}  \label{eq:PO_objfxn}}{\label{eq:PO_prob}}{}
    \addConstraint{\text{Equations \ref{eq:started_on_t} - \ref{eq:init_cond_init}}}{}{\label{eq:thermalPO_constraints}}{\qquad \qquad \forall u\in U^{th}_{pf}}
    \addConstraint{\text{Equations \ref{eq:optDisp_windPV1} - \ref{eq:optDisp_windPV2}}}{}{\label{eq:windPVPO_constraints}}{\qquad \qquad \forall u \in U^{w}_{pf}\cup U^{pv}_{pf}}
    \addConstraint{\text{Equations \ref{eq:optDisp_ndp1}}}{}{\label{eq:ndpPO_constraints}}{\qquad \qquad \forall u \in U^{ndp}_{pf} \cup U^{ndl}_{pf}}
    \addConstraint{\text{Equations \ref{eq:optDisp_flexLoad1} - \ref{eq:optDisp_flexLoad2}}}{}{\label{eq:flexLoadPO_constraints}}{\qquad \qquad \forall u \in U^{fl}_{pf}}
    \addConstraint{\text{Equations \ref{eq:optDisp_hydro1} - \ref{eq:optDisp_hydro2}}}{}{\label{eq:hydroPO_constraints}}{\qquad \qquad \forall u\in U^{h}_{pf}}
    \addConstraint{\text{Equations \ref{eq:optDisp_storage_init1} - \ref{eq:optDisp_storage_c7}}}{}{\label{eq:storagePO_constraints}}{\qquad \qquad \forall u\in U^{st}_{pf}}
    \addConstraint{\text{Equations \ref{eq:constraint_contract_auto} - \ref{eq:flat_reserve_constraint}}}{}{\label{eq:reservePO_constraints}}{\qquad \qquad \forall u \in U_{pf}}
\end{minie}

where 
\begin{align}
    \pi_{th}^{PO} &=  \sum_{u \in U^{th}_{pf}} \left( \pi_{th_u}^{OpCost} - \pi_{th_u}^{PotProf} \right)\\ 
    \pi_{h}^{PO} &=  \sum_{u \in U^{h}_{pf}} \left( \pi_{h_u}^{OpCost} - \pi_{h_u}^{PotProf} \right) \\ 
    \pi_{st}^{PO} &= \sum_{u \in U^{st}_{pf}} \left(  - \pi_{st_u}^{PotProf} \right) \\ 
    \pi_{fl}^{PO} &= \sum_{u \in U^{fl}_{pf}} \left( \pi_{fl_u}^{OpCost} - \pi_{fl_u}^{PotProf} \right) \\ 
    \pi_{pf}^{PO} &=  \pi_{pf}^{ImbCost} + \pi_{pf}^{UnprRes}
\end{align}

The pieces of the objective function corresponding to thermal and storage units have the same formulation as in Equations \ref{eq:dao_objfxn_th_OpCost} - \ref{eq:dao_objfxn_th_UnprRes} and Equations \ref{eq:dao_objfxn_storage_PotProf} - \ref{eq:dao_objfxn_storage_UnprRes}, respectively. 
Flexible load--in this case Power-to-Gas units--have the following objective function pieces: 
\begin{align}
    \pi_{fl_u}^{OpCost} &= p^{forecast,elec}_{t,t^{ex}} \times P_{u,t} \\
    \pi_{fl_u}^{PotProf} &= p^{forecast,gas}_{t,t^{ex}} \times P_{u,t} \\
\end{align}
Essentially, their profit is defined by the forecast gas price and their cost is defined by the forecast electricity price.

Finally, the imbalance cost and unprovided reserve penalties for the portfolio are defined as follows. 
\begin{align}
    \pi_{pf}^{ImbCost} &= p_t^{smallImb,up}I^{small,up} + p_t^{smallImb,down}I^{small,down} + \nonumber \\  
    & \qquad \qquad \qquad \qquad p_t^{largeImb,up}I^{large,up} + p_t^{largeImb,down}I^{large,down} \\
    \pi_{pf}^{UnprRes} &=  \sum_{u \in U_{pf}} \Big( \sum_{t \in T^{opt}} ( \eta^{manu} * (\Delta R_{u,t}^{manu,up} - \Delta R_{u,t}^{manu,dn}) + \nonumber \\
    & \qquad \qquad  \qquad \qquad \eta^{auto} * (\Delta R_{u,t}^{auto,up} - \Delta R_{u,t}^{auto,dn}) \Big) 
\end{align}
Each portfolio solves an optimal dispatch problem, taking into account the quantities previously bought or sold on prior markets. The goal of each portfolio is therefore to minimize its overall costs, while taking into account the most recent forecast and ensuring a feasible production plan for all its units. The portfolio also looks at a forecast of the imbalance price that they would have to pay if their overall balance changes, arbitrating between a change in overall dispatch and the cost of the imbalance.

\chapter{Order Creation Modules} \label{ch:OrderCreation}

\section{Order Types}

\begin{figure}[!ht]
    \centering
    \includegraphics[width=\textwidth]{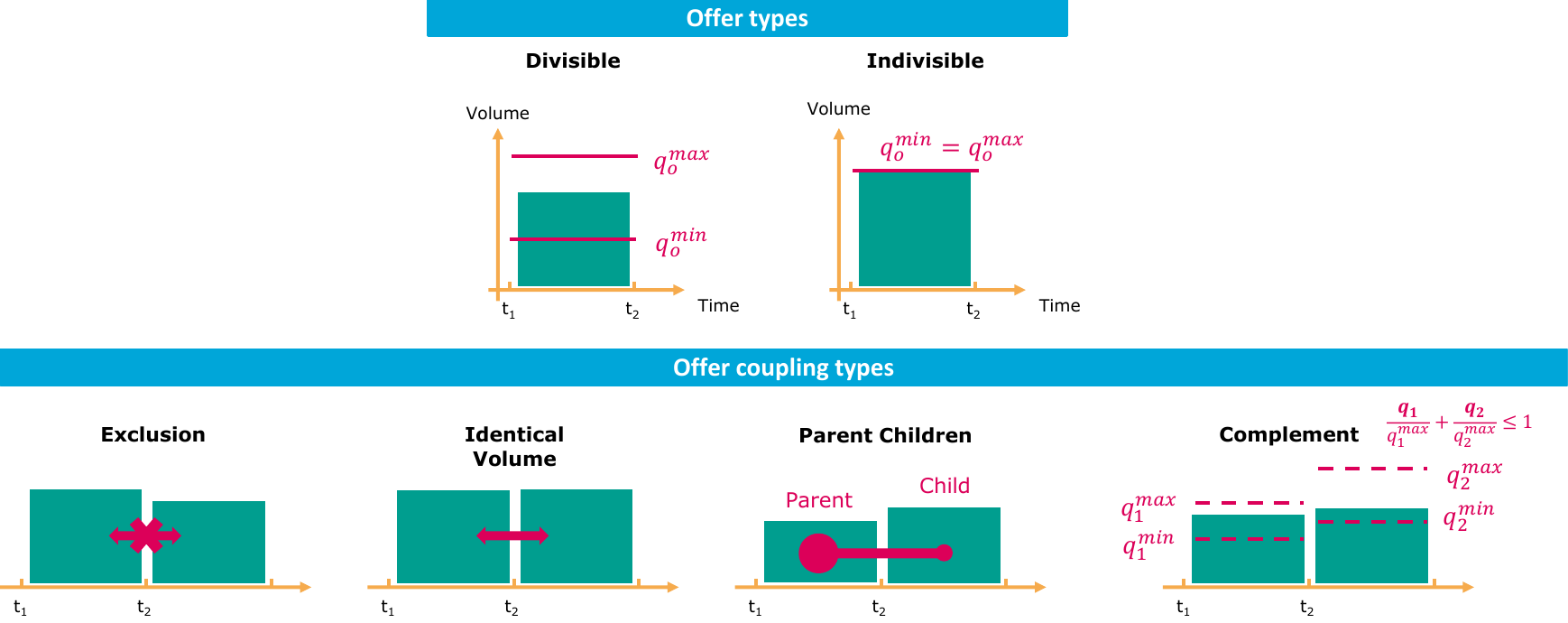}
    \caption{ATLAS Order Types}
    \label{fig:orderTypes_all}
\end{figure}

Figure \ref{fig:orderTypes_all} shows the different order types used in the ATLAS model. Some simplifications were made. Currently ATLAS does not have offers that last more than one time step. This capacity can easily be reproduced using an identical volume or identical ratio link between multiple offers. 

\section{Day Ahead Orders}
In this module, the offers are created for each unit for the day ahead market clearing. The offers are created differently based on the type of technology. For thermal units, some basic actor strategies have been included as well. 

The main goals are to reduce economic losses, ensure satisfaction of as many of the technical constraints as possible and take into account uncertainty at two levels: actor behavior and system state. The current version of ATLAS does not yet take into account the uncertainty based on the actions of other market players.

We can separate the units into two groups. The first set generate orders in a single phase based on their technical constraints and costs. This first set includes all unit types except thermal and storage units. 

\begin{figure}[H]
    \centering
    \includegraphics[width=\textwidth]{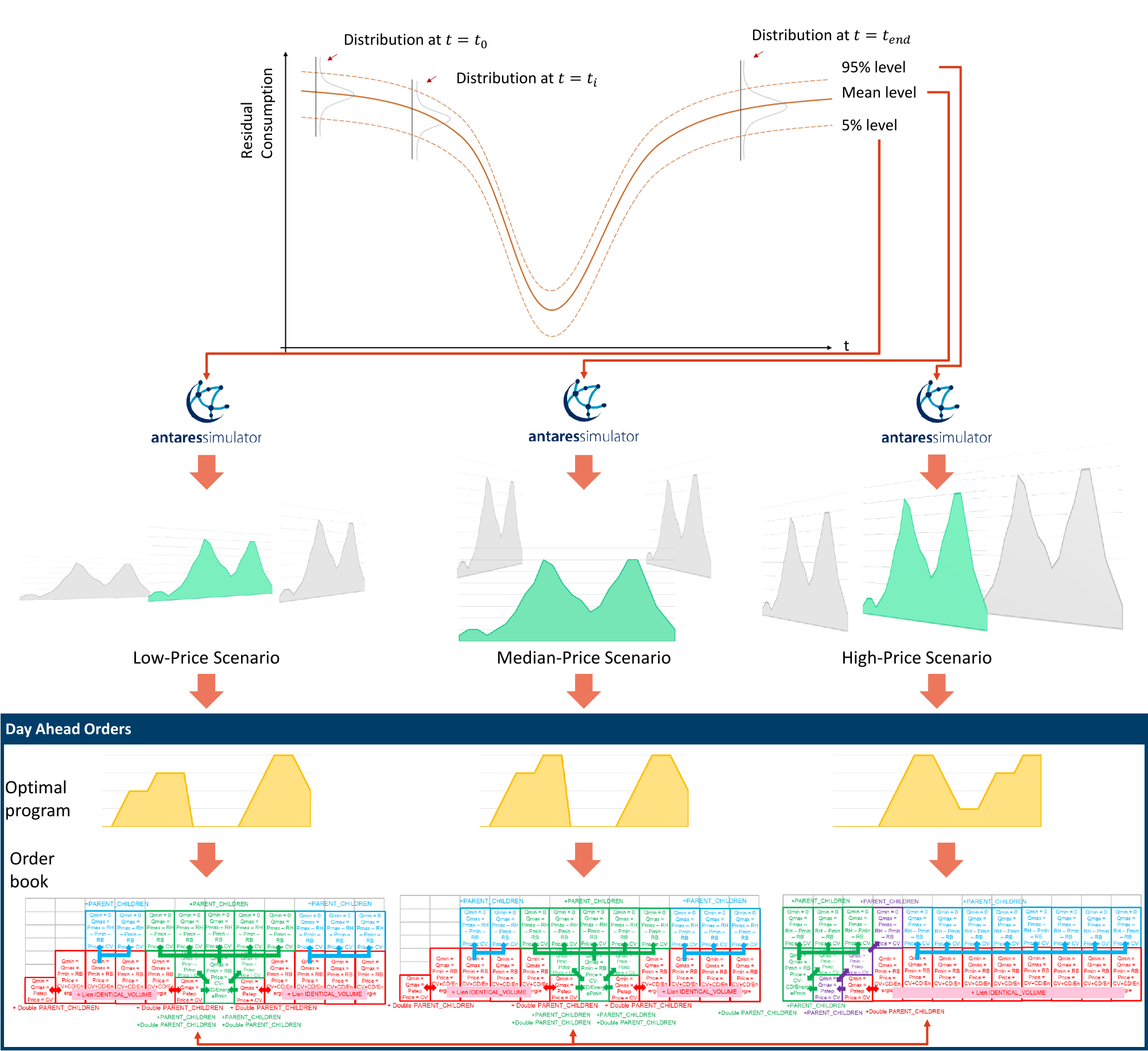}
    \caption{Day Ahead Order Process for Units with Two-Phase Order Generation}
    \label{fig:dao_processOverview}
\end{figure}

The second set solve a more complex problem. Based on the different price scenarios calculated in previous modules (see Section \ref{sec:PriceForecast}), these units generate their market orders in two stages. First, the module calculates the optimal program for each of these unit. Then it translates this into different order books, as shown in Figure \ref{fig:dao_processOverview}. 

The next few sub-sections will describe the different order books created for each of the different technologies of the ATLAS model.

\subsection{Thermal Units} \label{subsec:DAO_thermal}

% \begin{figure}[!ht]
%     \centering
%     \includegraphics[width=0.9\textwidth]{images/TechnicalConstraints.png}
%     \caption{Technical Constraints of a Thermal Power Plant}
%     \label{fig:thermalConstraints}
% \end{figure}

% \subsubsection{Stage }

% \subsubsection{Thermal Plants}

% In the ATLAS model, we consider 12 main constraints for thermal power plants, shown in Figure \ref{fig:thermalConstraints}.

% \begin{figure}[!ht]
% \centering
% \begin{subfigure}{.32\linewidth}
%     \centering
%     \includegraphics[width=\linewidth]{images/TechnicalConstraints_text.png}
%     % \caption{}\label{fig:thermalConstraints_text}
% \end{subfigure}
%     % \hfill
% \begin{subfigure}{.67\linewidth}
%     \centering
%     \includegraphics[width=\linewidth]{images/TechnicalConstraints.png}
%     % \caption{}\label{fig:thermalConstraints_img}
% \end{subfigure}

% \RawCaption{\caption{Constraints of a Thermal Power Plant modeled in ATLAS}
% \label{fig:thermalConstraints}}
% \end{figure}

Thermal plants have three different potential strategies: Base, Peak and Intermediate. Only those with the Intermediate strategy use the optimal dispatch problem to determine their order books. The Base and Peak strategies are applied to specific types of units (for instance nuclear power plants are often classified as Base, and OCGT units are often classified as Peak), and result in heuristics, rather than optimization problems, to determine their order books.

Base units are slow units with relatively low marginal cost, such as nuclear power plants, and are usually online when available. For this kind of unit, being shut down can have important economic consequences because of their significant dynamic constraints (for instance Minimum time off or Startup duration), so it is often better to stay online for some periods by selling electricity at a price below their marginal cost, to avoid having to pay startup costs again. 

On the other hand, Peak units are flexible and have higher marginal costs. They are often offline and only generate electricity when the demand cannot be covered by Base or Intermediate units. Consequently, the heuristic used creates bids over a single time unit, and their orders include their possible startup costs.

% \vspace{-30pt}
\begin{wrapfigure}[40]{r}{7cm}
    \centering
	\includegraphics[width = 6.5cm]{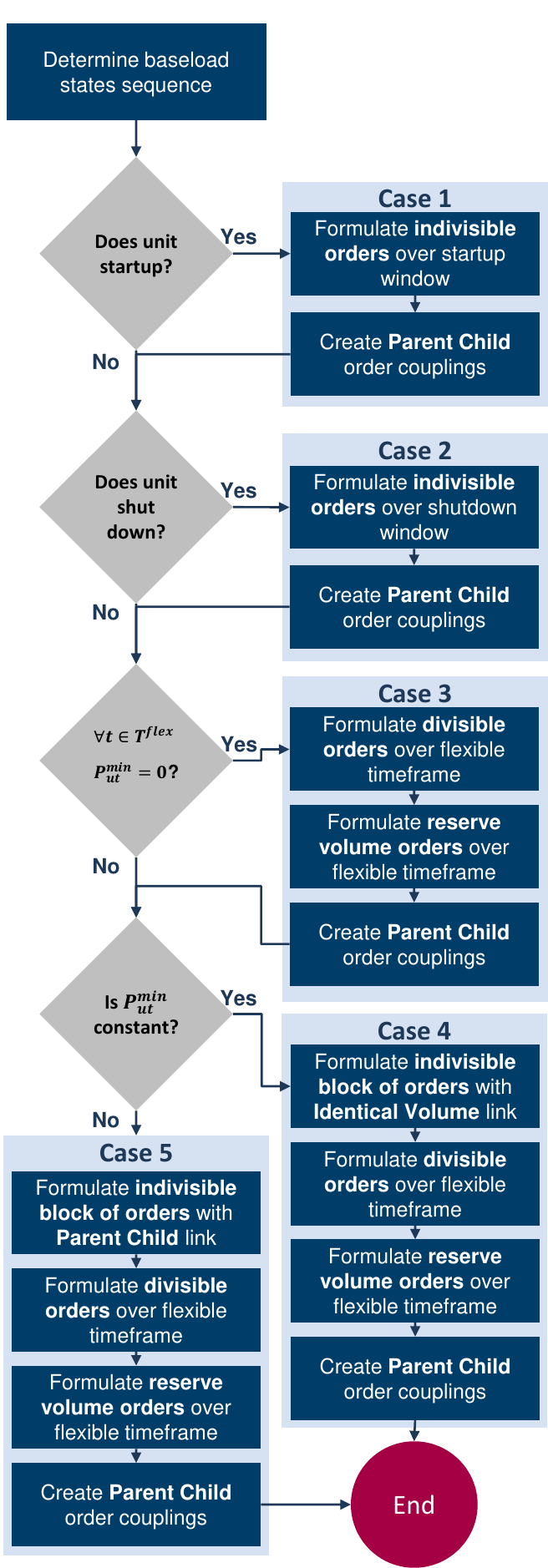}
	\caption[Day Ahead Orders - Base Strategy]{Day Ahead Orders - Base Strategy} 
 
    %\footnotemark{}}
	
	\label{fig:DAO_algoBase}
    % \vspace{-35pt}
\end{wrapfigure}

Intermediate units are not assumed to be mostly online or mostly offline. Instead, each one has its optimal program determined by the full optimal dispatch problem, described in Section \ref{subsec:OptDisp_DAO}.

These heuristics used for the Base and Peak strategies do involve certain rather strong simplification assumptions, but allow for significantly reduced calculation time. They also ensure that these units will offer available capacity at every moment, whereas a strategy based on a dispatch optimization assuming certain price forecasts does not. The choice of strategy is a parameter of each unit and therefore the use of these assumptions depends on the scope of each study. The Intermediate strategy has the advantage of a much more realistic distribution of startup costs. 

\subsubsection{Base Strategy} \label{subsec:BaseStrategy}

As mentioned above, units employing a Base strategy generally have longer start up durations and lower marginal costs. The Base strategy order books are formulated as shown in Figure \ref{fig:DAO_algoBase}.  

The first step in the process to generate these offers is to assess the forecast production plan. This production plan is translated into a vector of state variables, referred to in Figure \ref{fig:DAO_algoBase} as the "baseload states sequence". If there are periods where the unit is offline, it is assumed that these represent maintenance periods and thus the unit cannot offer at these times.

\begin{figure}[!ht]
\centering
\begin{subfigure}{0.8\textwidth}
    \centering
    \includegraphics[width=\textwidth]{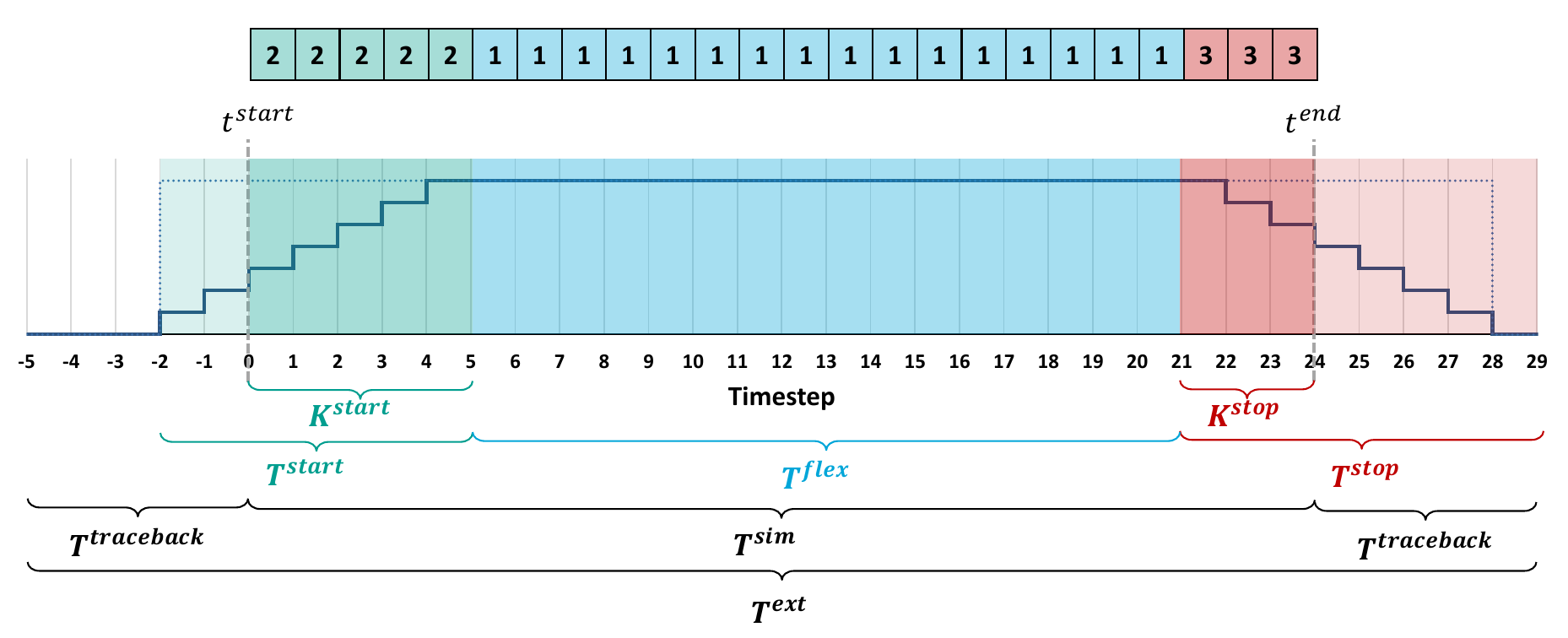}
    \caption{Production Example 1}\label{fig:timeFrames1}
\end{subfigure}
    % \hfill
\begin{subfigure}{0.8\textwidth}
    \centering
    \includegraphics[width=\textwidth]{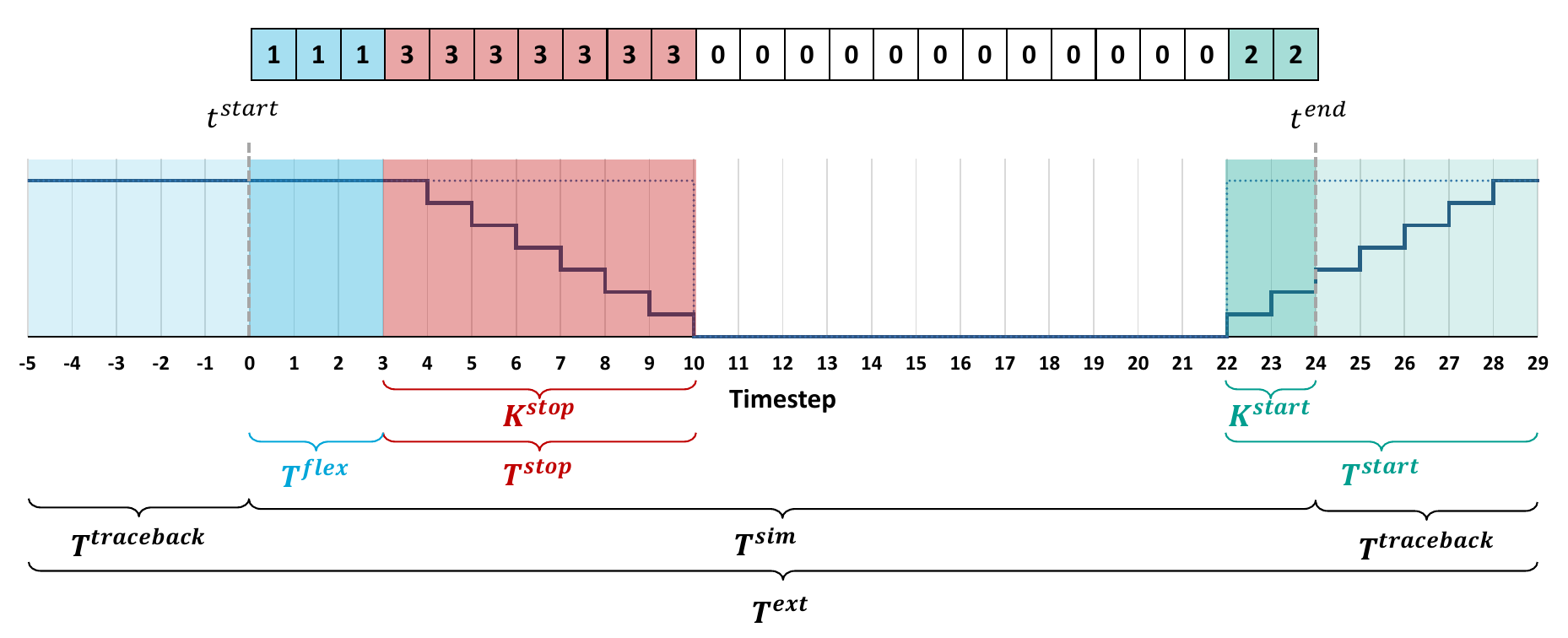}
    \caption{Production Example 2}\label{fig:timeFrames2}
\end{subfigure}
\caption{Two Base State Sequence Examples}
\label{fig:timeFrames_all}
\end{figure}

The online time period is deduced based on the values of different startup and shutdown phase durations. This online period is then split into several blocks. For instance, if the unit changes state, there will be a startup or shutdown phase as well as a time frame referred to as the \textit{flexible time frame} ($T^{flex}$) where the unit is at a stable power level. Two examples of different dispatches and the different related time periods are shown in Figures \ref{fig:timeFrames1} and \ref{fig:timeFrames2}. These figures also show the corresponding state vector representation, where offline periods are represented as 0, online stable as 1, startup as 2 and shutdown as 3. 

From these state assessments, there are several different potential order books created, based on this initial assessed state. Essentially, these translate into a combination of indivisible and divisible orders combined with either Parent Child or Identical Volume Couplings depending on the different Cases shown in Figure \ref{fig:DAO_algoBase}.

\begin{figure}[!ht]
\centering
\begin{subfigure}{0.3\textwidth}
    \centering
    \includegraphics[width=\textwidth]{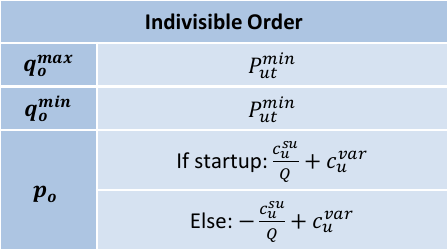}
    \caption{Indivisible Orders}\label{fig:DAO_baseOffers_indivisible}
\end{subfigure}
    % \hfill
\begin{subfigure}{0.3\textwidth}
    \centering
    \includegraphics[width=\textwidth]{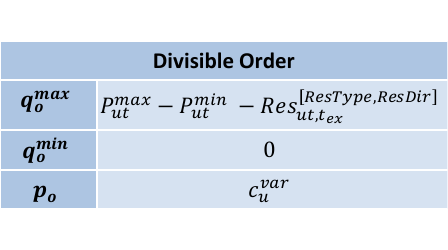}
    \caption{Divisible Orders}\label{fig:DAO_baseOffers_divisible}
\end{subfigure}
\begin{subfigure}{0.3\textwidth}
    \centering
    \includegraphics[width=\textwidth]{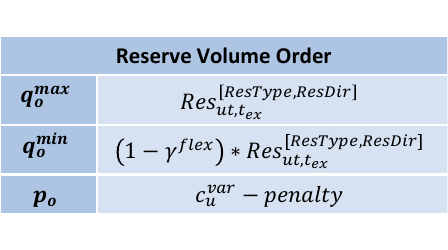}
    \caption{Reserve Volume Orders}\label{fig:DAO_baseOffers_resVol}
\end{subfigure}
\caption{Order Types - Thermal Plants} \label{fig:DAO_baseOffers_all}
\end{figure}

These orders are defined as shown in Figure \ref{fig:DAO_baseOffers_all}. 

\vspace{10pt}

\vspace{15pt}

For these offers, we define a value, $Q$, as follows: 
\begin{equation}
\centering
    Q =  K_{start}* \left( T_u^{SU} + \frac{3-K_{start}}{2} \right) * \frac{P^{min}_{ut}}{T_u^{SU}+1} + K_{stop}* \left( T_u^{SD} + \frac{3-K_{stop}}{2}* \frac{P^{min}_{ut}}{T_u^{SD}+1} + \right) \sum_{t\in T^{flex}} P_{ut}^{min} 
\end{equation}

The reserve volume offers are separated into automated and manual volumes and up and down directions. These orders are created at a price equal to the variable cost for the unit minus a user-defined penalty. This penalty corresponds to the amount the unit would pay for not providing these reserves. This allows the unit to arbitrate between the day ahead market price and the procured reserve market prices.

\begin{figure}[!ht]
    \centering
    \includegraphics[width=\textwidth]{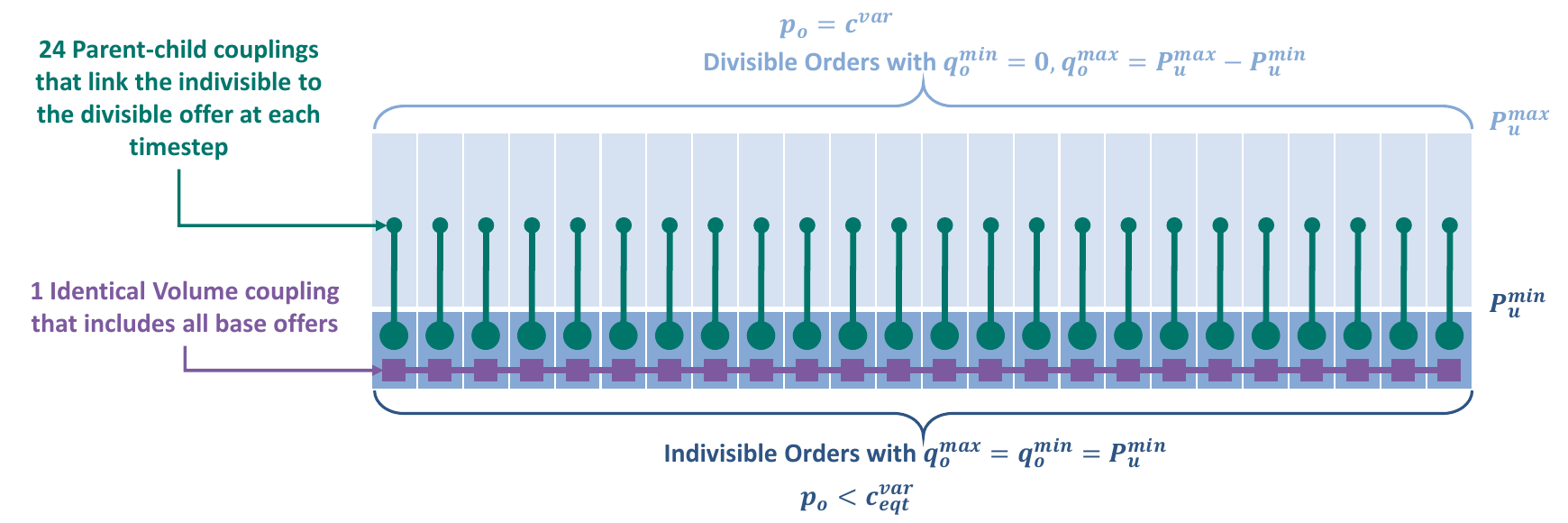}
    \caption{Base Strategy Orders - Constant State}
    \label{fig:baseStrategy_baseOrders}
\end{figure}

For instance, a base load unit with $ 0 < P_u^{min} < P_u^{max}$ that has a constant state (i.e. no maintenance planned) and no previously procured reserves would create the set of orders shown in Figure \ref{fig:baseStrategy_baseOrders}. 

\subsubsection{Peak Strategy} \label{subsec:Peak}

\textcolor{red}{TO BE COMPLETED}

\subsubsection{Intermediate Strategy} \label{subsec:IntermediateStrategy}

Thermal units with an intermediate offer strategy generate their offers in two stages. First, they perform a unit-based optimal dispatch, as described in Section \ref{subsec:OptDisp_DAO}. Each unit aims to minimize its per unit cost. 

Following the optimal dispatch phase, it is necessary to translate the optimal production plan into a coherent order book. This is done in a method similar to that of the Base and Peak strategies. 

\textcolor{red}{TO BE COMPLETED}

\subsection{Wind, Photovoltaic, Load and Non-Dispatchable Production}
Load, wind and photovoltaic units are all treated according to the same principles. They share the same type of property, called MaximumPowerForecast, a matrix that provides load or generation forecasts done at every hour of the year, over the next n hours. An example of the MaximumPowerForecast matrix of a wind generation unit with graphical representations can be found in Figure \ref{fig:forecastUncertainty}. 

% \begin{figure}[H]
%     \hspace*{-0.4in}
%     \includegraphics[width = 0.9 \paperwidth]{images/power_forecast_matrix_and_curves.png}
%     \caption{Subset of a MaximumPowerForecast matrix from a wind unit (right side), and a graphical representation of this table (left side)}
%     \label{fig:power_forecast_matrix_example}
% \end{figure}

In the matrix on the left side of the figure, each column is indexed by a forecasting date, and contains forecasts for the next n hours. As this figure shows, load units (resp. wind and photovoltaic units) have at their disposal a constantly updated view of their power demand (resp. generation) that includes forecasts errors and uncertainties.

An order $o$ created by a load unit for the time $t \in T$ has the following characteristics:
\begin{gather*}
    \forall u \in U^{l}
\end{gather*}
\begin{equation} \label{eq:qmax_load}
    \centering
    q^{max}_{o} = P^{for}_{u,t}
\end{equation}
\begin{equation} \label{eq:qmin_load}
    \centering
    q^{min}_{o} = 0
\end{equation}
\begin{equation} \label{eq:price_load}
    \centering
    p_{o} = p^{load}
\end{equation}
Where $p^{load}$ is an exegenous load price defined as a parameter of the Day-Ahead Orders module.

A specificity of wind and photovoltaic units is that they can be curtailed up to $Curt_{u}$ of their MaximumPowerForecast, $Curt_{u}$ being a parameter provided by the user of ATLAS. In addition, the price of their orders is given by their variable cost (usually set at 0, but that can be modified to replicate phenomena such as feed-in-tariffs).  Consequently, an order o created by one of these units has the following characteristics:
\begin{gather*}
    \forall u \in U^{w} \cup U^{pv}
\end{gather*}
\begin{equation} \label{eq:qmax_wind_pv}
    \centering
   q^{max}_{o} = P^{for}_{u,t}
\end{equation}
\begin{equation} \label{eq:qmin_wind_pv}
    \centering
    q^{min}_{o} = Curt_u * P^{for}_{u,t}
\end{equation}
\begin{equation} \label{eq:price_wind_pv}
    \centering
    p_{o} = c^{var}_{u}
\end{equation}

The last type of unit is called \textbf{OtherNonDispatchable}, and regroups all units that are required to produce and cannot be dispatched, such as run-of-river hydraulic units. The formulation of an order $o$ is then simply:

\begin{equation} \label{eq:qmax_qmin_ond}
    \centering
    q^{max}_{o} = q^{min}_{o} = P^{max}_{u,t}
\end{equation}
\begin{equation} \label{eq:price_ond}
    \centering
    p_{o} = c^{var}_{u}
\end{equation}
\subsection{Flexible load (Power-to-Gas Units)}

\textcolor{red}{TO BE COMPLETED}

\subsection{Storage} \label{subsec:DAO_storage}

Storage units regroup units that have a reservoir and can both produce energy or consume it to store it. For this kind of unit, an optimization problem is yet again necessary, and is solved over the market effective period plus an additional period to avoid having an outcome generation plan heavily influenced by having a short-term vision of a single day. 

Simply put, based on a price forecast, each storage unit will try to dispatch its buying and selling times by maximizing selling in times when the price is forecasted to be high, and buying when it is low. The optimal dispatch problem is described in Section \ref{subsec:OptDisp_DAO}.

Similarly to thermal units, the generation plan obtained after the resolution of the optimization problem is then translated into market orders. The price $p_{o}$ of these orders is computed based on price forecasts.

\textcolor{red}{TO BE COMPLETED}

\subsection{Hydraulic}

Hydraulic units have a reservoir of energy and can only produce (meaning that they can’t consume energy, which excludes pumped hydraulic storage that are treated as storage units). They offer all their available capacity, divided into 7 orders to avoid an effect of all or nothing on hydraulic units. The price of all of these orders is defined according to the water value of the unit, which gives an indication of the value of the current energy stored in the unit reservoir by estimating the opportunity cost of using it later in the year. The method used to compute water values for every hydraulic unit of a power system is based on Bellman’s algorithm \cite{bellman1966dynamic}.

\section{Intraday Orders} \label{ch:IntradayOrders}

The Intraday Orders module is the first of the Intraday market chain. The level of abstraction of the ATLAS model allows either to simulate successive hourly-framed markets (similar to the continuous market in place at the date of this publication) or to experiment with new regulations, such as a few daily-framed markets.

The Intraday Orders module formulates orders for the Intraday market, which aims at :
\begin{itemize}
    \item Taking into account the up-to-date forecast of load and VRE production and computing a new price forecast ;
    \item Based on that new forecast, optimizing once again the production plan of each portfolio and taking opportunities to achieve better coordination between actors ;
    \item Bridging the gap between consumption and production inherited from the last market due to portfolios exposing themselves to imbalance penalties.
\end{itemize}

To achieve these three goals, the module involves three corresponding ghost modules which are executed consecutively. 

\subsection{Intraday Price Forecast} \label{IDPriceForeast}

The first ghost module generates a new price forecast for each market area \(\displaystyle  z \in Z \). 

\subsubsection{Nomenclature}

For each market area, we define the following nomenclature.

\begin{longtable}[!ht]{!{\color{Grey}\vrule}>{\centering\arraybackslash}p{2.5cm} !{\color{Grey}\vrule} p{12.5cm}!{\color{Grey}\vrule} }
    \arrayrulecolor{Grey} \hline
    \rowcolor{Grey} \multicolumn{2}{|c|}{\large{\textbf{General temporal parameters}}} \\ 
    \textbf{Notation} & \textbf{Meaning}\\ \hline
        $T_{ID}$ & Complete time frame of the Intraday market \\ 
        \hline
        \(\displaystyle  t \in T_{ID} \) & Date within the market time frame \\ 
        \hline
        $t^{ex}_{ID}$ & Date at which the Intraday market is cleared. \\ 
        \hline
        $t^{ex}_{DA}$ & Date at which the former Day-ahead market was cleared. \\ 
        \hline
\end{longtable}

\begin{longtable}[!ht]{!{\color{Grey}\vrule}>{\centering\arraybackslash}p{2.5cm} !{\color{Grey}\vrule} p{12.5cm}!{\color{Grey}\vrule} }
    \arrayrulecolor{Grey} \hline
    \rowcolor{Grey} \multicolumn{2}{|c|}{\large{\textbf{Inputs and outputs}}} \\ 
    \textbf{Notation} & \textbf{Meaning}\\ 
        \hline
        $\tilde{\lambda}_{z, t}^{\bullet}$ & Long term price forecast in the market area $z$ at time $t$, with $\bullet$ standing for either the $LOW$, $MEDIUM$ or $HIGH$ scenario. Historical data or prospective simulations based on dispatch models such as Antares can be used. These scenarios are clarified in Figure \ref{fig:dao_processOverview} \\ 
        \hline
        $\tilde{\lambda}_{z,t,t^{ex}}$ & Price forecast in the market area $z$ at time $t$, seen from $t^{ex}$ \\
        \hline
        $\lambda^{DA}_{z,t}$ & Cleared Day-Ahead price in the market area $z$ at time $t$ \\
        \hline
        $P_{z,t}^{max, \bullet}$ & Total baseload consumption forecast in the market area $z$ at time $t$, either in the $LOW$ or $HIGH$ scenario. These scenarios are clarified in Figure \ref{fig:dao_processOverview} \\  
        \hline
        $P^{max}_{u,t,t^{ex}}$ & Maximum production forecast of unit $u$ at time $t$, seen from $t^{ex}$ \\ 
        \hline
        $\Delta RD_{z,t,t^{ex}}$ & The variation of residual demand in market area $z$ at time $t$, seen from $t^{ex}$ using the last cleared market as reference\\ 
        \hline
\end{longtable}

\subsubsection{Process}

The price forecast $\tilde{\lambda}$ is obtained as the sum of the Day-ahead cleared price and the expected price evolution, product of the evolution of the residual demand and the price sensitivity to the demand. The following equation illustrates this process :

\begin{align}
    \centering
    \forall t \in T_{ID},\forall z \in Z,\quad \tilde{\lambda}_{z,t,t^{ex}_{ID}} = \lambda^{DA}_{z,t} + \Delta RD_{z,t,t^{ex}_{ID}} * \frac{\tilde{\lambda}_{z, t}^{HIGH} - \tilde{\lambda}_{z, t}^{LOW}}{P_{z,t}^{max,HIGH} - P_{z,t}^{max,LOW}} 
\label{eq:idpriceforecast}
\end{align}

where :

\begin{align}
    \centering
    \forall t \in T_{ID},\forall z \in Z,\quad \Delta RD_{z,t,t^{ex}_{ID}} = \sum_{u \in U^{l}}{P^{max}_{u,t,t^{ex}_{ID}} - P_{u,t,t^{ex}_{DA}}^{max}} - \sum_{u \in U^{w}}{P^{max}_{u,t,t^{ex}_{ID}} - P^{max}_{u,t,t^{ex}_{DA}}} - \sum_{u \in U^{pv}}{P^{max}_{u,t,t^{ex}_{ID}} - P^{max}_{u,t,t^{ex}_{DA}}}
\end{align}

\subsubsection{Special use case: several intraday markets}

In the use case where several Intraday markets are simulated consecutively on overlapping time frames (but with different execution dates), the price forecast is not computed as an affine function of the DayAhead price $\lambda^{DA}$ but rather as an affine function of the mean value of known prices at any given time. This method is used to grasp the overall price tendency instead of being influenced by a single market which might introduce a bias due to a lack of coordination between actors with restrained vision. \newline

For instance, if the Intraday market is executed twice, the second price forecast computation is given by:

\begin{align}
    \centering
    \forall t \in T_{ID},\forall z \in Z,\quad \tilde{\lambda}_{z,t,t^{ex}_{ID2}} = \frac{\lambda^{DA}_{z,t} + \lambda^{ID1}_{z,t}}{2} + \Delta RD_{z,t,t^{ex}_{ID2}} * \frac{\tilde{\lambda}_{z,t}^{HIGH} - \tilde{\lambda}_{z,t}^{LOW}}{P_{z,t}^{max,HIGH} - P_{z,t}^{max,LOW}} 
\end{align}

where :

\begin{align}
    \centering
    \forall t \in T_{ID},\forall z \in Z,\quad \Delta RD_{z,t,t^{ex}_{ID2}} = \sum_{u \in U^{l}}{P^{max}_{u,t,t^{ex}_{ ID2}} - P^{max}_{u,t,t^{ex}_{ID1}}} - \sum_{u \in U^{w}}{P^{max}_{u,t,t^{ex}_{ID2}} - P^{max}_{u,t,t^{ex}_{ID1}}} - \sum_{u \in U^{pv}}{P^{max}_{u,t,t^{ex}_{ID2}} - P^{max}_{u,t,t^{ex}_{ID1}}}
\end{align}

\subsection{Portfolio Optimization} \label{IDPO}

The process of taking into account the new price forecast is the same used in the Portfolio Optimization module presented in Chapter \ref{ch:OptimalDispatch} (but based on a forecast rather than a cleared market price). The same module is thus used here as a ghost module to generate an optimal production plan $Q^{OptimalDispatch}$ for each unit  (from the portfolio perspective and given the uncertainty of the forecast), used as a reference by the next ghost module to create ID market orders. The module requires running the optimization of the production on all units to match overall consumption, yet the only units that use the result of this optimization for the Intraday market are thermic and storage units. An optimization period longer than the studied time frame is used for both storage and thermic units to address a short-term bias; however, the price previously forecasted may not cover the entire extended timeframe. In the remaining time steps, the price forecast $\tilde{\lambda}^{MEDIUM}$ is used.

\subsection{Intraday Orders Formulation} \label{IDOrders}

The last module formulates orders to the Intraday market using the optimization results computed previously.

In all sections, the following notations are used to indicate the sum of previously cleared quantities on market $\bullet$:

\begin{equation}
    Qcleared^{\bullet}_{u,t} = \sum\limits_{o \in O^{\bullet}_{u,t}} - \sigma_{o} * q^{acc}_{o}
\end{equation}
Where $O^{\bullet}_{u,t}$ is the set of orders with the start date $t^{start}_{o} = t$, formulated on market $\bullet$ by unit $u$.

\subsubsection{Thermic units}

\paragraph{Base and Intermediate strategy}

For both the Base and the Intermediate strategy, the bidding strategy is the following :
\begin{itemize}
    \item For each thermic unit, the optimal production plan computed in the last ghost module is compared to the sum of all previously cleared quantities (Day Ahead market and potential Intraday market on the same timeframe) ;
    \item The delta planning is transcripted as hourly sequences according to the current state of the unit. Possible sequences are: starting or shutting down the unit, extending or reducing the operating period (anticipated or delayed startup, anticipated or delayed shut down) or modulating the output power in flexible operation.
    \item Corresponding orders are formulated to the market using appropriate order coupling and including the startup cost if necessary, similar to the process used in Day Ahead Orders in Section \ref{subsec:DAO_thermal}.
\end{itemize}

\paragraph{Peak strategy}

For the Peak strategy, the result of the optimization is not used; similar to the Day Ahead Orders module in Section \ref{subsec:Peak}, all available power is submitted using both inflexible and flexible orders for each hour $t$ of the time frame, corresponding prices and order couplings.

\subsubsection{Hydraulic units}

Hydraulic units submit orders based on the results of all previous markets. 

The logic used in Day Ahead Orders stays the same: they are provided with a reservoir of energy and can only produce. Available capacity is divided into 7 fragments to avoid an effect of all or nothing. The price of all of these orders is defined according to the water value of the unit.

All fragments that were accepted in the former markets are offered as Buy orders at the corresponding price. Fragments that were refused previously are offered as Sell orders on the market at the corresponding price. Fragments that were partially accepted are separated into the two previous cases using the same logic.\\

\subsubsection{Storage units}

For storage units, the optimization computed in the second ghost module (Section \ref{IDPO}) uses the price forecast from the first ghost module (Section \ref{IDPriceForeast}) to dispatch buying and selling times to maximize profit, similar to the strategy used in Section \ref{subsec:DAO_storage}.
In this submodule, the results are used to compute maximal buying and minimum selling prices assuring economical break-even for the unit when taking into account efficiency. These are then used as prices respectively for buying and selling orders. Then, we compare the computed quantity from the optimal dispatch problem, $Q^{OptimalDispatch}$, to the previously cleared quantities to send orders to the market.

An order $o$ created by a storage unit at $t^{ex}$, for the time $t \in T_{ID}$ has thus the following characteristics:
\begin{gather*}
    \forall u \in U_{pf}^{st}
\end{gather*}
\begin{equation} \label{eq:qmax_storage_ido}
    \centering
    q^{max}_{o} = \abs{Qcleared_{u,t}^{Day Ahead} + Qcleared_{u,t}^{Previous Intraday} - Q_{u,t,t^{ex}}^{OptimalDispatch}}
\end{equation}
\begin{equation} \label{eq:qmin_storage_ido}
    \centering
    q^{min}_{o} = 0
\end{equation}
\begin{equation} \label{eq:price_storage_ido}
    \centering
    p_{o} = \left\{
                \begin{array}{ll}
                 MinimalSellPrice & \mbox{for a sell order} \\
                 MaximalBuyPrice & \mbox{for a buy order} 
                \end{array}
                \right.
\end{equation}

Furthermore, electric vehicles use order coupling to formulate an overall COMPLEMENT order (meaning that individual order volumes may be changed by the market but not the sum itself). This is used both to make sure that the energy used for vehicle displacement is provided and to make up for partial information of the actor by delegating to the market the right to modify selling and buying volumes (under the hypothesis that the market's interest and the actor's are aligned).

\subsubsection{Load, wind and photovoltaic units}
Similar to Day Ahead Orders, load, wind and photovoltaic units are all treated according to the same principles. They formulate orders to the Intraday market based on the evolution of either production or consumption forecast. The Day Ahead market results are used as a reference.

If several Intraday markets are simulated for the same time frame, ATLAS keeps track of previous Intraday market results. 

Hence an order $o$ created by a load unit at $t^{ex}$, for the time $t \in T_{ID}$ has the following characteristics:
\begin{gather*}
    \forall u \in U_{pf}^{l}
\end{gather*}
\begin{equation} \label{eq:qmax_load_ido}
    \centering
    q^{max}_{o} = \abs{Qcleared_{u,t}^{Day Ahead} + Qcleared_{u,t}^{Previous Intraday} - Qforecast_{u,t,t^{ex}}}
\end{equation}
\begin{equation} \label{eq:qmin_load_ido}
    \centering
    q^{min}_{o} = 0
\end{equation}
\begin{equation} \label{eq:price_load_ido}
    \centering
    p_{o} = p^{load}
\end{equation}

Wind and photovoltaic units formulate two orders: one to update the production forecast and one curtailment order.
A forecast order $o$ created by one of these units has the following characteristics :
\begin{gather*}
    \forall u \in U_{pf}^{w} \cup U_{pf}^{pv}
\end{gather*}
\begin{equation} \label{eq:qmax_wind_pv_ido}
    \centering
    q^{max}_{o} = \abs{Qcleared_{u,t}^{Day Ahead} + Qcleared_{u,t}^{Previous Intraday} - Qforecast_{u,t,t^{ex}}}
\end{equation}

\begin{equation} \label{eq:qmin_wind_pv_ido}
    \centering
    q^{min}_{o} = 0
\end{equation}

\begin{equation} \label{eq:price_wind_pv_ido}
    \centering
    p_{o} = \left\{
                \begin{array}{ll}
                 c^{var}_u & \mbox{for a sell order} \\
                 c^{var}_u + p^{largeImbal,down}_t*\tilde{\lambda}_{z,t, t^{ex}} & \mbox{for a buy order} 
                \end{array}
                \right.
\end{equation}

If the new forecast is lower than the up-to-date cleared quantity, the unit may not want to buy the lacking energy if market prices are high and uses the forecasted imbalance price as a decision criterion when formulating buy orders in Equation \ref{eq:price_wind_pv_ido}.

A curtailment order $o$ created by one of these units has the following characteristics:

\begin{gather*}
    \forall u \in U_{pf}^{w} \cup U_{pf}^{pv}
\end{gather*}
\begin{equation} \label{eq:qmax_wind_pv_ido_curt}
    \centering
    q^{max}_{o} = P^{for}_{u,t}
\end{equation}
\begin{equation} \label{eq:qmin_wind_pv_ido_curt}
    \centering
    q^{min}_{o} = 0
\end{equation}
\begin{equation} 
    \centering
    p_{o} = c^{var}_u
\end{equation}

Note that the $Curt_u$ is not yet used in Equation \ref{eq:qmax_wind_pv_ido_curt} and will be the subject of a future update. This means that, in the Intraday market, it is assumed that such units can be entirely curtailed.

\subsubsection{Other Non-dispatchable Units}
All units regrouped in this category have a production considered fatal and certain (MaximumPowerForecast is the same at any given $ExecutionDate$). Thus, their production plan stays the same and they do not formulate orders on the Intraday market. The inherent hypothesis is that ATLAS favors photovoltaic and wind units for curtailment.

\chapter{Market Clearing} \label{ch:MarketClearing}

The Market Clearing module was created through a merge of the original market clearing algorithm of the Optimate project and the TERRE algorithm at the time of the merge (2017).

% \subsection{Offer Types}

% To mimic the clearing of the different electricity markets, the ATLAS Market Clearing module accepts XX different order types. A market order is defined by the following characteristics: 

% !TeX root = ../main.tex
% Add the above to each chapter to make compiling the PDF easier in some editors.
\begin{wrapfigure}{r}{8cm}
    \centering
	\includegraphics[width = 7cm]{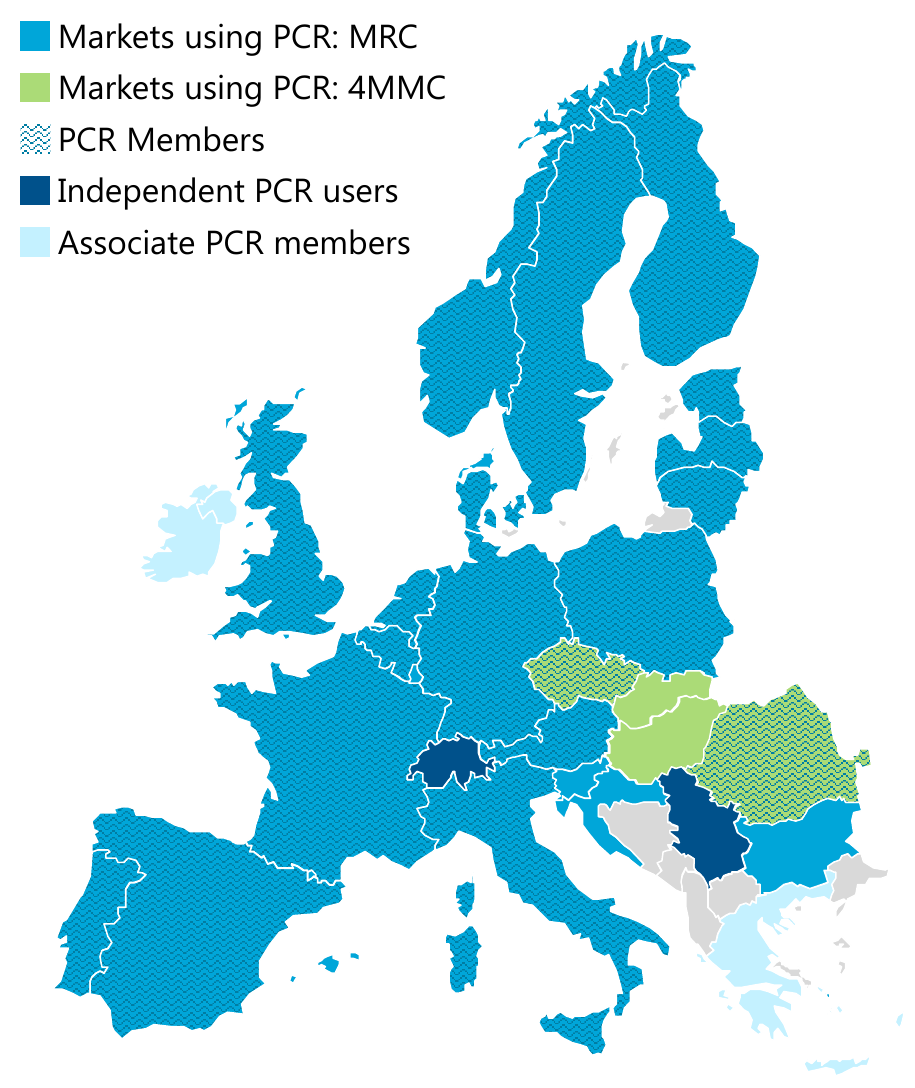}
	\caption[PCR Region.]{PCR Region} %\footnotemark{}}
	% \vspace{-30pt}
	\label{figure:pcrmap}
\end{wrapfigure}

Since February 2014, a single price coupling algorithm has been used across Europe known as EUPHEMIA (Pan-European Hybrid Electricity Market Integration Algorithm). EUPHEMIA is currently used across 23 European countries (see Figure \ref{figure:pcrmap}), enabling trade of more than EUR 200 million.\footnote{https://www.n-side.com/pcr-euphemia-algorithm-european-power-exchanges-price-coupling-electricity-market/} It processes the orders from seven power exchanges, deciding which offers are accepted or rejected in order to maximize the social welfare and remain within the physical network constraints. The social welfare is a term that represents the sum of the profits of all market players. The market coupling model is a mixed integer linear program based on a simplified version of the EUPHEMIA algorithm with some differences.\footnote{The type of order is less complex for the model discussed here which allows, for one, the objective function of this model to be linear, rather than quadratic as it is in EUPHEMIA.}

\begin{figure}[H]
    % \vspace{-15pt}
    \centering
    \includegraphics[width=0.6\textwidth]{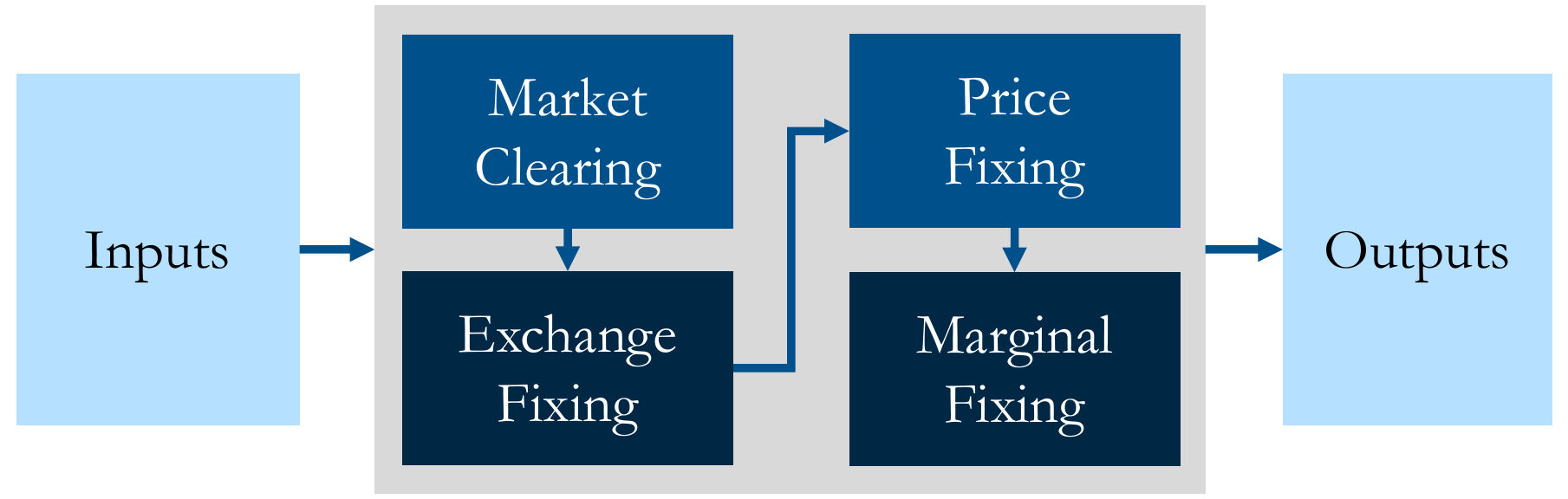}
    \caption{Model Structure}
    \label{fig:model_desc}
    % \vspace{-10pt}
\end{figure}

Both algorithms rely on the concept of duality in order to determine the quantity of accepted market offers as well as the market clearing price. The ATLAS Market Clearing model is broken down into four sub-problems (see Figure \ref{fig:model_desc}): 1) the Clearing phase, 2) the Exchange Fixing phase, 3) the Price Fixing phase, and 4) the Marginal Fixing phase. The Clearing phase is the primal problem, with the Price Fixing as its dual. The other two modules are subproblems, used to improve the solution generated by the Clearing module. 

The model takes as an input the orders to market from each market area, as well as the market border constraints, whether this is the ATC method or Flow-based Method (see \autoref{app:CapCalc} for a full explanation of the differences between the two). From these inputs, the algorithm matches purchase and sale offers to maximize the overall social welfare. It also calculates the market prices and other values and market indicators, including the cross-border capacity flow and the congestion rents. 

In the following sections, these modules will be discussed in detail.

%\end{tikzpicture}
\section{Inputs}\label{sec:Inputs}

The inputs to the algorithm come in the form of a packet of data in the PROMETHEUS platform (discussed in Chapter \ref{ch:Introduction}. The complete structure of the data packet can be seen in Figure \ref{fig:ATLASmodel}, while the specific pieces of the data shell relevant to the Market Coupling algorithm are summarized in the following sections. The exhaustive list of terms and notations that will be used in the following pages is included in Section \ref{sec:Nomenclature}. The key inputs as discussed are the data associated with each offer, the physical network limits and the geographical market zones. 

\subsection{Order Data}\label{subsec:OrderInputData}

\begin{figure}[!ht]
    \centering
    \includegraphics[width=0.5\textwidth]{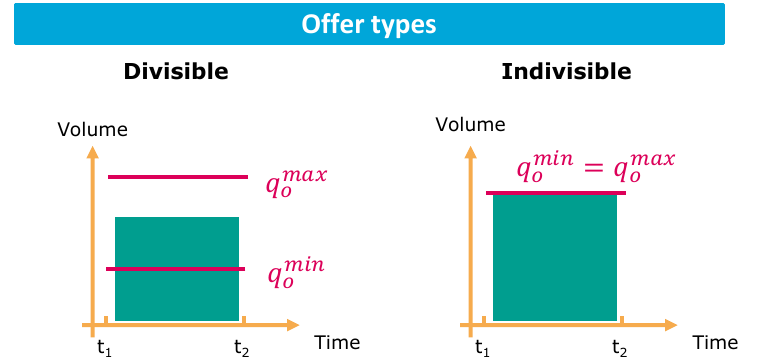}
    \caption{Basic Order Types}
    \label{fig:basicOrderTypes}
\end{figure}

The model takes as a main input all the offers to market. Each order is defined by the characteristics shown in Table \ref{tab:orderData_mc}, as well as by one to two optimization variables associated with it, depending on the type of offer. First off, orders can be divisible or indivisible, as shown in Figure \ref{fig:basicOrderTypes}.  

In the ATLAS model, orders are almost exclusively generated for a specific time step. In order to simulate the various order types--as well as orders defined across multiple time steps--found in the different European operational market clearing algorithms, market orders can be coupled in several ways. Currently the following coupling types are implemented (see Figure \ref{fig:orderCouplings}):

\begin{figure}[!ht]
    \centering
    \includegraphics[width=\textwidth]{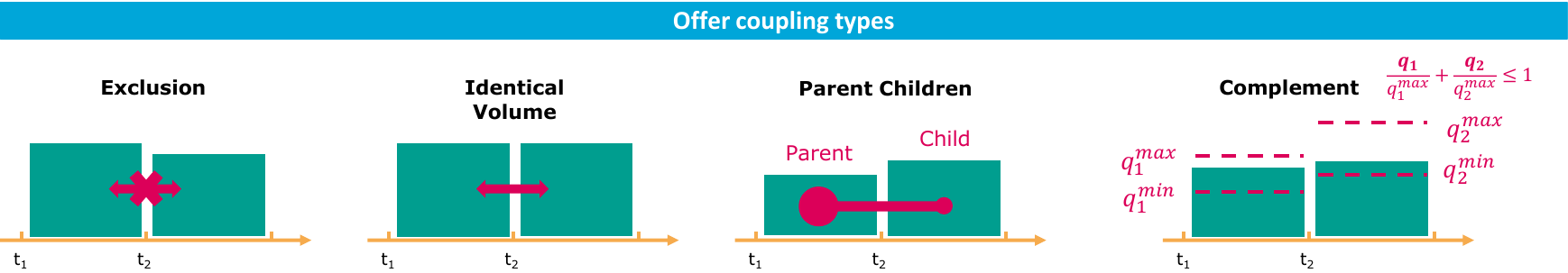}
    \caption{Order Couplings}
    \label{fig:orderCouplings}
\end{figure}

\begin{itemize}
    \item \textbf{\textit{Exclusion}}: A single order among the coupled orders can be accepted.
    \item \textbf{\textit{Parent Child}}: The orders linked by a parent-child coupling can only be accepted if the order defined as the parent is accepted.
    \item \textbf{\textit{Identical Volume}}: The quantity accepted of each order must be identical.
    \item \textbf{\textit{Identical Ratio}}: The ratio of the accepted quantity of each order must be identical.
    \item \textbf{\textit{Complement}}: The sum of accepted quantities of all orders must be less than a given value $E^{COMP}_{co}$.
\end{itemize}

The second entry in the table, \bm{$\delta_{o}$}, is a binary variable that represents the state of acceptance of the offer. It is only present for offers that have a minimum power. It is especially important for indivisible offers--offers that have a single feasible power and thus can not be partially accepted. These offers are characterized by a minimum power equal to their maximum power. This comes into play in the clearing module within the constraints that limit the total accepted quantity to within the operating region specified in the order. The offer data is generated based on price forecasts for each zone and the technical constraints of each equipment. This process is described in more detail in \cref{ch:OrderCreation}.
% \autoref{ch:OrderCreation}. 
% Chapter~\ref{ch:OrderCreation}. 

\begin{longtable}[!ht]{!{\color{Grey}\vrule}>{\centering\arraybackslash}p{2.5cm} !{\color{Grey}\vrule} p{12.5cm}!{\color{Grey}\vrule} }
    \arrayrulecolor{Grey} \hline
    \rowcolor{Grey} \multicolumn{2}{|c|}{\textbf{Order characteristics}} \\ 
    \hline
    \rowcolor{Grey} 
    \textbf{Notation} & \textbf{Meaning}\\ \hline
    $\boldsymbol{q_o}$ & \textbf{Variable} - Amount of power accepted for order $o$ \\
    $\boldsymbol{\delta_o}$ & \textbf{Variable} - Binary unit commitment variable for order $o$ \\
    $p_o$ & Price of order $o$ \\
    $q_o^{min}$ & Minimum quantity of power offered for order $o$ \\
    $q_o^{max}$ & Maximum quantity of power offered for order $o$ \\
    $t_o^{start}$ & Start time of order \\
    $t_o^{exec}$ & Creation date of order \\
    $t_o^{end}$ & End time of order \\
    $d_{o}$ & Duration of order \\
    $\sigma_{o}$ & Sale/Purchase indicator, $\sigma = 1$ for purchase, $-1$ for sale \\ 
    \hline 
    \caption{Order data} \label{tab:orderData_mc}
\end{longtable}

\subsection{Geographic Data}\label{subsec:GeographicalData}

The geographical data is split first into different control blocks. Each control block is overseen by a single control entity and can contain multiple market areas or zones. During the market coupling algorithm--specifically during the pricing module--a market price will be determined for each zone. 

\begin{longtable}[!ht]{!{\color{Grey}\vrule}>{\centering\arraybackslash}p{2.5cm} !{\color{Grey}\vrule} p{12.5cm}!{\color{Grey}\vrule} }
    \arrayrulecolor{Grey} \hline
    \rowcolor{Grey} \multicolumn{2}{|c|}{\large{\textbf{Zonal characteristics}}} \\ 
    \hline
    \rowcolor{Grey} 
    \textbf{Notation} & \textbf{Meaning}\\ \hline
    $c_z$ & Control block $c$ owning zone $z$ \\
    $z_l^U$, $z_l^D$ & Upstream and downstream zone defined for market border $mb_l$, respectively \\
    $(\Delta q)^{bal,ref}_{z,t} $ & Supply/demand reference balance in zone $z$ at time $t$ \\
    $\boldsymbol{(\Delta q)^{bal}_{z,t} }$ & \textbf{Variable (FB mode)} - Supply/demand balance in zone $z$ at time $t$ \\
    \hline 
    \caption{Zonal data} \label{tab:zonalData_mc}
\end{longtable}

\subsection{Cross-border Data}\label{subsec:CrossBorderData}
Depending on the type of network representation desired, different input data will be used. ATC constrains the cross-border quantity in terms of market borders. For each market border, there is a maximum and minimum Net Transfer Capacity (NTC). Each market border is represented once, so the minimum term represents the maximum flow in the opposite direction.

The Flow-based mechanism constrains the cross-border quantity for specific critical branches in the system. For each critical branch, there is a maximum flow allowed, a reference flow--representing any flow that has already been planned--and a Flow Reliability Margin (FRM). The FRM is a value set by the TSO for each branch and each time-step that gives a safety margin on the amount of cross-border capacity. Flow-based mode also requires the input of a Power Transfer Distribution Matrix (PTDF) which essentially denotes the impact of every possible trade between each pair of zones on each critical branch in the system (see Appendix \ref{app:CapCalc} for more information).

The variables optimized in each network representation mode are somewhat different, as shown in Tables \ref{tab:zonalData_mc}, \ref{tab:cbData_mc}, and \ref{tab:mbData_mc}.

\begin{longtable}[!ht]{!{\color{Grey}\vrule}>{\centering\arraybackslash}p{2.5cm} !{\color{Grey}\vrule} p{12.5cm}!{\color{Grey}\vrule} }
    \arrayrulecolor{Grey} \hline
    \rowcolor{Grey} \multicolumn{2}{|c|}{\large{\textbf{Critical branch characteristics (Flow-based Mode)}}} \\ 
    \hline
    \rowcolor{Grey} 
    \textbf{Notation} & \textbf{Meaning}\\ \hline
    $\boldsymbol{q_{k,t}^{CB}}$ & Flow of critical branch $cb_k$ at time $t$ \\
    $q_{k,t}^{CB,max}$ & Max flow on critical branch $cb_k$ at time $t$ \\
    $q_{k,t}^{CB,FRM}$ & Flow reliability margin set by TSO on critical branch $cb_k$ at time $t$ \\
    $q_{k,t}^{CB,ref}$ & Reference flow of critical branch $cb_k$ at time $t$ \\
    $D_{t,k,z}^{Z}$ & Zonal power transfer distribution factor for zone $z$ on critical branch $cb_k$ at time $t$ \\
    \hline
    \caption{Critical branch data} \label{tab:cbData_mc}
\end{longtable}

\begin{longtable}[!ht]{!{\color{Grey}\vrule}>{\centering\arraybackslash}p{2.5cm} !{\color{Grey}\vrule} p{12.5cm}!{\color{Grey}\vrule} }
    \arrayrulecolor{Grey} \hline
    \rowcolor{Grey} \multicolumn{2}{|c|}{\large{\textbf{Market border characteristics}}} \\ 
    \hline
    \rowcolor{Grey} 
    \textbf{Notation} & \textbf{Meaning}\\ \hline
    $\boldsymbol{(\Delta q)_{l,t}^{MB}}$ & Net exchange of traded power going through border $mb_l$ (defined from upstream zone $z_l^{U}$ to downstream zone $z_l^D$) at time $t$ \\
    $\boldsymbol{(\Delta q)_{l,t}^{MB,exp}}$ & Power exported on border $mb_l$ (defined from  upstream zone $z_l^{U}$ to downstream zone $z_l^D$) at time $t$ \\   
    $\boldsymbol{(\Delta q)_{l,t}^{MB,imp}}$ & Power imported on border $mb_l$ (defined from  upstream zone $z_l^{U}$ to downstream zone $z_l^D$) at time $t$ \\  
    $(\Delta q)_{l,t}^{MB,max}$ & Maximum power allowed to pass on border $mb_l$ from upstream zone $z_l^{U}$ to downstream zone $z_l^D$ at time $t$ \\   
    $(\Delta q)_{l,t}^{MB,min}$ & Minimum power allowed to pass on border $mb_l$ from upstream zone $z_l^{U}$ to downstream zone $z_l^D$ at time $t$ \\
    $\sigma_{lz}$ & $\sigma_{lz} = 1$ if $z = z_l^{U}$, $\sigma_{lz} = -1$ else\\
    $a_l$ & Technical parameter used for modeling losses on a DC border \\
    $\nu_{l,t}$ & Coefficient used to model losses on DC borders (see Section \ref{ch:MarketClearing}) \\
    $\xi_{l,t}$ and $\xi^{aux}_{l,t}$ & Helper variables for linearizing constraints on DC borders (see Section \ref{ch:MarketClearing}) \\    
    \hline 
    \caption{Market border data} \label{tab:mbData_mc}
\end{longtable}

\section{Market Clearing Algorithm}
\subsection{Clearing}\label{subsec:Clearing}

The market clearing is the first phase of the coupling algorithm and the primal problem. In this phase, purchase and sale offers are associated, while maximizing the overall social welfare--in other words, finding the best solution for the most market players. Equation \ref{objfxn_clearing} shows the full objective function for the clearing, where the bold terms are variables. 
\begin{multline} \label{objfxn_clearing}
\max \sum_{o\in O}\sigma_{o}d_{o}\boldsymbol{q_{o}}(p_o + \sigma_o\lambda_1) - \lambda_2 \sum_{t\in T,mb_l\in MB} \lvert\boldsymbol{(\Delta q)^{MB}_{l,t}}\rvert -\\ \lambda_3 \sum_{t\in T,mb_l\in MB} \lvert\boldsymbol{(\Delta q)^{MB}_{l,t}} - (\Delta q)^{MB,max}_{l,t}\rvert - \lambda_4 \sum_{t\in T,mb_l\in MB} \lvert\boldsymbol{(\Delta q)^{MB}_{l,t}} - (\Delta q)^{MB,min}_{l,t}\rvert
\end{multline}
\vspace{-20pt}
\begin{figure}[H]
    \centering
    \includegraphics[width=0.7\textwidth]{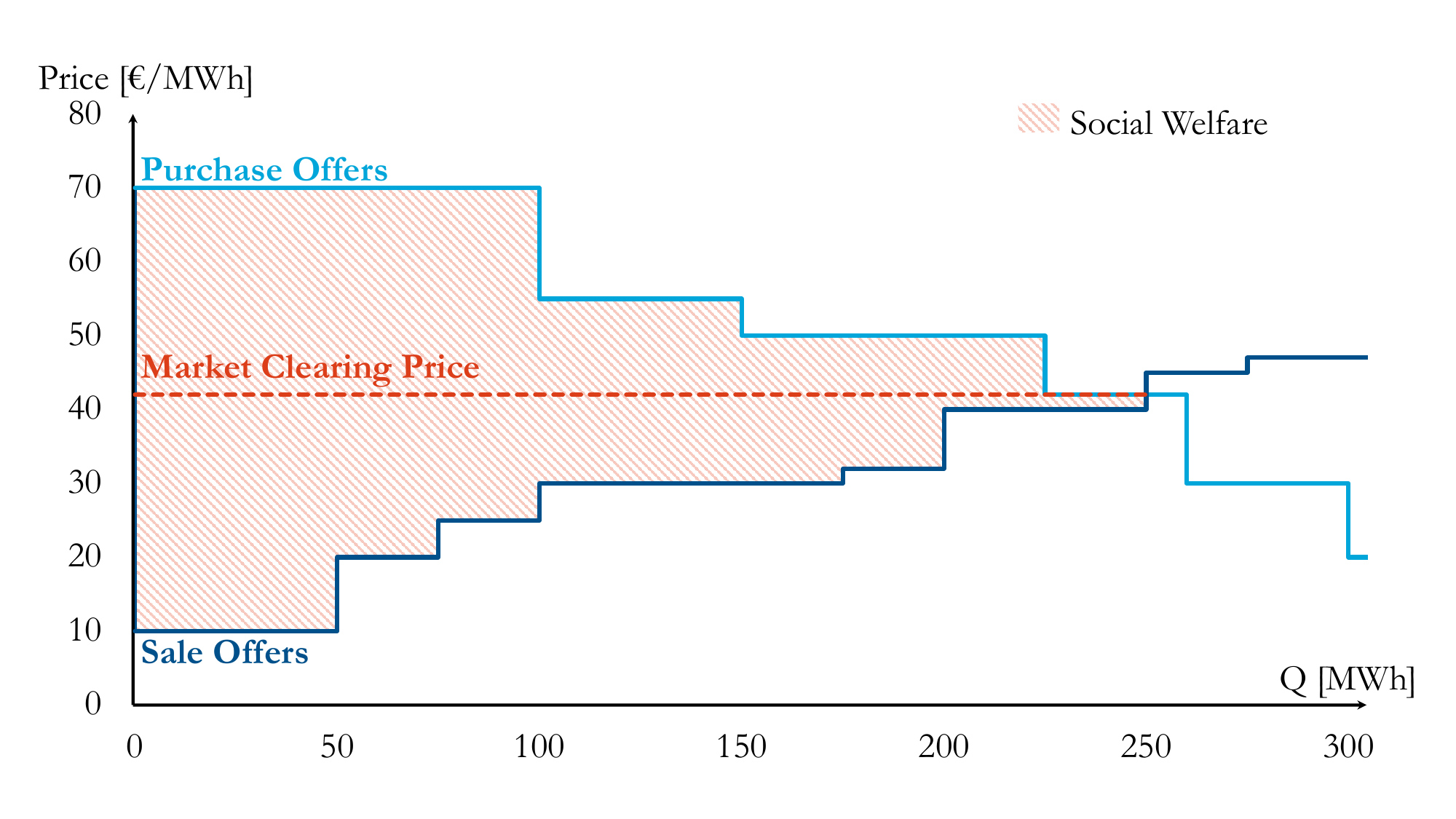}
    \caption{Social Welfare}
    \label{fig:socialwelfare}
\end{figure}
This equation calculates the sum of the social welfare for each product. The $\sigma_{o}$ term is positive for purchase offers and negative for sale offers. Looking at a single product for one zone at one time step, the social welfare is equivalent to the shaded region of the plot in Figure \ref{fig:socialwelfare}.

There are three main types of constraints in the Market Clearing:
\begin{enumerate}
    \item \textbf{Balance constraints:} Everything sold is purchased.
    \item \textbf{Order constraints:} The total quantity accepted for each order remains within the operating range.
    \item \textbf{Cross-border constraints:} The flow (related to intra-zonal commercial exchanges) remains feasible given the physical constraints of the network.
\end{enumerate}
    
\subsubsection{Balance Constraints}\label{subsubsec:BalanceConstraints}
These constraints guarantee that all purchased offers correspond to sale offers and are ensured as a whole across all zones connected by a common market border. This means that if cheaper offers are found in other zones, they can be accepted, assuming the border constraints are not saturated. These constraints are described by Equations \ref{eqn:balance1_mc}-\ref{eqn:balance2_mc}. Any terms not described explicitly in the previous section can be found in Section \ref{sec:Nomenclature}. 

$\forall z\in Z, t\in T:$
\begin{center}
    \begin{align}
    	\sum_{o\in O_{z,t}}-\sigma_{o}q_o = &\Delta q_{z,t}^{bal} \label{eqn:balance1_mc}  \\ where \text{  } &\Delta q_{z,t}^{bal} = \sum_{mb_{l}\in MB_{z}} \sigma_{lzt}\Delta q_{l,t}^{mb} \label{eqn:balance2_mc}
    \end{align}
\end{center}

In the case that there are DC borders where we consider losses, Equation \ref{eqn:balance2_mc} is modified to: 

$\forall z\in Z, t\in T:$
\begin{equation}
    \Delta q_{z,t}^{bal} = \sum_{mb_{l}\in MB_{z} - MB_{z}^L} \sigma_{lzt}\Delta q_{l,t}^{mb} + \sum_{mb_{l}\in MB_{z}^L} \left(\frac{\sigma_{lzt}+1}{2}\Delta q_{l,t}^{mb,exp} +\frac{\sigma_{lzt}-1}{2}\Delta q_{l,t}^{mb,imp}\right) \label{eqn:balance2_losses_mc} 
\end{equation}

Here, the model essentially takes the sum of the orders within each zone at each time step to calculate the local balance and set it equal to the quantity of exchanges exercised in and out of each particular zone.\footnote{The negative sign in the balance calculation was added to maintain the convention that exports are positive and imports are negative.} As briefly mentioned in the previous section, the optimized variables are represented slightly different for flow-based and ATC modes. For ATC, the zonal balances--the $q^{bal}_{z,t}$ terms--are not in fact variables in the optimization, while the market border exchanges--the $\Delta q^{mb}_{l,t}$ terms--are. For flow-based, since the market border terms do not come into play in the cross-border constraints, it is possible to simplify Equations \ref{eqn:balance1_mc} and \ref{eqn:balance2_mc} in this way:\\ \\
$\forall  t\in T:$
\begin{align}
\vspace{-10pt}
    \sum_{z\in Z}\Delta q_{z,t}^{bal} &= 0 \label{eqn:zoneSum_energy} \\
\end{align}

\subsubsection{Order constraints} \label{subsubsec:OrderConstraints}
The order constraints ensure that the entire quantity accepted for each offer remains feasible.

\textbf{Divisible offers}: 

$\forall  o\in O_D:$
\begin{equation}
    0 \leq q_o \leq q_o^{max}
\end{equation}

\textbf{Fully indivisible or partially indivisible offers}: 

$\forall  o\in O_D:$
\begin{equation}
    \delta_o q_o^{min} \leq q_o \leq \delta_o q_o^{max}
\end{equation}

\textbf{Exclusive orders:}

$\forall co \in CO^{E},  :$

\begin{equation}
    \sum_{o_{co} \in \{o_1, ... ,o_n\}} \delta_o \leq 1
\end{equation}

% \textbf{Complementary:}

% $\forall co \in CO^{C},  :$

% Proposition avec la définition dans ATLAS:
% \begin{equation}
%     \sum_{o_{co} \in \{o_1, ... ,o_n\}} q_{o_{co}} * \frac{\Delta t}{60} * \leq E^{COMP}_{co}
% \end{equation}

% \begin{equation}
%     \sum_{o_{co} \in \{o_1, ... ,o_n\}} \frac{q_{o_{co}}}{q^{max}_{o_{co}}} \leq 1
% \end{equation}

% For divisible or partially divisible complement orders, this becomes:
% \begin{equation}
%     \sum_{o_{co} \in \{o_1, ... ,o_n\}} \frac{q_{o_{co}}}{(q^{max}_{o_{co}}-q^{min}_{o_{co}})} \leq 0
% \end{equation}

% \textcolor{red}{Do we want this?}
\textbf{Complementary:}

$\forall co \in CO^{C},  :$
\begin{equation}
    \sum_{o_{co} \in \{o_1, ... ,o_n\}} \frac{q_{o_{co}}}{q^{max}_{o_{co}}} \leq 1
\end{equation}

For divisible or partially divisible complement orders, this becomes:
\begin{equation}
    \sum_{o_{co} \in \{o_1, ... ,o_n\}} \frac{q_{o_{co}}}{(q^{max}_{o_{co}}-q^{min}_{o_{co}})} \leq 0
\end{equation}

We also allow them to be limited in energy through the addition of a value, $E^{COMP}_{co}$:
% Proposition avec la définition dans ATLAS:
\begin{equation}
    \sum_{o_{co} \in \{o_1, ... ,o_n\}} q_{o_{co}} * \frac{\Delta t}{60} \leq E^{COMP}_{co}
\end{equation}

\textbf{Identical Volume:}

$\forall co \in CO^{IV}, o_{co} \in \{o_1, ... ,o_n\} :$

\begin{equation}
    q_{o_1} = q_{o_2} = \cdots = q_{o_n}
\end{equation}

\textbf{Identical Ratio}:

$\forall co \in CO^{IR}, o_{co} \in \{o_1, ... ,o_n\} :$

\begin{equation}
    \frac{q_{o_1} - q^{min}_{o_1}}{q^{max}_{o_1} - q^{min}_{o_1}} = \frac{q_{o_2} - q^{min}_{o_2}}{q^{max}_{o_2} - q^{min}_{o_2}} = \cdots = \frac{q_{o_n} - q^{min}_{o_n}}{q^{max}_{o_n} - q^{min}_{o_n}}
\end{equation}

\textbf{Parent Child:}

$\forall co \in CO^{PC}, o^{child}_{co} \in \{o^{child}_1, ... ,o^{child}_n\} :$
\begin{equation}
    \delta_{o^{child}_{co}} \leq \delta_{o^{parent}_{co}}
\end{equation}
   
\subsubsection{Cross-border constraints} \label{subsubsec:CrossborderConstraints}
Here, the model supports two options: the Available Transfer Capacity (ATC) mechanism and the Flow-based (FB) mechanism. ATC constraints are ensured across each direction of border, while Flow-based constrains the capacity across 'critical branches' of the network.\footnote{Critical branches represent all lines that are "heavily impacted" by cross-border exchanges.} 

The ATC constraints take the upstream and downstream market flows and constrain them separately by the minimum and maximum NTC limits, respectively. This is shown in Equation \ref{eqn:ATC_mc}.

$\forall t\in T, mb_{l}\in MB : $
\begin{equation}
    NTC_{l,t}^{min} \leq \Delta q_{l,t}^{mb} \leq NTC_{l,t}^{max} \label{eqn:ATC_mc}
\end{equation}

To model current losses for zones connected by DC lines, we add in the following constraints :

$\forall t\in T, mb_{l}\in MB^L : $
\begin{gather}
    \Delta q^{mb}_{l,t} = \frac{1}{2}\left(\Delta q^{mb,exp}_{l,t} + \Delta q^{mb,imp}_{l,t}\right) \label{eqn:losses1_mc}\\
    \Delta q^{mb,imp}_{l,t} = \left((1-a_l)\nu_{l,t} + \frac{1-\nu_{l,t}}{1-a_l}\right)\Delta q^{mb,exp}_{l,t}  \label{eqn:losses2_mc} \\
    \Delta q^{mb,exp}_{l,t} \left(\nu_{l,t}-\frac{1}{2}\right) \leq 0 \label{eqn:losses3_mc}
\end{gather}

Since both $\nu_{l,t}$ and $\Delta q^{mb,exp}_{l,t}$ are variables, we can see that these equations are non-linear. They are thus linearized using two (for each market border and time step) auxiliary variables: $\xi_{l,t}$ and $\xi_{l,t}^{aux}$. We can then rewrite these equations in the following manner: 

$\forall t\in T, mb_{l}\in MB^L : $
\begin{gather}
    \Delta q^{mb,imp}_{l,t} = \left((1-a_l) - \frac{1}{1-a_l}\right)\xi_{l,t}  + \frac{\Delta q^{mb,exp}_{l,t}}{1-a_l}  \label{eqn:lossesLin2_mc} \\
    \xi_{l,t} - \frac{\Delta q^{mb,exp}_{l,t}}{2} \geq 0 \\
    \xi_{l,t} = \Delta q^{mb,exp}_{l,t} - \xi_{l,t}^{aux} \\
    \nu_{l,t}\Delta q^{mb,min}_{l,t} \leq \xi_{l,t} \leq \nu_{l,t}\Delta q^{mb,max}_{l,t} \\
    (1-\nu_{l,t})\Delta q^{mb,min}_{l,t} \leq \xi^{aux}_{l,t} \leq (1-\nu_{l,t})\Delta q^{mb,max}_{l,t} \label{eqn:lossesLin5_mc}
\end{gather}

The flow-based constraints are shown in Equation \ref{eqn:energyFB}.

$\forall t\in T, cb_{k}\in CB : $
\begin{equation}
    q_{k,t}^{CB,ref} + \sum_{z\in Z}D_{t,k,z}^{PT}\left(\Delta q_{z,t}^{bal} - \Delta q_{z,t}^{bal,ref}\right)\leq q_{k,t}^{CB,max}-q_{k,t}^{CB,FRM} \label{eqn:energyFB}
\end{equation}
This equation includes a reference flow for each critical branch and each zone--denoted by $q_{k,t}^{CB,ref}$ and $\Delta q_{z,t}^{bal,ref}$ respectively--as well as the flow reliability margin for each critical branch, $q_{k,t}^{CB,FRM}$. The term, $D_{t,k,z}^{PT}$, represents the PTDF value for each zone, critical branch and time step. It can be positive or negative depending on the direction of the critical branch flow that it is representing. This equation limits the sum of the impacts of each zone's import-export balance, $\Delta q_{z,t}^{bal}$, to the maximum allowed on each critical branch, $q_{k,t}^{CB,max}$.

\subsection{Exchange Fixing}\label{subsec:ExchangeFixing}

The outputs of the optimal solution of the clearing module then move into the second phase of the algorithm: the Exchange Fixing module. Since the clearing module concentrates on associating sellers and buyers, it may not accurately represent the physical flows through the network. The exchange fixing is thus in place to better model these real physical flows. It does this by minimizing the exchanges through the market borders while retaining the solution found by the clearing module. The objective function is laid out in Equation \ref{eqn:exchangefixingobjfxn}:
% \begin{equation}
%     \min \sum_{mb_{l}\in MB,t\in T}\left(\kappa^e \left\lvert\Delta q_{l,t}^{mb,e}\right\rvert+ \kappa^r\left(q_{US_{l,t}}^{mb,r_{u}} +  q_{US_{l,t}}^{mb,r_{d}} + q_{DS_{l,t}}^{mb,r_{u}} + q_{DS_{l,t}}^{mb,r_{d}}\right)\right)
%     \label{eqn:exchangefixingobjfxn}
% \end{equation}

\begin{equation}
    \min \sum_{mb_{l}\in MB_{z} - MB_{z}^L} \lvert\Delta q_{l,t}^{mb}\rvert + \sum_{mb_{l}\in MB_{z}^L} \lvert\frac{\Delta q_{l,t}^{mb,exp} + \Delta q_{l,t}^{mb,imp}}{2}\rvert \label{eqn:exchangefixingobjfxn} 
\end{equation}

\begin{wrapfigure}[24]{r}{7cm}
    % \centering
	\includegraphics[width = 6.5cm]{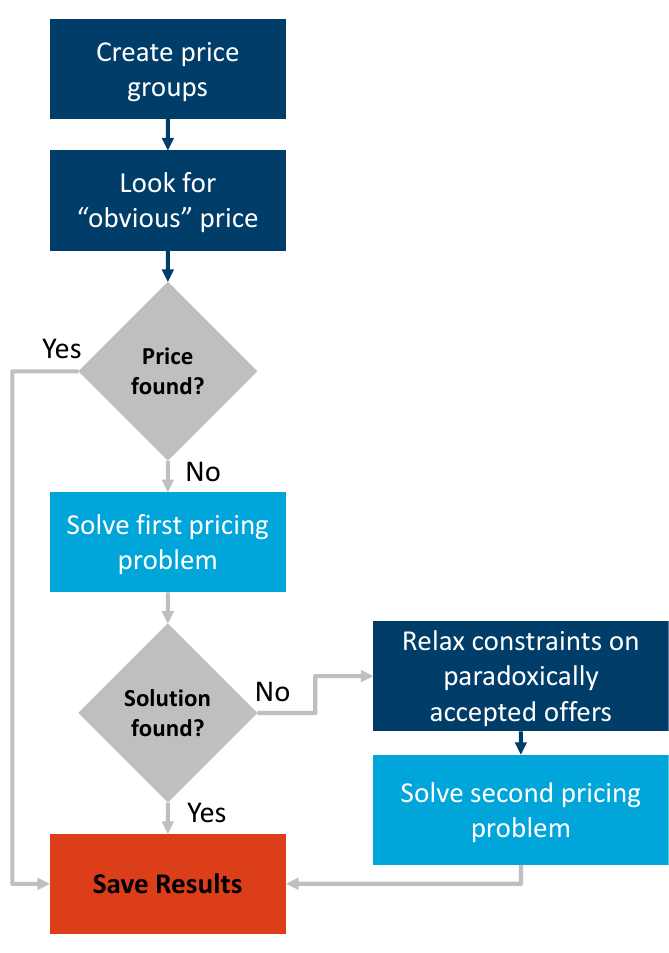}
	\caption[ATLAS Pricing Algorithm]{ATLAS Pricing Algorithm} %\footnotemark{}}
	% \vspace{-30pt}
	\label{figure:pricingAlgo_ATLAS}
\end{wrapfigure}

The constraints for the exchange fixing are similar to those of the clearing. The balance constraint is identical to Equation \ref{eqn:balance2_losses_mc}, although $\Delta q_{z,t}^{bal}$ is now a constant rather than a variable. Then, if ATC mode is active, the cross-border constraints are as shown in Equations \ref{eqn:ATC_mc}, \ref{eqn:losses1_mc} and \ref{eqn:lossesLin2_mc}-\ref{eqn:lossesLin5_mc}. If, on the other hand the flow-based mode is active, this step is not strictly necessary. It is maintained however in order to extract the corrected market border exchanges. This is useful for instance if a subsequent market will use ATC constraints. 

\subsection{Pricing}\label{subsec:Pricing}

In the price fixing module of the algorithm, the market clearing prices are set for for each zone and each time step of the simulation. The algorithm used in the ATLAS model is based on the dual of the clearing module.\footnote{Note: One important thing to note is that the duality theory is only applicable to linear problems. This means that the introduction of integer variables (such as the binary term, $\delta_{o}$) in the primal problem negates the relationship between the two problems. However, we overcame this issue by finding the dual of a slight variant of the primal problem. This is possible for one reason: the result of the primal would be identical if we removed the orders that are not cleared (i.e. the offers where $\delta_{o} = 0$). By ignoring these orders, and treating $\delta_{o}$ as a constant equal to 1, we now have a linear problem that adheres to duality theory. In this manner, the relationships and interpretations between the primal and the dual problem (the clearing and the pricing modules) remain somewhat valid, although it does raise some issues. Many sources discuss this problem and will not all be cited here. One reference specific to the European market clearing problem can be found in Annex B of \cite{nemo_committee_euphemia_2020} for clarification on this issue.} 

The ATLAS pricing algorithm is shown in Figure \ref{figure:pricingAlgo_ATLAS}. At this point, the accepted and rejected offers have already been determined. The goal of the price fixing problem is to fix the price for each zone that respects certain rules. This phase is necessarily a compromise between the accepted market players. The breakdown of profits between market actors depends on this phase. Unlike several operational pricing algorithms (see \cite{entsoe2023terre}, \cite{entsoe2023mari}, or \cite{nemo_committee_euphemia_2020} for instance), there is no iteration between the clearing problem and the pricing problem. This means that introducing a hard constraint on paradoxically accepted offers (as done in many market clearing algorithms, again, see \cite{entsoe2023terre} or \cite{nemo_committee_euphemia_2020}) can render the problem infeasible. The strategy used in ATLAS is therefore to attempt to solve a first optimization problem with a hard constraint on paradoxically accepted orders. Then, if there is no feasible solution, to relax the constraints and add them to the objective function. In this way, we minimize the paradoxically accepted orders. Any necessary make-whole payments can then be calculated ex-post. 

% Prior to solving the first pricing problem, several heuristics are used to attempt to find the optimal market price quickly. These--as well as the details of each optimization problem--are described in the next sections. 

\subsubsection{Price Groups}\label{subsubsec:PriceGroups} %\vspace{5pt}  \\
\begin{wrapfigure}[11]{r}{0.27\textwidth}%[h]
    \vspace{-20pt}
    \centering
    \includegraphics[width=\textwidth]{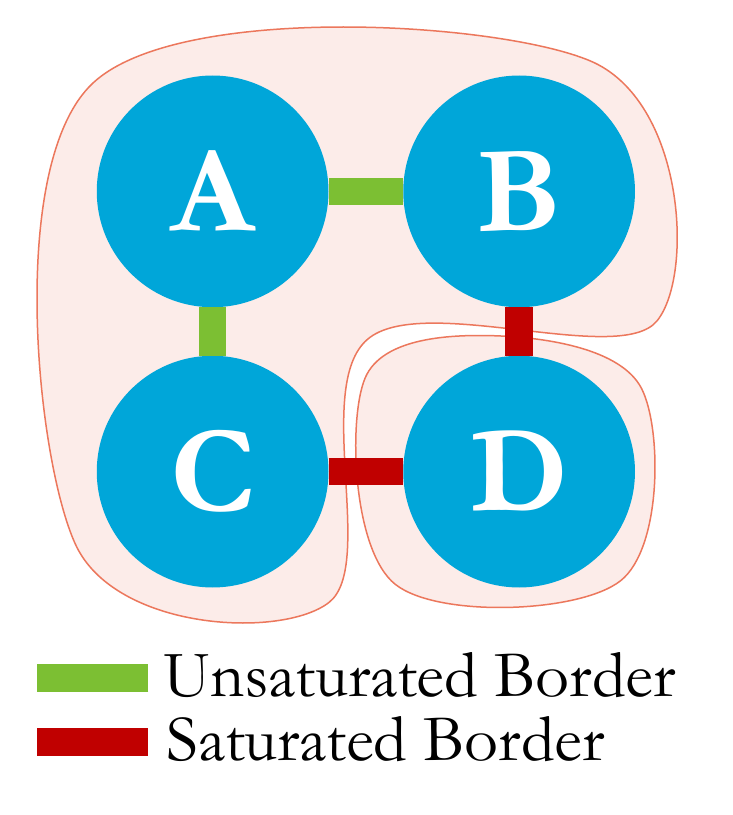}
    \vspace{-10pt}
    \caption{ATC Price Group Demonstration}
    \label{fig:pricegroups_atc}
    \vspace{-10pt}
\end{wrapfigure}
From a combination of the pricing constraints and the complementary slackness conditions, an interesting relationship between the market clearing prices of different zones emerges. For ATC, this creates a link between all zones connected by unsaturated borders; they all have the same market clearing price. This is demonstrated in Figure \ref{fig:pricegroups_atc}. The groups connected by unsaturated market borders end up with the same market clearing price while zone D has its own price. An unsaturated border allows for free trade, whereas a saturated border blocks trade, thus leading to a price delta--and corresponding congestion rent--between zones. The first of the pricing algorithm when using ATC mode is thus to calculate these price groups.

The search for price groups is quite straightforward. The algorithm loops over the zones and applies the following steps:

\begin{enumerate}
    \item If the current zone $z$ does not belong to any price group, create a new one. 
    \item Loop over all unsaturated borders of $z$ (borders for which Equation \ref{eqn:ATC_mc} holds strictly. Then:
    \begin{enumerate}
        \item Place the neighbor zone $z'$ in the same price group
        \item Apply step 2 to $z'$ (recursion)
    \end{enumerate}
    \item Iterate to the next zone
\end{enumerate}

For flow-based, this relationship expresses itself a bit differently. Equation \ref{eqn:energyFB} shows the flow-based constraint. It can be seen that each critical branch constraint takes into account the imports and exports from each zone. This essentially translates into a relationship, not just between the nearest critical branches, but all the critical branches at the same time. This means that a single saturated branch in the system can indicate a different market price for every zone.\footnote{This property has been discussed at length with regard to nodal markets. See for instance \cite{wu_folk_1996}.} For flow-based mode, the previous price group search is therefore not performed. Each zone remains in its own price group. 

\subsubsection{Obvious Price Search} \label{subsubsec:ObviousPrice}
If there is an order over a single time step that has been partially accepted ($q_o < q_o^{max}$), then the price of this price group can be immediately determined as the order price, without entering into an optimization problem. This search looks specifically at orders fulfilling the following requirements: 
\begin{enumerate}
    \item $d_o \leq \Delta t$: The duration of the order is not longer than the current market time step.
    \item $q_o^{min} < q_o < q_o^{max}$: The order has been marginally accepted.
    \item $o \notin co, \forall co \in CO^{IV} \cup CO^{IR} \cup CO^{C}$: The order is not part of an Identical Volume, Identical Ratio or Complement coupling.
\end{enumerate}

Then, in most cases, the obvious price for the price group is equal to the price of the order. If, however, the order is part of a Parent Child coupling and has a price strictly less than the parent order in the coupling, then the market clearing price is set to the average of the other orders in the coupling that have been accepted, as shown in Algorithm \ref{alg:obvPrice}.

\begin{figure}[ht]
\centering
\begin{minipage}{.45\linewidth}
\begin{algorithm}[H]
\caption{Obvious Price Calculation}\label{alg:obvPrice}
\begin{algorithmic}
\If{$o \in co, co \in CO^{PC}$}
    \If{$ p_o \geq p_{o^{parent}_{co}}$}
        $p^{obvious}_g = p_o$\;
    \Else
        % $p^{obvious}_g=\frac{\sum_{o'\in co,q_{o'}>0} p_{o'}}{\sum_{o'\in co,q_{o'}>0} q_{o'}}$
        \For{$o' \in co$}
            \If{$q_{o'} > 0$}
                $p^{*}_{co} = p_{o'} + p^{*}_{co}$\
                $q^{*}_{co} = q_{o'} + q^{*}_{co}$\
            \EndIf
        \EndFor
        $p^{obvious}_g = \frac{p^{*}_{co}}{q^{*}_{co}}$\
    \EndIf
\EndIf
\State $p^{obvious}_g = p_o$\
\end{algorithmic}

\end{algorithm}
\end{minipage}
\end{figure}

\subsubsection{First Pricing Problem} \label{subsubsec:FirstPricing}
If no obvious price is found, the algorithm passes to the first pricing problem. The objective function is\footnote{This objective function was determined during the Optimate project. Currently a study is underway comparing different pricing algorithms and objective functions to assess their impacts on market outcomes. This therefore has the potential to change in the near future.}: 
\begin{equation}
    \sum_{\substack{t \in T \\ (g_1,g_2)\in G_t \\ (g_1,g_2) neighbors}} \lvert p_{g_2t} - p_{g_1t} \rvert + \alpha \sum_{\substack{t \in T \\ g \in G_t}} p_{gt} + \beta \sum_{\substack{t \in T \\ g \in G_t}} \lvert p_{gt} \rvert
    \label{eqn:firstPricing_objfxn}
\end{equation}
where $\alpha$ and $\beta$ are user-defined parameters that set the focus either on price convergence or lowering market prices. The prices are bound by the following constraints: 

$\forall t \in T, g\in G_t $ :
\begin{equation}
    p^{min}_{gt} \leq p_{gt} \leq p^{max}_{gt}
\end{equation}
where 
\begin{align}
    p^{min}_{gt} &\coloneq \max_{z\in Z_g}\left(p^{min}_{z,t},\max_{\substack{o\in O^{'+}_{z,t} \\ \delta_o =1}}(p_o),  \max_{\substack{o\in O^{'-}_{z,t} \\ \delta_o =0}}(p_o) \right) \label{eqn:priceBoundmin} \\
    p^{max}_{gt} &\coloneq \min_{z\in Z_g}\left(p^{max}_{z,t},\min_{\substack{o\in O^{'-}_{z,t} \\ \delta_o =1}}(p_o),  \min_{\substack{o\in O^{'+}_{z,t} \\ \delta_o =0}}(p_o) \right) \label{eqn:priceBoundmax}
\end{align}
where $O^{'+}_{z,t}$ and $O^{'-}_{z,t}$ are subsets of purchase and sale (respectively) offers in the zone $z$ at time $t$. The subset excludes:

\begin{itemize}
    \item orders with a duration larger than the current time step
    \item fully indivisible orders
    \item partially indivisible orders that are rejected or accepted at their minimum power
    \item and orders belonging to Individual Volume, Individual Ratio, Complement or Parent Child couplings
\end{itemize}

The price bounds are then calculated as in Equations \ref{eqn:priceBoundmin} and \ref{eqn:priceBoundmax}, which look as follows when simplified:
\begin{align}    
    p^{min}_{gt} &\coloneq \max(\text{\textcolor{A3}{Max. Accepted Sale Price}, \textcolor{C4}{Max. Rejected Purchase Price}}) \\
    p^{max}_{gt} &\coloneq \min(\text{\textcolor{C2}{Min. Rejected Sale Price}, \textcolor{D3}{Min. Accepted Purchase Price}}) 
\end{align}

\begin{figure}[ht!]
    \centering
    \includegraphics[width=0.6\textwidth]{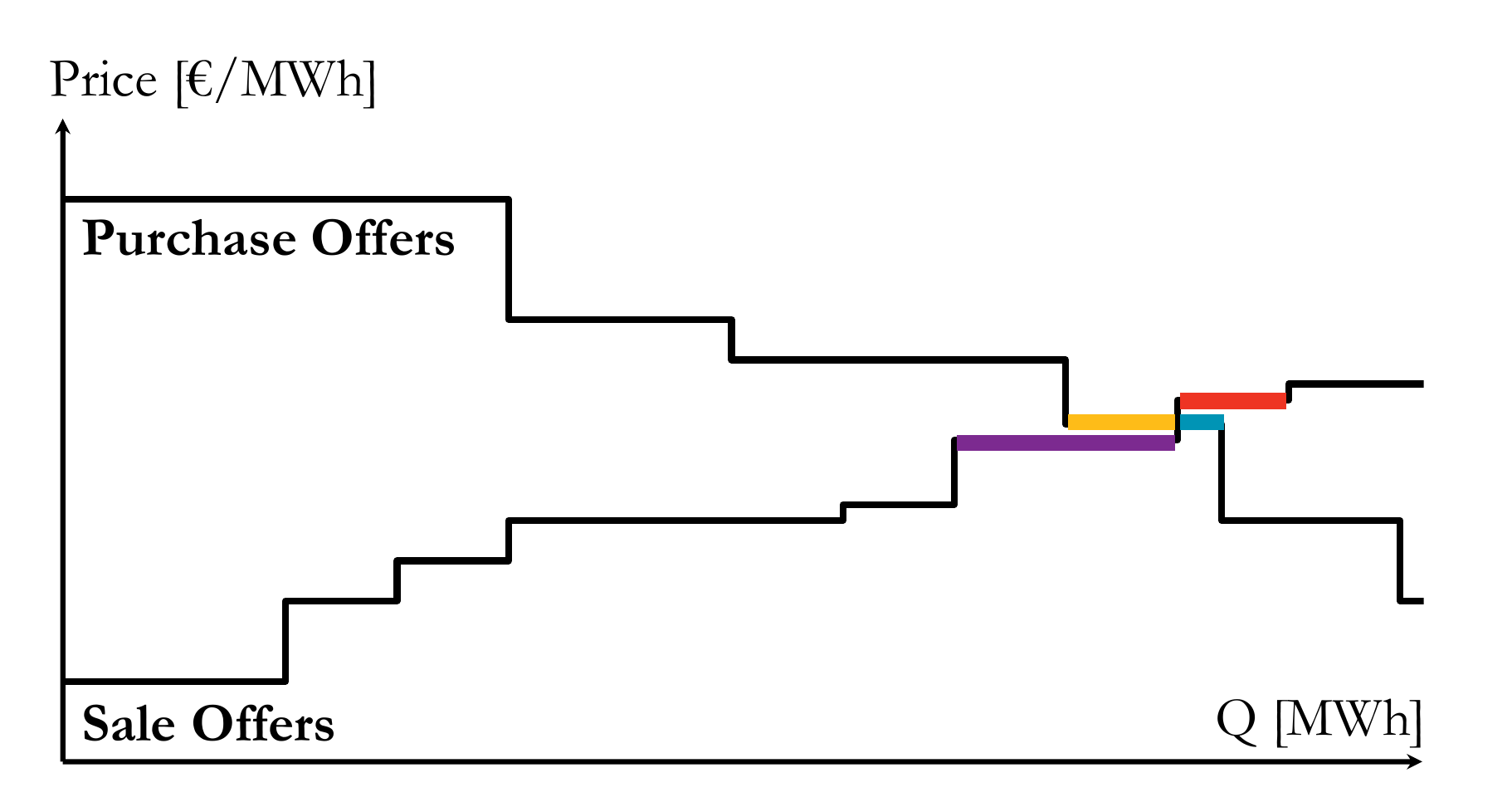}
    \caption{Upper and Lower Price Bounds}
    \label{fig:priceBounds}
\end{figure}

The colors here correspond to those offers shown in Figure \ref{fig:priceBounds}.

During this phase, no paradoxically accepted orders are allowed. 
% $\forall o \in O^{+}, d_o > \De$
% \begin{equation}
    
% \end{equation}

% \textcolor{red}{[XXX - Some things to verify. PAB constraint does not seem to be there for FB --> Run study case from before and see if it goes through first pricing or second (First means that there is no constraint for PAB in FB mode). JK, seems to be dealt with correctly in surplus variable definition. Question remains as to how to show the many difference between the ATC and FB problems...]}

\subsubsection{Second Pricing Problem} \label{subsubsec:SecondPricing}
If the first pricing problem fails, the algorithm enters a second optimization problem that is identical to the first except the constraints on the paradoxically accepted offers are relaxed. We therefore remove Equations \ref{eqn:mc_paroxicallyRej} from the problem and add the following to the objective function:

\begin{equation}
    M\times \left( \max \left( \sum_{o\in O^{+}} \delta_o \left(p_o - \frac{\Delta t}{d_o} \sum_{t \in [t_o^{start},t_o^{end}]} p_{g_{z_o}t} \right), 0 \right)  + \max \left( \sum_{o\in O^{-}} \delta_o \left(\frac{\Delta t}{d_o} \sum_{t \in [t_o^{start},t_o^{end}]} p_{g_{z_o}t} - p_o \right), 0 \right) \right)
\end{equation}

% \textcolor{red}{Duration??}

Then any necessary make-whole payments are stored and passed on to the next steps. 

\subsection{Marginal Fixing}\label{subsec:MarginalFixing}
\begin{figure}[h]
    \centering
    \vspace{-10pt}
    \includegraphics[width = 0.74\textwidth]{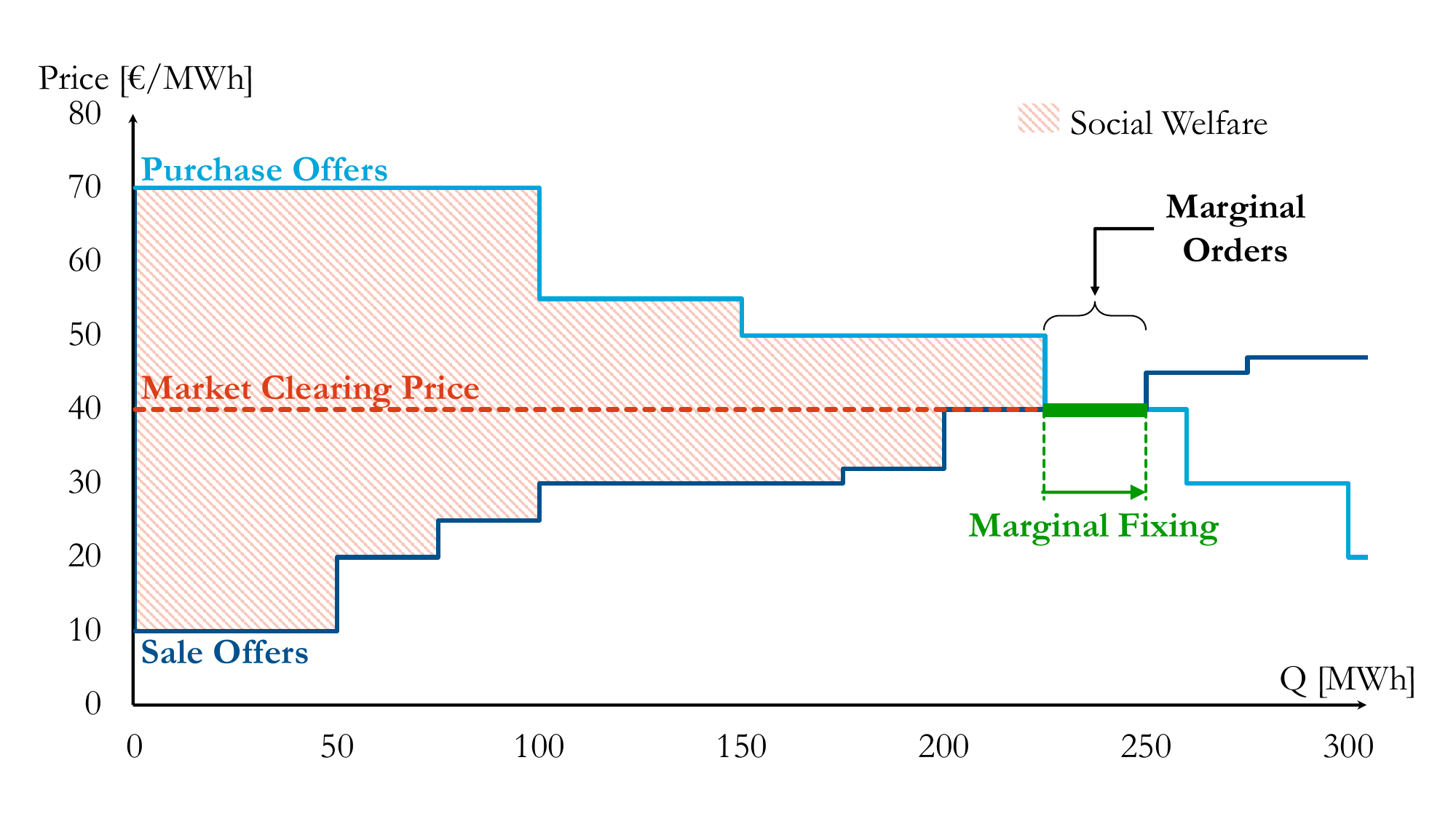}
    \caption[Marginal Fixing Example.]{Marginal Fixing Example}
    \label{fig:marginalfixing}
    \vspace{-15pt}
\end{figure}

The final subproblem of the market clearing algorithm is the Marginal Fixing. An offer is deemed 'marginal' if its offer price is the same as the market price. This module thus comes into play only in specific situations. Some of these marginal orders might not be fully accepted as the acceptance of these offers does not change the social welfare. The idea here is to maximize the number of offers accepted given the situation described above and shown in Figure \ref{fig:marginalfixing}. This step finds these marginal orders and accepts the available volume at the market price.

% \section{Outputs}

% \textcolor{red}{TO BE COMPLETED}

% \section{Platform and Solver}\label{sec:PlatformSolver}

% \subsection{PROMETHEUS}\label{subsec:Prometheus}
% The model was developed using an online platform known as PROMETHEUS. PROMETHEUS is an integrated platform developed at RTE for simulating energy systems. It allows for flexible modeling across any time frame and uses a variety of visualization tools to facilitate efficient analysis. It enables the user to create multiple prototypes and processes and link them together to simulate a variety of market designs. \autoref{fig:examplewf} shows a workflow that was used in this study and that will be explained in further detail in \autoref{chapter:MarketStudies}.

% \begin{figure}[h]
%     \centering
%     \includegraphics[width=\textwidth]{images/exampleWF.pdf}
%     \caption{PROMETHEUS Workflow for Sequential Market Design (No Pre-cooptimization)}
%     \label{fig:examplewf}
% \end{figure}

% Each circle in \autoref{fig:examplewf} represents a packet of data with all or some of the data described in \autoref{app:DataSchema}. The data is carried through a variety of different prototypes represented by the square boxes. Each workflow can also integrate different conditions and loops to follow a variety of paths and studies.

% The model was run using one of two solvers integrated with the PROMETHEUS Platform:
% \begin{enumerate}
% 	\item GLPK (GNU Linear Programming Kit)\footnote{For more information on GLPK: https://www.gnu.org/software/glpk/\#documentation}
% 	\item PNE: an in-house solver developed by RTE. 
% \end{enumerate}

\chapter{Example Studies} \label{ch:ExampleStudies}

\section{A Brief History of the Model} \label{sec:history}

The ATLAS model is a direct descendant of the OPTIMATE model, developed in a European project in the early 2010s (https://cordis.europa.eu/project/id/239456). 

\textcolor{red}{TO BE COMPLETED}
\chapter{Conclusion and Further Work} \label{ch:Conclusion}
\textcolor{red}{TO BE COMPLETED}

\appendix

\chapter{EUPHEMIA Algorithm}\label{app_EUPHEMIA}
\begin{figure}[ht!]
    \centering
    \includegraphics[width=\textwidth]{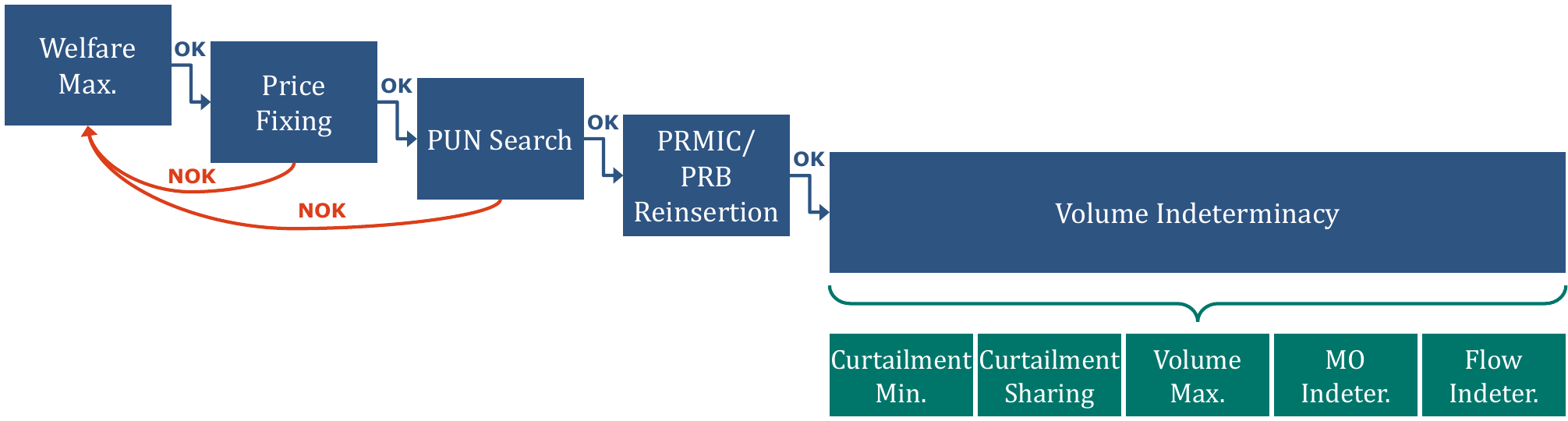}
    \caption{EUPHEMIA Algorithm}
    \label{fig:EuphemiaAlgo}
\end{figure}

EUPHEMIA uses a branch-and-cut approach to solve a mixed integer quadratic program (MIQP). The program itself--shown in Figure \ref{fig:EuphemiaAlgo}--is a single master problem (market welfare maximization, see Section \ref{subsubsec:4_welfareMax}) with several sub-problems, two of which serve to generate the relevant cutting planes. The initial point is the solution to the welfare maximization with integer variables relaxed. Then, the problem branches by analyzing the 0/1 values of binary variables linked to violated constraints.  

We will not cover the PUN search in this section as it is not modeled in the ATLAS project and requires quite a bit of explanation.\footnote{Note that these offers are specific to Italy.} 

\section{Offer Types} \label{subsubsec:4_offerTypes}
In order to represent the complex technical constraints of various power plants in the form of bids, there are several types of offers permitted in EUPHEMIA, some of which render the problem much less well-behaved than theoretical models. 

The simplest orders are either linear or stepwise, all of which are ultimately aggregated into a single curve, either following Figure \ref{fig:offerTypes_stepCurve} or Figure \ref{fig:offerTypes_linearCurve}.

\begin{figure}[ht!]
    \centering
    \begin{subfigure}{0.45\linewidth}
        \centering
        \includegraphics[width=\linewidth]{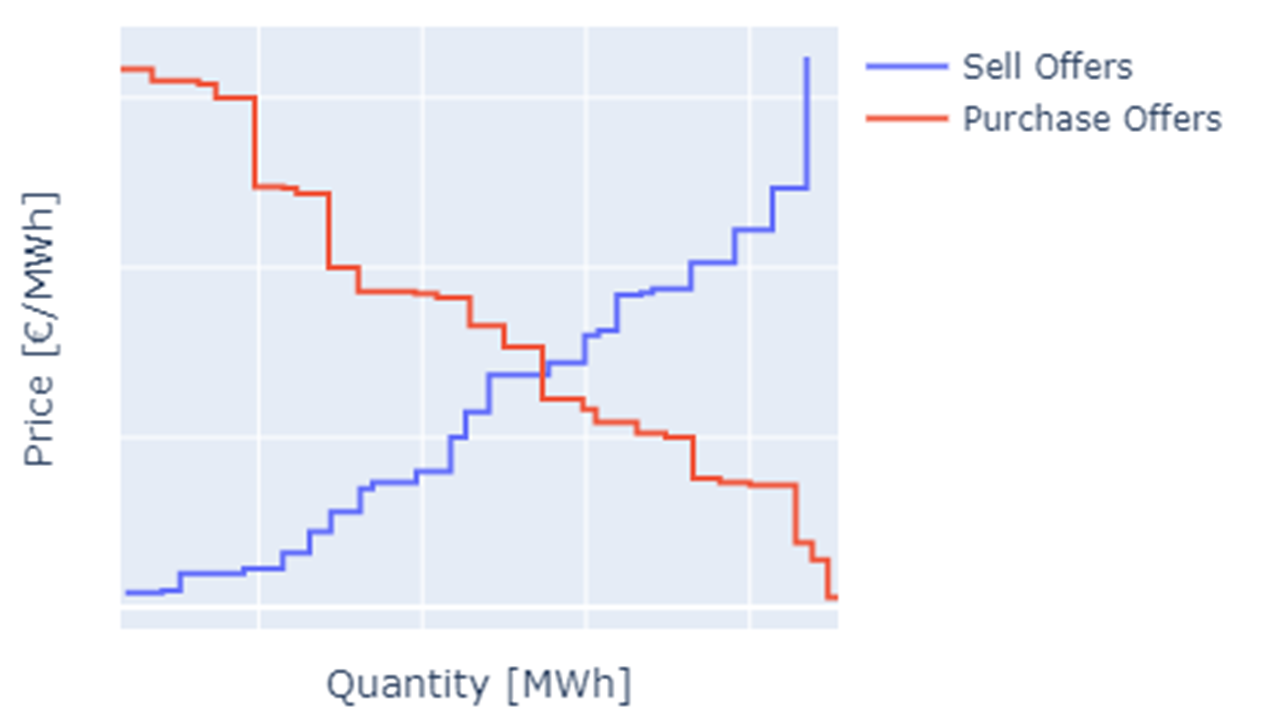}
        \caption{Step Curve} \label{fig:offerTypes_stepCurve}
    \end{subfigure}
    \begin{subfigure}{0.45\linewidth}
        \centering
        \includegraphics[width=\linewidth]{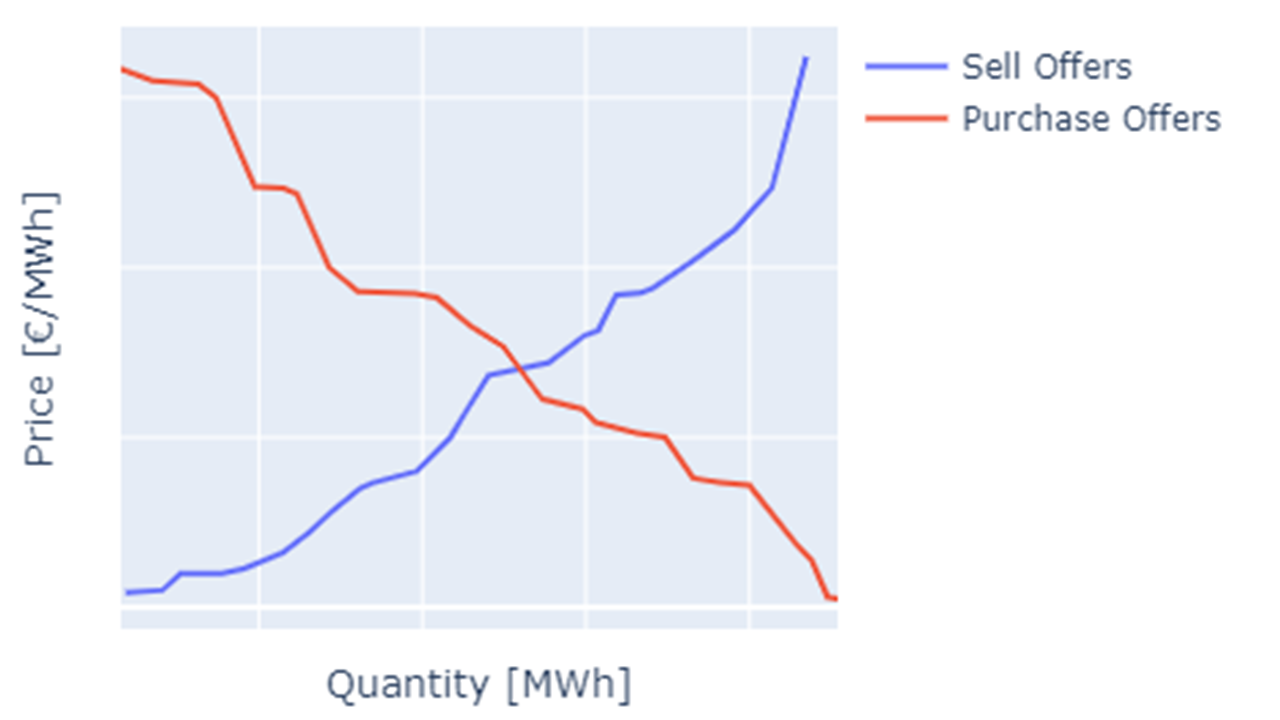}
        \caption{Linear Curve} \label{fig:offerTypes_linearCurve}
    \end{subfigure}
    \caption{Offer Curve Types}
    \label{fig:offerTypes}
\end{figure}
\begin{figure}[ht!]
    \centering
    \begin{subfigure}{0.45\linewidth}
        \centering
        \includegraphics[width=\linewidth]{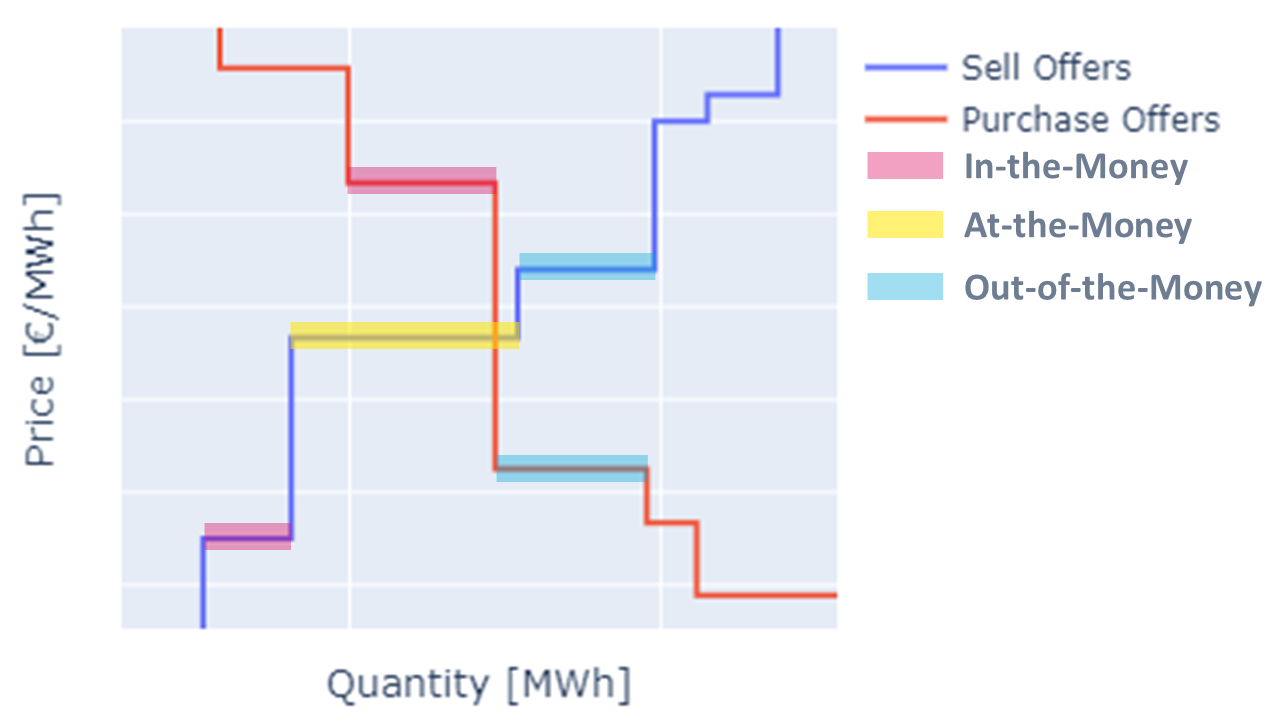}
        \caption{Step Curve} \label{fig:offerTypes_stepCurve_zoomed}
    \end{subfigure}
    \begin{subfigure}{0.45\linewidth}
        \centering
        \includegraphics[width=\linewidth]{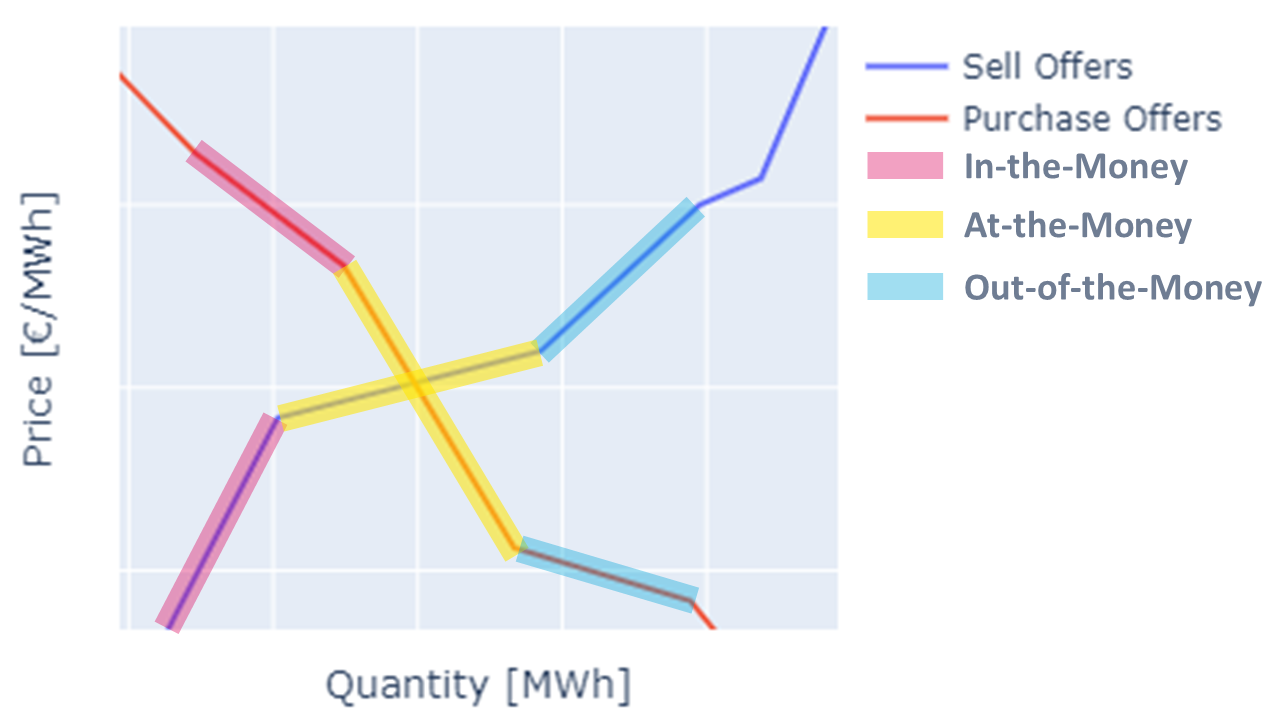}
        \caption{Linear Curve} \label{fig:offerTypes_linearCurve_zoomed}
    \end{subfigure}
    \caption{Offer Acceptance Status}
    \label{fig:offerTypes_zoomed}
\end{figure}

In reality, the full aggregated curves are a hybrid combination of these types of offers. Figure \ref{fig:offerTypes_zoomed} shows which offers should be accepted (assuming a single market zone).

Since the linear curve offers have a price that depends on the quantity, the resulting objective function will be quadratic. In order to somewhat simplify the problem resolution, for this work, we have ignored this offer type in the ATLAS model and in the studies tested in EUPHEMIA. 

Next we have two sets of more complicated order types: 
\begin{enumerate}
    \item Complex Orders 
    \item Block Orders 
\end{enumerate}

\setlength\leftskip{\paragraphmargin}

\textbf{Complex Orders} are sets of simply hourly piece-wise orders that have an additional constraint applying to the overall set. Currently there are three types of additional constraints permitted: 1) Minimum Income constraints (MIC), 2)  Scheduled Stop constraints, and 3) Load Gradient constraints. The first is described as follows: 

\begin{displayquote}
    \emph{"Generally speaking, the Minimum Income economical constraint means that the amount of money collected by the order in all periods must cover its production costs, which is defined by a fixed term (representing the startup cost of a power plant) and a variable term multiplied by the total assigned energy (representing the operation cost per MWh of a power plant)."}\footnote{\citeay{nemo_committee_euphemia_2020}}
\end{displayquote}

This means that the revenue for the total set of offers must be sufficient to cover fixed and variable costs: 
\begin{equation}
    \sum_{o,t\in \mathscr{O}_{set}} \lambda_{t,z} q_o^{acc} \geq C_o^{fix} + C_o^{var}\sum_{o\in \mathscr{O}_{set}} q_o^{acc}
\end{equation}
where $\lambda_{t,z}$ is the market price for the zone and time of the offer and $q_o^{acc}$ is the accepted quantity of each offer in the MIC set, respectively. $\mathscr{O}_{set}$ represents the set of offers included in a single Minimum Income Condition. $C_o^{fix}$ and $C_o^{var}$ are the terms of the offer, representing a fixed (start-up) cost and a variable cost. While this type of offer seems quite helpful, it can also increase the risk of rejection. If a single offer price in the set is out-of-the-money, the whole set is rejected. However, this can be avoided in the MIC is combined with a Scheduled Stop constraint. This is generally used if a power plant is already running during the previous day. 

Finally, the Load Gradient constraint can be added to a set of offers. This means that the volume that can be accepted in a time, $t$, is contingent upon the volume accepted in previous time steps. The difference in volume accepted between time steps must respect the increment given, in either direction:
\begin{equation}
    Grad^{dec} \leq  q_{o,t}^{acc} - q_{o,t-1}^{acc} \leq Grad^{inc}
\end{equation}
\textbf{Block Orders} are unique for two reasons: they are defined over multiple time periods and they have a minimum acceptance ratio (MAR) (a $q_{min}$, in other words). Because of this, these offers are inherently associated with binary variables. The minimum acceptance ratio of these offers -- relative to their given maximum power -- can vary between 0.01 and 1 (commonly referred to as a fill-or-kill offer). Additionally, blocks can be \emph{linked}, \emph{flexible} or \emph{exclusive}. Linked offers are sets of parent and child offers whose acceptance adheres to the following principles: 
\begin{quote}
    \emph{ \vspace{-20pt}
    \begin{inparaenum}
            \item "The acceptance ratio of a parent block is greater than or equal to the highest acceptance ratio of its child blocks (acceptance ratio of a child block can be at most the lowest acceptance ratio among own parent blocks)
            \item (Possibly partial) acceptance of child blocks can allow the acceptance of the parent block when:
            \begin{inparaenum}
                \item the surplus of a family is non-negative
                \item leaf blocks (block order without child blocks) do not generate welfare loss
            \end{inparaenum}
            \item A parent block which is out-of-the-money can be accepted in case its accepted child blocks provide sufficient surplus to at least compensate the loss of the parent.
            \item A child block which is out-of-the-money cannot be accepted even if its accepted parent provides sufficient surplus to compensate the loss of the child, unless the child block is in turn parent of other blocks (in which case rule 3 applies)."\footnote{EUPHEMIA DOC CITATION}
        \end{inparaenum}}
\end{quote}
Essentially, a parent offer can be accepted even if it is out-of-the-money if the child offer(s) generate enough additional welfare, but not the other way around. A parent offer might represent a start-up cost, for instance. If the hours after the start-up occurs bring about enough surplus, the start-up offer can be accepted. However, the other offers can definitely not be accepted if the start-up offer is rejected. 

Next we have flexible offers, which are defined for a single time step (and have a minimum acceptance ratio of 1, i.e. $q_{o}^{min} = q_o^{max}$), but can be accepted at any point of the day. Exclusive offers are a slightly more complex version of these, where they can be defined across multiple time steps and the sum of the acceptance ratios must be lower than one: 
\begin{equation}
    \sum_{o \in \mathscr{O}_{block}} \frac{q_o^{acc}}{q_o^{max}} \leq 1
\end{equation}
If the offers within a certain exclusive block have their own minimum acceptance ratio equal to 1, it means that among the block, only a single offer can be accepted. 

\setlength{\leftskip}{0cm}

For both complex and block offers, if they are out-of-the-money, they must be rejected, even if it leads to a lower market welfare. A block or complex offer that is accepted during the market clearing despite being out-of-the-money is referred to as a \emph{paradoxically accepted} offer. We will discuss these a bit more in Appendix \ref{app_EUPHEMIA}. There are two additional offer types: Merit Orders and PUN Orders, which we will skip over for the moment as they are not directly relevant to the work presented in this chapter.  

As a last note, in order for the solution to be perfectly reproducible, EUPHEMIA has developed criteria to differentiate between identical orders when necessary, first by timestep, then randomly by hash.

% \setlength{\leftskip}{0cm}

% \paragraph{Welfare Maximization}\hspace{0pt} \label{para:4_welfareMax}

% \setlength\leftskip{\paragraphmargin}
\section{Welfare Maximization} \label{subsubsec:4_welfareMax}

The clearing, or Welfare Maximization, is the classic economic dispatch problem. It has the constraints described above for the different offer types, zonal balance constraints and either the flow-based or ATC flow constraints. The objective function maximizes the total market welfare--the sum of the profits of all market players. It gives as an output the quantity accepted of each offer and the net positions of each bidding zone.

The full objective function, as given by \citeay{nemo_committee_euphemia_2020} is:
%  <flush right> & <flush left> & <right> & <left> &...
\begin{equation*}
    \max f(x) \text{    where: } 
\end{equation*}
\begin{alignat}{7}
    &&f(x)=  -\sum_{o\in \mathscr{O}_{step}}& x_o p_o q_o \label{eq:objfxn_clearing_step} \\
    && &-\sum_{o \in \mathscr{O}_{lin}} x_o q_o (p_o^{min} + &x_o \frac{p^{max}_o - p^{min}_o}{2}) \label{eq:objfxn_clearing_lin}  \\
    && & &-\sum_{o\in \{\mathscr{O}_{BO},\mathscr{O}_{CO},\mathscr{O}_{MO}\}} &&x_o p_o q_o& \label{eq:objfxn_clearing_bo} \\
    && & & && -\sum_{k\in \mathscr{K}_t}&Tariff_k FLOW_k  \label{eq:objfxn_clearing_hvdc} \\
    && & & && & -M\sum_{o\in \mathscr{O}_{PTD}} \lvert q_o \rvert (1-x_o)^2  \label{eq:objfxn_clearing_curt}
    % & && && && &&\nonumber
\end{alignat}

where $x_o$ ($\in [0,1]$) is the amount of the offered volume, $q_o$, accepted. The negative sign comes from the fact that $q_o$ is defined as negative for purchase orders and positive for production offers. Parts \ref{eq:objfxn_clearing_step} and \ref{eq:objfxn_clearing_lin} refer to the market welfare from the step and linear offers, while part \ref{eq:objfxn_clearing_bo} refers to the same for block offers, complex offers and merit orders. We can see that due to part \ref{eq:objfxn_clearing_lin}, the objective function is quadratic. Part \ref{eq:objfxn_clearing_hvdc} is a reduction of market welfare that occurs for certain HVDC lines operated by external companies who implement tariffs relative to the flow on the line. Finally, part \ref{eq:objfxn_clearing_curt} is an additional term that is equivalent to minimizing curtailment in a unit commitment model. It regards $\mathscr{O}_{PTD}$, the set of price-taking demand offers that might be rejected. This term comes back into play later in the Volume Indeterminacy sub-problem. 

% \setlength{\leftskip}{0cm}

% \paragraph{Price Fixing}\hspace{0pt} \label{para:4_priceFixing}

% \setlength\leftskip{\paragraphmargin}

\section{Price Fixing} \label{subsubsec:4_priceFixing}

\begin{figure}[ht!]
    \centering
    \includegraphics[width=0.6\textwidth]{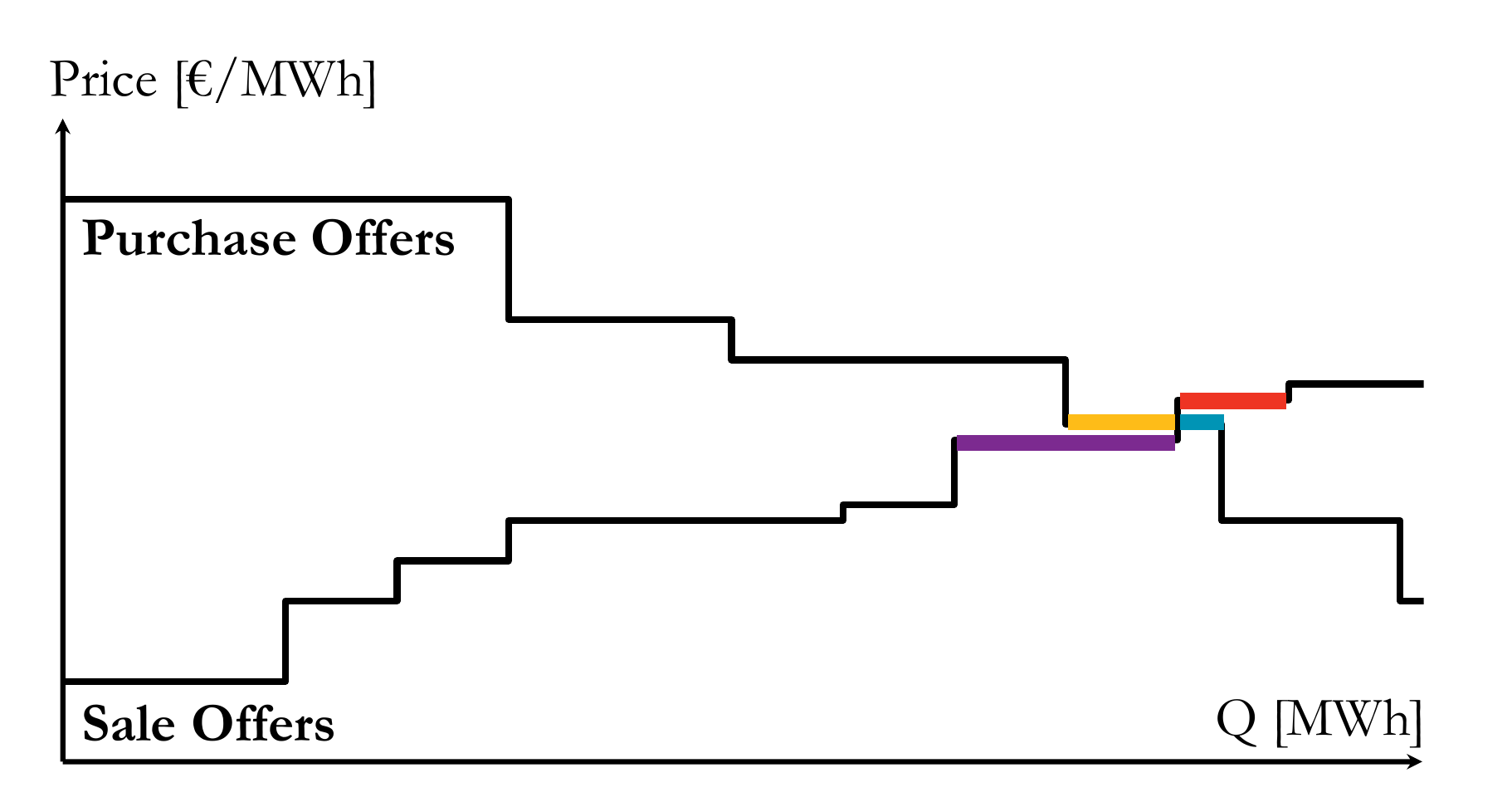}
    \caption{Upper and Lower Price Bounds}
    \label{fig:priceBounds_euphemia}
\end{figure}

In a perfect world, the pricing is the dual of the clearing problem, where the zonal prices are the dual variables associated with the zonal balance constraint (just as nodal prices would be the dual variable associated with the nodal balance constraint). However, certain additional constraints are added to the pricing phase, notably the concrete rejection of paradoxically accepted offers.\footnote{See Annex B of \citeay{nemo_committee_euphemia_2020} for clarification on the relation of the primal/dual problems.} The objective function is also modified to: 
\begin{equation}
    \min \sum_{\substack{z\in\mathscr{Z} \\ t\in \mathscr{T}}} \left(\lambda_{z,t} - \frac{p_{z,t}^{UB} + p_{z,t}^{LB}}{2}\right)^2
\end{equation}
where $p_{z,t}^{UB}$ and $p_{z,t}^{LB}$ are calculated as: 
\begin{align}
    p_{z,t}^{UB} &\coloneqq \min(\text{\textcolor{C2}{Min. Rejected Sale Price}, \textcolor{D3}{Min. Accepted Purchase Price}}) \\
    p_{z,t}^{LB} &\coloneqq \max(\text{\textcolor{A3}{Max. Accepted Sale Price}, \textcolor{C4}{Max. Rejected Purchase Price}})
\end{align}
The colors here correspond to those offers shown in Figure \ref{fig:priceBounds_euphemia}.

\begin{figure}[ht!]
    \centering
    \includegraphics[width=0.6\textwidth]{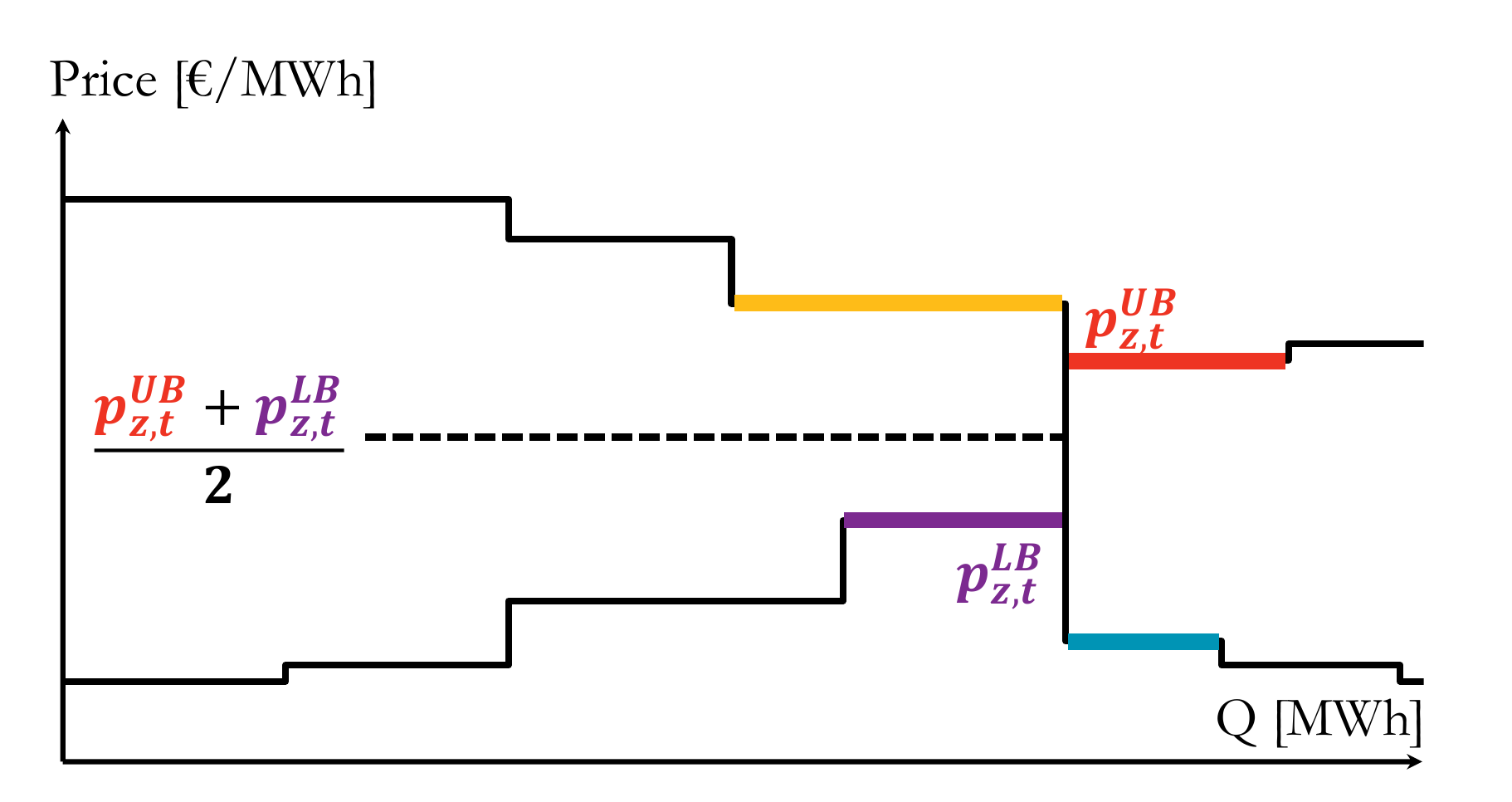}
    \caption{Reasoning for Pricing Objective Function}
    \label{fig:priceBounds_objfxn}
\end{figure}

In most cases, the objective function will go to zero, and is not necessary. However, it is crucial for points when the volume offered matches identically, as in Figure \ref{fig:priceBounds_objfxn}; otherwise, the price could theoretically be found at any point along the volume acceptance line. 

The price fixing tries to find a market clearing price for each zone that corresponds to the solution found in the welfare maximization stage AND that has no paradoxically accepted offers. If this is infeasible, the algorithm returns to the welfare maximization problem and introduces a cutting plane. The PUN search is similar--if the solution found in the first stage is not feasible with the additional constraints on PUN offers, the algorithm agains returns to the master problem. 

% \setlength{\leftskip}{0cm}

% \paragraph{PRMIC/PRB Reinsertion}\hspace{0pt}  \label{para:4_PRB Reinsertion}

% \setlength\leftskip{\paragraphmargin}
\section{PRMIC/PRB Reinsertion}  \label{subsubsec:4_PRB Reinsertion}

In this phase, EUPHEMIA takes the list of paradoxically rejected minimum income condition offers (PRMIC) and paradoxically rejected block offers (PRB) and iteratively tests them. Each offer is accepted, and the clearing and price fixing modules are rerun. If the market welfare is not reduced and the price fixing is feasible, the offer is accepted. Otherwise, it is rejected. 

% \setlength{\leftskip}{0cm}

% \paragraph{Volume Indeterminacy}\hspace{0pt}  \label{para:4_volIndeterminacy}

% \setlength\leftskip{\paragraphmargin}
\section{Volume Indeterminacy}  \label{subsubsec:4_volIndeterminacy}

The Volume Indeterminacy module aims to differentiate between solutions that have the same market welfare. It comprises five smaller optimization problems: 
\begin{enumerate}
    \item Curtailment Minimization
    \item Curtailment Sharing
    \item Volume Maximization
    \item Merit Order Indeterminacy
    \item Flow Indeterminacy
\end{enumerate}

\textbf{Curtailment Minimization} and \textbf{Curtailment Sharing} deal with times when the market clearing price of a zone is set at the maximum (or minimum) permitted price and the offers at this price are not fully accepted. This quantity is first minimized, then, if there remains some volume curtailed, it is shared as equally as possible amongst the bidding zones, while respecting network constraints.

\begin{figure}[ht!]
    \centering
    \vspace{-10pt}
    \includegraphics[width = 0.74\textwidth]{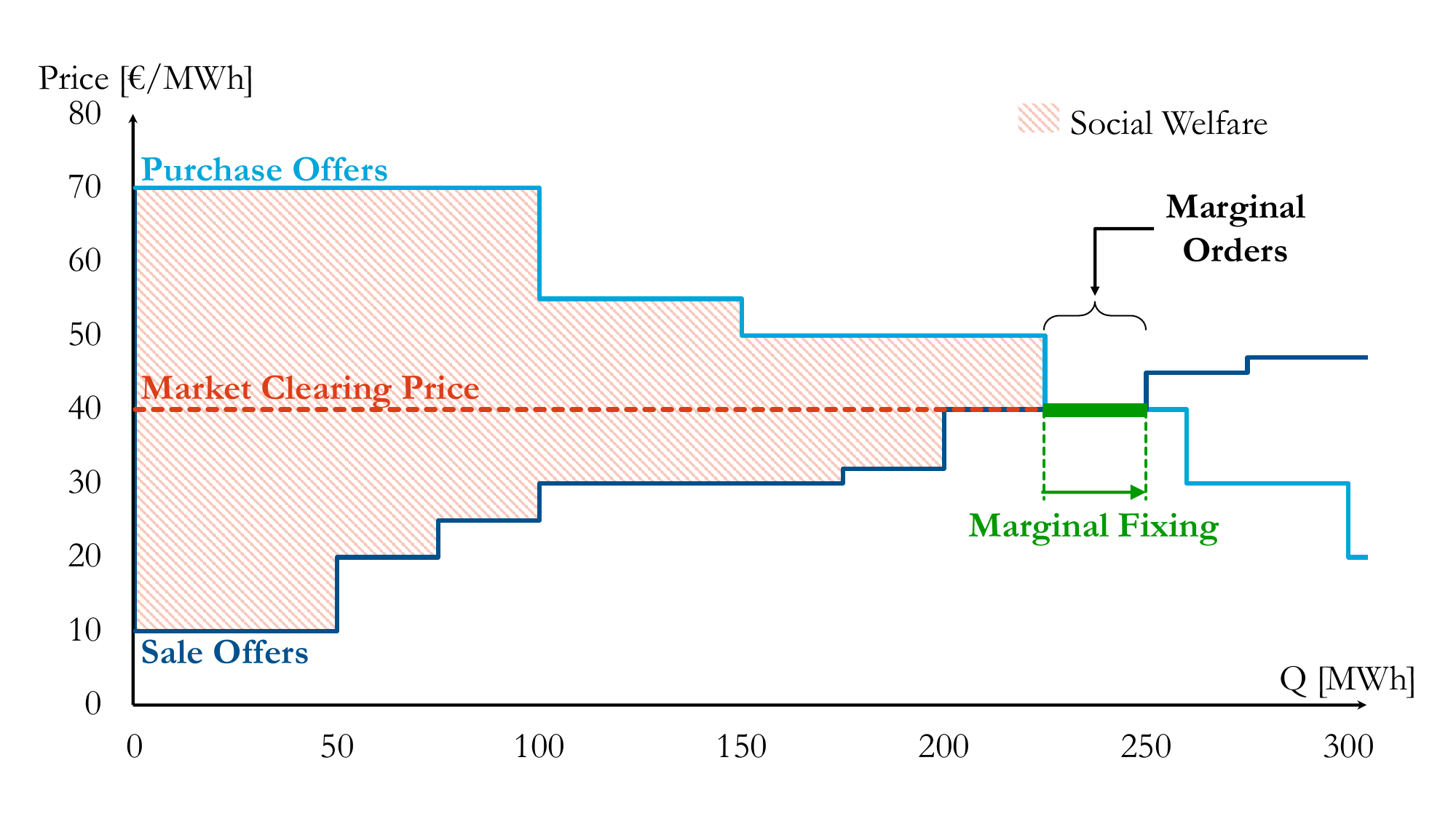}
    \caption[Marginal Fixing Example]{Marginal Fixing Example}
    \label{fig:marginalfixing_euphemia}
    \vspace{-15pt}
\end{figure}

The \textbf{Volume Maximization} problem is the next step. As shown in Figure \ref{fig:marginalfixing_euphemia}, some marginal offers might not be fully accepted as the acceptance of these offers does not change the market welfare. The idea here is to maximize the number of offers accepted given the previously-found solution. An example of a case where this is necessary is demonstrated in Figure \ref{fig:marginalfixing_euphemia}. This step finds these marginal orders and accepts the available volume at the market price.

\textbf{Merit Order Indeterminacy} deals with merit order numbers, which are out of the scope of this work. 

Finally, the \textbf{Flow Indeterminacy} stage occurs. Since the clearing module concentrates on associating sellers and buyers, it may not accurately represent the physical flows through the network. This phase is thus in place to better model these real physical flows. It does this by minimizing the exchanges through the market borders while retaining the solution found by the clearing module. The objective function is laid out in \ref{eq:exchangefixingobjfxn}:
\begin{equation}
    \min \sum_{\substack{mb_{l}\in MB \\ t_{i}\in T}}\alpha \left\lvert\Delta q_{li}^{mb,e}\right\rvert + \beta \left(\Delta q_{li}^{mb,e}\right)^2
    \label{eq:exchangefixingobjfxn}
\end{equation}
where $\Delta q_{li}^{mb,e}$ is the exchange on a market border, and $\alpha$ and $\beta$ are terms parametrizing the linear and quadratic terms. The constraints for this phase are essentially the same as the balance constraints and the cross-border constraints for the welfare maximization module, only now the computed local balances are parameters, rather than variables. 

\setlength{\leftskip}{0cm}

\chapter{Capacity Calculation}\label{app:CapCalc}

\textcolor{red}{TO BE COMPLETED}

\section{Acknowledgements}
\textcolor{red}{TO BE COMPLETED}

\bibliography{bibliography} %-->reference list is on the template.bib file

\end{document}